\documentclass[onecolumn]{aastex63}

\usepackage{array}
\usepackage{CJK}
\usepackage{amsmath}
\usepackage{multirow}
\usepackage{graphicx}  
\usepackage{float} 
\usepackage{subfigure} 
\usepackage{threeparttable}
\usepackage{captcont}
\usepackage{xcolor}
\usepackage{booktabs}
\usepackage{soul}
\usepackage{longtable}
\usepackage{appendix}
\usepackage{chngcntr}

\begin{document}

\newcommand*{\mkblue}[1]{\textbf{\color{blue}{#1}}}
\newcommand*{\LYJ}[1]{\textbf{\color{orange}{#1}}}

\newcommand{\ZJUT}{Institute for Theoretical Physics and Cosmology, Zhejiang University of Technology, Hangzhou, 310032, China.}

\newcommand{\PMO}{Key Laboratory of Dark Matter and Space Astronomy, Purple Mountain Observatory, Chinese Academy of Sciences, Nanjing 210033, People's Republic of China.}

\newcommand{\NJU}
{Department of Astronomy, Nanjing University, Nanjing 210023, People's Republic of China.}

\newcommand{\AEI}
{Max-Planck-Institut f{\"u}r Gravitationsphysik (Albert-Einstein-Institut), D-30167 Hannover, Germany}
\newcommand{\LUH}
{Leibniz Universit{\"a}t Hannover, D-30167 Hannover, Germany}

\newcommand{\UCGWP}
{United Center for Gravitational Wave Physics, Zhejiang University of Technology, Hangzhou, 310032, China}

%\title{The Net Spin sheds new light on Formation Channel of the Merging BBH Systems}
\title{The Heavier the Faster: A Sub-population of Heavy, Rapidly Spinning and Quickly Evolving Binary Black Holes}

\author{Wei-Hua Guo}
\affiliation{\ZJUT}
\affiliation{\UCGWP}

\author{Yin-Jie Li}
\affiliation{\PMO}
\author{Yuan-Zhu Wang}
\affiliation{\ZJUT}
\affiliation{\PMO}
\author{Yong Shao}
\affiliation{\NJU}
\author{Shichao Wu}
\affiliation{\AEI}
\affiliation{\LUH}
\author{Tao Zhu}
\affiliation{\ZJUT}
\affiliation{\UCGWP}
\author{Yi-Zhong Fan}
\affiliation{\PMO}

\correspondingauthor{Yin-Jie Li, Yuan-Zhu Wang and Yin-Zhong Fan}
\email{liyinjie@pmo.ac.cn, vamdrew@zjut.edu.cn, yzfan@pmo.ac.cn}

\date{\today}

\begin{abstract}
The spins of binary black holes (BBHs) measured from gravitational waves carry notable information of the formation pathways. Here we propose a quantity ``dimensionless net spin" ($\chi_{\rm N}$), which is related to the sum of angular momentum of component black holes in the system, to provide a novel perspective to study the origin(s) of BBHs. By performing hierarchical Bayesian inference on $\chi_{\rm N}$, we find strong evidence that the marginal distribution of this quantity can be better fitted by two Gaussian components rather than one: there is a narrow peak at $\chi_{\rm N} \sim 0.15$ and another extended peak at $\chi_{\rm N} \sim 0.47$. We also find that the rapidly spinning systems likely dominate the high-mass end of the population and they evolve with redshift much quicker. These findings bring new challenges to the field binary scenario, and suggest that dynamical process should plays a key role in forming high total mass BBHs.
\end{abstract}

\section{Introduction}\label{sec:intro}
Gravitational wave has become one of the most important messenger for studying the compact object in the Universe. More than 90 binary black hole (BBH) candidates have been released with the current catalogs of compact binary coalescences (GWTC-3) \citep{2023PhRvX..13d1039A}, and this number is expected to increase   several-fold after the 4th observing run (O4) of the LIGO/Virgo/KAGRA detector network \citep{2020LRR....23....3A}. The origins of these BBHs, however, remains a mystery. 

Researches on the observed events have revealed some broad features for the distribution of their measured properties, such as the BBH mass distribution has substructure beyond a smooth truncated power law \citep{2021ApJ...913...42W,2021ApJ...913L..19T,2021ApJ...917...33L,2021ApJ...922..258V,2022ApJ...924..101E}, the observed black hole spins are small \citep{2019PhRvD.100d3012W,2019MNRAS.484.4216R,2020ApJ...895..128M,2021PDU....3100791G,2021PhRvL.126q1103B} with some of their tilts misaligned to the orbital angular momentum \citep{2017PhRvD..96b3012T,2019ApJ...882L..24A,2023PhRvX..13a1048A,2024arXiv240409668L}, the mass-ratio distribution evolves with the primary mass \citep{2022ApJ...933L..14L}, and the merger rate evolution remains consistent with the star formation rate\citep{2023PhRvX..13a1048A}. Some other topics, like the existence of non-spinning sub-populations \citep{2021PhRvD.104h3010R,2021ApJ...921L..15G,2022ApJ...937L..13C,2022PhRvD.106j3019T,2023PhRvX..13a1048A} and the exact distribution shape for the tilt angle of component BHs remain subjects of ongoing debate \citep{2022A&A...668L...2V}. 

The statistics on the spins of BBHs plays an key role in identifying the formation pathways. Being the best measured spin parameter, the effective inspiral spin $\chi_{\rm eff}$ \citep{2011PhRvL.106x1101A,2010PhRvD..82f4016S}, which quantifies the mass-weighted average of the two component spins projected parallel to the binary’s orbital angular momentum, is widely studied among literature. \citet{2021ApJ...922L...5C} found an anti-correlation between $\chi_{\rm eff}$ and mass ratio ($q$). \citet{2020ApJ...894..129S} examined whether the effective spin distribution correlates with various mass parameters of the binary. \citet{2022ApJ...932L..19B} proposed that the $\chi_{\rm eff}$ distribution likely broadens with redshift. Yet these works have provided important clues, other spin parameters may carry additional useful information that can be extracted from current observations. Unfortunately, the effective precession spin $\chi_{\rm p}$, which describes the mass-weighted in-plane spin component that contributes to spin precession \citep{2014PhRvL.113o1101H,2015PhRvD..91b4043S}, is poorly measured for most events \citep{2023PhRvX..13d1039A,2024PhRvD.109b2001A,2021PhRvX..11b1053A,2019PhRvX...9c1040A}. Nevertheless, the ability to measure individual spins with the LIGO and Virgo detectors has also been limited \citep{2021PhRvL.126q1103B}. As shown in \citet{2021PhRvL.126q1103B}, for systems with mass ratios close to unity, the mass-sorted parameter estimation method (which is the method adopted to generate the publicity available posterior samples for hierarchical analysis) may yield misleading results, while for systems with significantly asymmetric mass ratios, only the spin for the primary BH can be well constrained. The choice of priors (e.g., some astrophysical priors motivated by simulations of stellar evolution \citep{2020ApJ...895L..28M,2021ApJ...922L..14M,2022ApJ...941..179Q}) also has great impact on the inference on individual spins. Additional works focusing on spin quantities other than $\chi_{\rm eff}$ and individual spins may bring new opportunities to understand the formation of BBHs. 

In our previous work, we found that the population of component BHs can be explained by the mixture of two sub-populations that have distinct mass and spin distributions, with one consist with the field binary scenario and the other consist with the dynamical formation scenario including different generations of mergers \citep{2022ApJ...941L..39W}. \cite{2023arXiv230302973L} further found a sub-population of higher-generation BHs using an semi-parametric methods. While these works focused on the properties of individual BHs, in this work we investigate the properties describing the whole BBH system: total mass, redshift, and the ``dimensionless net spin" ($\chi_{\rm N}$) which we will define in Sec.\ref{sec:chi_N}. We build simple models that can describe the net spin, and search for sub-populations in the joint $(M_{\rm tot}, q, \chi_{\rm N}, z)$ distribution. The rest of the paper is arrange as follows: in Sec.\ref{sec:chi_N} we propose the definition for the dimensionless net spin, and demonstrate the posterior distributions of $\chi_{\rm N}$ inferred from 69 high significance events; in Sec.\ref{sec:method} we introduce the details of hierarchical Bayesian inference; The analysis focusing on the marginal distribution of $\chi_{\rm N}$ is carried out in Sec.\ref{sec:spinonly}, and we study the mass dependency and redshift evolution of $\chi_{\rm N}$ in Sec.\ref{sec:joint}. We summarize the paper and give some discussions in Sec.\ref{sec:summary}.

\section{Dimensionless Net Spin From Component Black Holes}\label{sec:chi_N}

For a binary black hole system, the total angular momentum can be expressed as $\mathbf{J} = \mathbf{J}_{\rm orbit} + \mathbf{J}_1 + \mathbf{J}_2$, in which $\mathbf{J}_{\rm orbit}$ is the orbital angular momentum, $\mathbf{J}_1$ and $\mathbf{J}_2$ are the angular momentum for the primary and secondary black hole respectively. In this work, we focus on the net angular momentum from the two individual black holes: $\mathbf{J}_{\rm N} = \mathbf{J}_1 + \mathbf{J}_2$. This quantity is relevant to the spins and masses of the BHs.

\begin{figure*}
	\centering  
\includegraphics[width=0.48\linewidth]{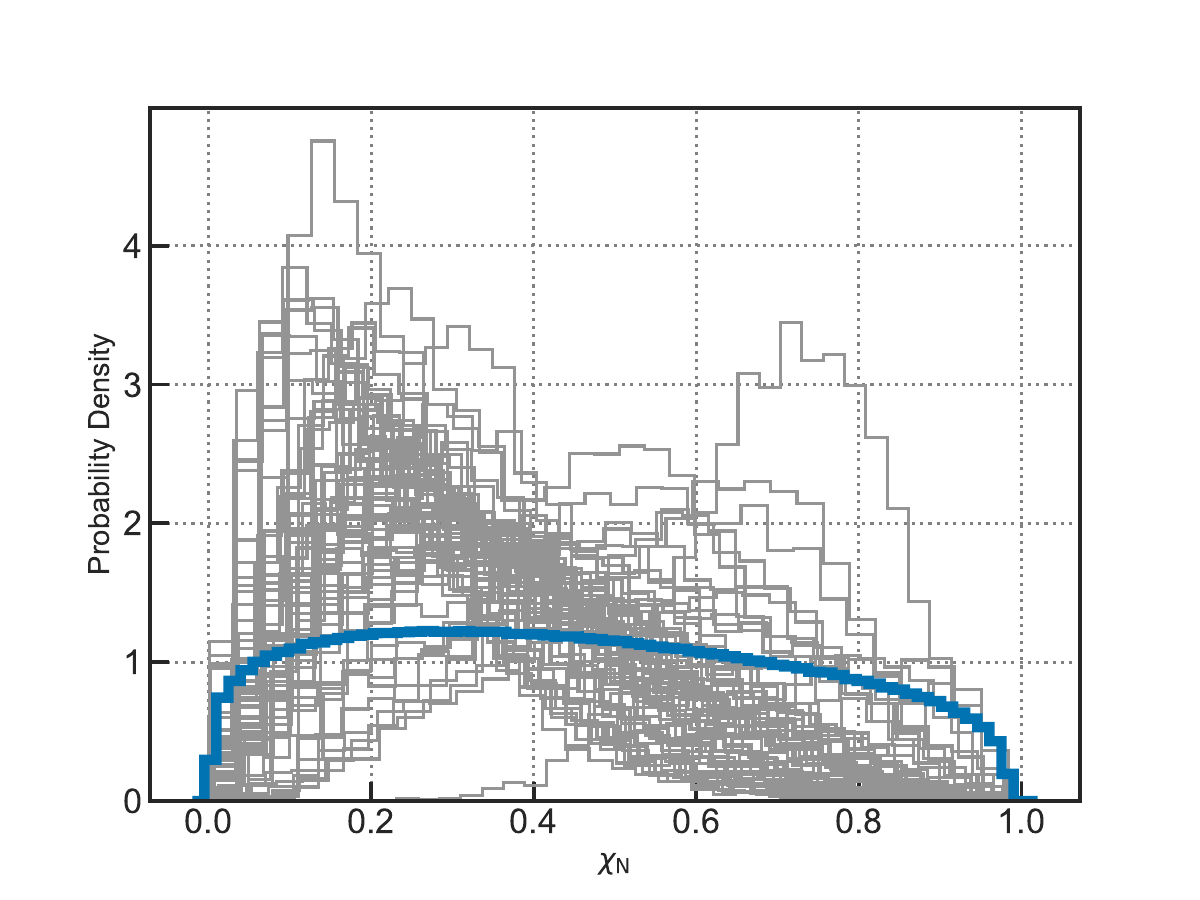}
\caption{Marginal prior distribution (blue) for $\chi_{\rm N}$ in the parameter estimation and the converted posterior distributions (grey) of each events.}
\label{fig:chiN}
\end{figure*}

The dimensionless component spin parameter for an individual BH with mass $m$ is defined as \citet{2023PhRvX..13d1039A}
\begin{equation}\label{eq:1}
\chi_i = \frac{\left | \mathbf{J}_i \right |  c}{Gm^2},
\end{equation}
where $\mathbf{J}_i$ is the spin angular momentum, and $Gm^2/c$ is the maximum angular momentum allowed by the third law of black hole mechanics. Analogy with this definition, here we define the dimensionless net spin as 
\begin{equation}\label{eq:2}
\chi_{\rm N} = \frac{ \left |\mathbf{J}_1+\mathbf{J}_2 \right | c}{G(m^2_1+m^2_2)},
\end{equation}
Note that the value of $\chi_{\rm N}$ will always in the range of (0,1). The dimensionless net spin contains information from component masses (or mass ratio) as well as the individual spin magnitudes and directions. In particular, when considering two component BHs with identical masses and maximum spin magnitudes, we get $\chi_{\rm N} = 1$ if their spins are aligned; if they have opposite spin direction, $\chi_{\rm N} = 0$. We use the ``C01:Mixed" parameter estimation result released by \citet{2023PhRvX..13d1039A} to calculate the posterior distribution of $\chi_{\rm N} = 1$ for each event. Giving a posterior sample $(m_1,m_2,s_{1x},s_{1y},s_{1z},s_{2x},s_{2y},s_{2z})$, the net spin at a giving direction $k$ is
\begin{equation}\label{eq:3}
\mathbf{J}_k = \frac{G}{c}(s_{1k} m^2_1 + s_{2k} m^2_2)\widehat{\mathbf{L}}_k,
\end{equation}
where $\widehat{\mathbf{L}}_k$ is the unit vector in the $k$ direction. According to Eq.~(\ref{eq:2}), $\chi_{\rm N}$ can be calculated as
\begin{equation}\label{eq:4}
\chi_{\rm N} = \frac{ c\sqrt{\sum_{k}^{}\mathbf{J}^2_k}}{G(m^2_1+m^2_2)}.
\end{equation}
In Fig.\ref{fig:chiN} we present in grey the histograms for the converted $\chi_{\rm N}$ samples from each of the 69 events included in this work, and the details of the constraints are also listed in Tab.\ref{tab:long_table} of Appendix.\ref{app:notzero}. Note that while the $\chi_{\rm N}$ posterior samples are calculated from the projections of individual spin ($s_{ik}$), in general the individual spins are not well constrained and there are degeneracies among them in the parameter estimation \citep{2021PhRvL.126q1103B}. As shown in Fig.\ref{fig:strain}, for a binary with identical $\chi_{\rm N}$ and nearly equal component masses, different combinations of individual spin projections can produce GW with similar inspiral signals. Whether a binary has one BH with negligible spin, as a result of efficient transportation of angular momentum in stellar evolution scenario, is of particular interest in many studies \citep{2016MNRAS.462..844K,2017ApJ...842..111H,2018A&A...616A..28Q,2020A&A...635A..97B}. For each event, we estimate the probability that its $\chi_{\rm N}$ exceed the maximum value under the assumption that the primary or the secondary black hole has zero spin (see Appendix.\ref{app:notzero} for more details). We find that up to $\sim 40\%$ of the events are more likely (with a probability $>50\%$) to have $\chi_{\rm N}$ larger than the maximum value if only the secondary BH is allowed to spin, and $\sim 10\%$ of the events have probability of $>10\%$ that their $\chi_{\rm N}$ larger than the case in which only the primary BH is allowed to spin.

We derive the prior distribution of $\chi_{\rm N}$ implemented in the parameter estimation in order to compare with the posteriors. With the default prior as introduced in \citet{2023PhRvX..13d1039A}, we randomly generate $10^7$ prior samples. Then we convert the sampled parameters to $\chi_{\rm N}$ samples, and the resulting $\chi_{\rm N}$ distribution is the marginal prior distribution, $P(\chi_{\rm N})$. In Fig.\ref{fig:chiN} we present the marginal prior in blue. When constructing a mass-correlated population model, we also need to obtain the conditional prior distribution, $P(\chi_{\rm N} | q)$. Similar to the marginal distribution, we use simulations to obtain the prior for any giving mass ratio.

Different models for the formation pathways would have diverse predictions for $\chi_{\rm N}$. For demonstration, we adopt the simulation results released by \citep{2021ApJ...910..152Z} and show the corresponding marginal $\chi_{\rm N}$ distributions in Fig.\ref{fig:simulation_chis}. The simulation contains results for five independent formation channels: the common envelope evolution (CEE), chemically homogeneous evolution (CHE), stable mass transfer (SMT), the dynamical processes in globular clusters (GC) and in neuclear star clusters (NSC). For comparison, the predicted $\chi_{\rm eff}$ distributions and $\chi_{\rm p}$ distributions are also shown. The $\chi_{\rm N}$ reflects the relative magnitude of the net angular momentum at its maximum direction, and the characteristics of different channels could be better distinguished by analysing both the observed $\chi_{\rm N}$ and $\chi_{\rm eff}$ distributions.

\section{Hierarchical Bayesian Inference}\label{sec:method}
We perform hierarchical Bayesian inference to constrain our model parameters. By choosing a uniform-in-log prior for the total merger rate, the likelihood under hyper-parameters $\mathbf{\Lambda}$ can be written as \citep{2019PASA...36...10T}
\begin{equation}\label{eq:5}
\mathcal{L}(\vec{d}\mid \mathbf{\Lambda}) \propto \prod_{i}^{N_{\rm obs}}\frac{1}{n_i \eta(\mathbf{\Lambda})}\sum_{k}^{n_i}\frac{p(\theta_{i}^k\mid \mathbf{\Lambda})}{p(\theta_{i}^k\mid \varnothing)},
\end{equation}
 where $\eta(\mathbf{\Lambda})$ is the detection efficiency, following the procedures described in \citet{2023PhRvX..13a1048A}, we use the injection campaign released in \citet{2023PhRvX..13a1048A} to estimated this quantity. The $n_i$ posterior samples for the $i$-th event and the default prior $\pi(\theta_k \mid \varnothing)$ are obtained from the released data accompanying with \citet{2023PhRvX..13d1039A,2024PhRvD.109b2001A,2021PhRvX..11b1053A,2019PhRvX...9c1040A}. 
 The estimation of likelihood is approximated with
 Monte Carlo summations over samples, which will bring statistical error. Therefore, we follow \cite{2023PhRvX..13a1048A} and constrain the prior of hyperparameter to ensure $N_{{\rm eff},i} > N_{\rm obs}$, where $N_{{\rm eff},i}$ is the effective numbers of samples for $i$-th event. We use the same criteria that define the detectable events as \citet{2023PhRvX..13a1048A}, i.e., ${\rm FAR} < 1/{\rm yr}$, and 69 BBH events passed the threshold cut. We use the python package {\sc Bilby} \citep{2019ApJS..241...27A} and the {\sc Nessai} sampler \citep{2021PhRvD.103j3006W} to obtain the Bayesian evidence and posteriors of the hyper-parameters for each model.

\section{The Marginal Distribution of $\chi_{\rm N}$: One Component Versus Two}\label{sec:spinonly}

We first consider the marginal distribution of $\chi_{\rm N}$. In this case, the selection bias is neglected since it is mainly induced by the masses of BBHs. We will further compare the results obtained in this section with those in the following section which includes the joint distributions and selection effects.

Our analysis starts with describing the marginal distribution with a single truncated Gaussian distribution, $\mathcal{G}(\chi_{\rm N}|\mu,\sigma)$
\begin{equation}\label{eq:6}
\mathcal{G}(\chi_{\rm N}|\mu_\chi,\sigma_\chi)= \left\{\begin{matrix}
\mathcal{N}(\chi_{\rm N}|\mu_\chi,\sigma_\chi) \ \ \ \ 0 < \chi_{\rm N} < 1, 
 \\
0 \ \ \ \ \ \ \  \ \ \ \ \ \ \ \ otherwise.
\end{matrix}\right.
\end{equation}
Note that $\mathcal{N}(\chi_{\rm N}|\mu_\chi,\sigma_\chi)$ is normalized in the interval $(0,1)$, and Eq.~(\ref{eq:6}) is labelled with model ``M1" in the following context. We also introduce a model consists of two truncated Gaussians for $\chi_{\rm N}$:
\begin{equation}\label{eq:7}
P(\chi_{\rm N}) =  (1-\xi)\mathcal{G}(\chi_{\rm N}|\mu_{\rm \chi,I},\sigma_{\rm \chi,I}) + \xi\mathcal{G}(\chi_{\rm N}|\mu_{\rm \chi,II},\sigma_{\rm \chi,II})
\end{equation}
where the parameter $\xi$ controls the fraction of the second Gaussian component, and the model is labelled with ``M2". In order to avoid degeneracy in the inference, we set additional constraint that $\mu_{\rm \chi,II}-\mu_{\rm \chi,I}>0 $ in the hierarchical inference. In Table \ref{tab:models} we list the priors and posteriors for the hyper-parameters in both models. The full corner plot for the inference are demonstrated in Fig.\ref{fig:one_corner} and Fig.\ref{fig:two_corner} of Appendix.\ref{sec:app_post}. When inferring the marginal distribution of $\chi_{\rm N}$ with model M1, we obtain a distribution with mean $\sim 0.29$ and standard deviation $\sim 0.1$. However, the Bayes factor of model M2 compared to M1 is 22, suggesting a strong preference of two components against one by the data. The inferred results for model M2 reveal two distinct Gaussian components: one peaks at $0.22^{+0.03}_{-0.02}$ and has a narrow width of $0.03^{+0.02}_{-0.01}$, another peaks at $0.55^{+0.26}_{-0.20}$ and could be possibly much more extended ($\sigma_{\rm II} \sim 0.15$). 

Although the constraint for the second component is weak with current observations when only considering the one-dimensional data of $\chi_{\rm N}$, the above analysis indicates that the marginal $\chi_{\rm N}$ distribution have structures beyond a single Gaussian.

\section{The Mass Dependency and Redshift Evolution}\label{sec:joint}

We take a further step to study if there is mass dependency or redshift evolution behind the two components identified above. Our model is designed base on the possibility that the over-all BBH population could be the result of superposition of multiple populations formed via independent channels. The model consists of sub-populations labeled with ``I'' and ``II" respectively. Each of them has its own $\chi_{\rm N}$ distribution described by truncated Gaussian $\mathcal{G}(\chi_{\rm N}|\mu_i,\sigma_i)$ (where $i\in \left \{ {\rm I,II} \right \}$). According to the chain rule, the joint likelihood for a system with parameter ($M_{\rm tot},q, \chi_{\rm N}, z$) is written as :
\begin{equation}\label{eq:8}
P(M_{\rm tot},q, \chi_{\rm N}, z) = P(M_{\rm tot},q)\Gamma(z|M_{\rm tot}) 
\left \{ 
\left [ 1-\xi(M_{\rm tot},z) \right ] 
\mathcal{G}(\chi_{\rm N}|\mu_{\rm I},\sigma_{\rm I}) +
\xi(M_{\rm tot},z) 
\mathcal{G}(\chi_{\rm N}|\mu_{\rm II},\sigma_{\rm II}) 
\right \},   \end{equation}
where $P(M_{\rm tot},q)$ is the over-all joint distribution for the total mass and mass ratio marginalized over other parameters, $\Gamma(z|M_{\rm tot})$ is the over-all redshift distribution conditioned on $M_{\rm tot}$. These distributions include contributions from both sub-populations. $\xi(M_{\rm tot},z)$ is the fraction (branch ratio) of the ``II" sub-population given $M_{\rm tot}$  and $z$. Assuming the merger rate of the two sub-populations evolve with redshift as $\mathcal{R}_i = \mathcal{R}_{0,i}(1+z)^{\gamma_i}$, and let $\mathcal{F}_0(M_{\rm tot}) = \mathcal{R}_{\rm 0,II}(M_{\rm tot})/[\mathcal{R}_{\rm 0,I}(M_{\rm tot})+\mathcal{R}_{\rm 0,II}(M_{\rm tot})]$, the redshift distribution $\Gamma(z|M_{\rm tot})$ can be written as:
\begin{equation}\label{eq:9}
\Gamma(z|M_{\rm tot}) \propto
\left [ (1-\mathcal{F}_0)(1+z)^{\gamma_{\rm I}} \\
+ \mathcal{F}_0(1+z)^{\gamma_{\rm II}} \right ] \\
\frac{\mathrm{d} V_{\rm c}}{\mathrm{d} z} \\
\frac{1}{1+z}.   
\end{equation}
The branch ratio $\xi(M_{\rm tot},z)$ satisfies:
\begin{equation}\label{eq:10}
\xi(M_{\rm tot},z) = \frac{\mathcal{F}_0(1+z)^{\gamma_{\rm II}}}{ \\
(1-\mathcal{F}_0)(1+z)^{\gamma_{\rm I}} + \mathcal{F}_0(1+z)^{\gamma_{\rm II}}}.    
\end{equation}

We employ the Power-law+Peak model as introduced in \citet{2018ApJ...856..173T} and \citet{2021ApJ...913L...7A} for the total mass and a Power-law distribution for the mass ratio. To investigate the dependency between $\mathcal{F}_0$ and $M_{\rm tot}$, we introduce a modified logistic function:
\begin{equation}\label{eq:11}
\mathcal{F}_0(M_{\rm tot}) = f_{\rm L} + \frac{f_{\rm H}-f_{\rm L}}{1+\exp[-k(M_{\rm tot}-m_{\rm ref})]}, 
\end{equation}
in which the value of $\mathcal{F}_0$ changes from $f_{\rm L}$ at low-mass edge to $f_{\rm H}$ at high-mass edge. The fraction reach $\mathcal{F}_0 = (f_{\rm H}-f_{\rm L})/2$ at a reference mass $m_{\rm ref}$, and $k$ controls the rapidness of the evolution. The model described in this section is label as the ``\textit{JOINT}" model, the analytical functions used in the model, their parameters and the corresponding priors are summarized in Tab.~\ref{tab:models}. 

The full corner plot for the parameters in the \textit{JOINT} model obtained from the hierarchical inference is shown in Fig.\ref{fig:full}. The shape of astrophysical distributions for $M_{\rm tot}$ recovered from the posteriors of hyper-parameters are very similar to that of the primary mass distribution in \citet{2023PhRvX..13a1048A}, with approximately doubled $m_{\rm min}$, $m_{\rm max}$, and $\mu_{\rm m}$. This result can be understood because the binaries tend to pair with symmetric masses ($\beta \sim 2$ as inferred). Comparing with the results for Model M2 in Sec.\ref{sec:spinonly}, the \textit{JOINT} model also reveals two distinct $\chi_{\rm N}$ distributions. The constraint on  $\sigma_{\rm \chi,II}$ benefits from the relations of BBH parameters embedded in the \textit{JOINT} model and is consistent with the peak of the corresponding posterior for Model M2.

\begin{figure*}
	\centering  
\includegraphics[width=0.5\linewidth]{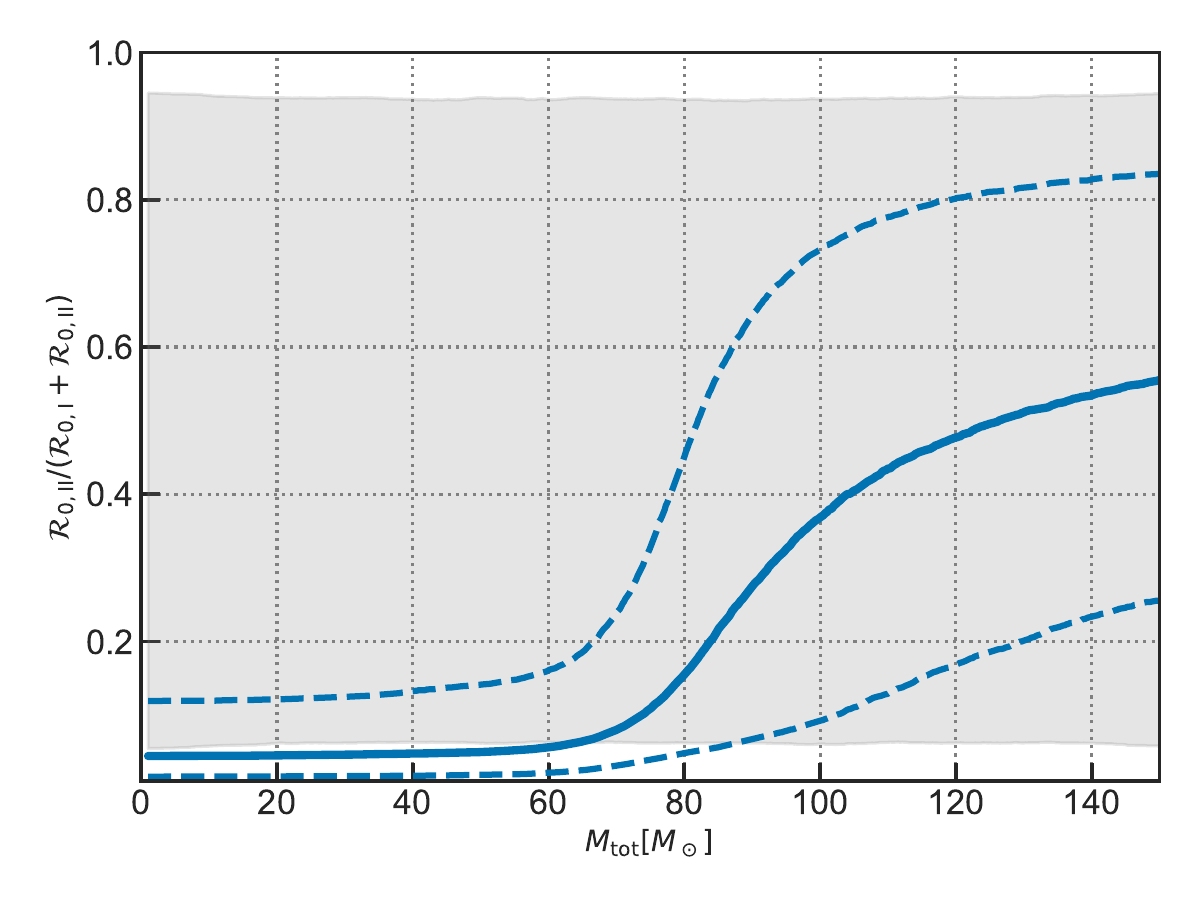}
\caption{The evolution of branch ratio ($\mathcal{F}_0$) in the local Universe ($z=0$) across total mass for the large $\chi_{\rm N}$ component. The grey area is the $90\%$ region for the prior used in the inference, and the blue dashed lines show the $68\%$ posterior credible region.}
\label{fig:F0}
\end{figure*}

For the mass dependence of $\mathcal{F}_0$, there is a clear hint in the posterior distributions: the distribution of $f_{\rm L}$ rails against $f_{\rm L}=0$, which reflects the small $\chi_{\rm N}$ sub-population may dominant the low-mass edge; though $f_{\rm H}$ is not well-constrained, its posterior disfavor the region of $f_{\rm H}<0.2$, and we find that $f_{\rm H}>f_{\rm L}$ at $96.8\%$ credibility. For a better demonstration, we show the evolution of $\mathcal{F}_0$ on different total masses in Fig.\ref{fig:F0}. The upper left prior region (representing the fraction of high $\chi_{\rm N}$ component $>0.2$ at low total mass end) is excluded by the posterior. In addition, we find that the fraction increase rapidly at $80-100 M_\odot$.

Besides the mass evolution of $\mathcal{F}_0$, we also find distinct redshift evolutions for the two sub-populations. The small $\chi_{\rm N}$ BBHs evolve much slower than the large $\chi_{\rm N}$ ones, with $\gamma_{\rm II}>\gamma_{\rm I}$ at $92.2\%$ credibility. 

To avoid model misspecification \citep{2022PASA...39...25R}, we perform posterior predicted checks following the procedures described in \citet{2021ApJ...913L...7A}. As shown in Fig.~\ref{fig:ppcheck}, the observed data well-match the prediction of the \textit{JOINT} model.

\begin{figure*}
	\centering  
\includegraphics[width=1.\linewidth]{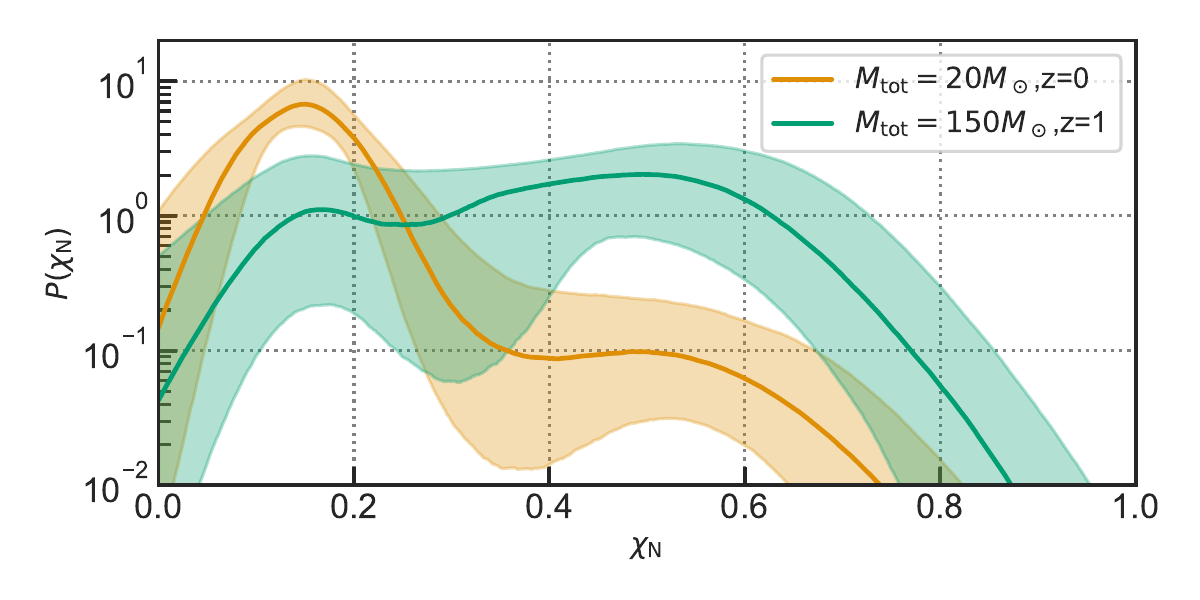}
\caption{The over-all $\chi_{\rm N}$ posterior distributions conditioned on different total masses and redshifts. The shaded areas are the $68\%$ credible regions, and the solid lines are the medians of the posterior probability density function.}
\label{fig:chiN_at_MZ}
\end{figure*}

The above results suggest we are more likely to observe high $\chi_{\rm N}$ BBHs with high masses and high redshifts. In Fig.\ref{fig:chiN_at_MZ} we show the inferred over-all $\chi_{\rm N}$ distribution (contributed by both sub-populations) at different total masses and redshifts. These trend, if true, can be validated by comparing the net spins of the smallest and nearest BBHs with the heaviest and furthest ones in future LIGO/Virgo/KAGRA observations.

\begin{table*}[htpb]
\centering
\caption{Summary of BBH parameters, the hyper-parameters 
 as well as their priors in the \textit{JOINT} population model.}\label{tab:models}
\begin{tabular}{c|c|c|c}
\hline
\hline
\multirow{2}{*}{property}  & {BBH parameter \&} & {hyper-parameter \&} & \multirow{2}{*}{prior} \\
& distribution & definition &\\
\cline{1-4}
\multirow{16}{*}{mass} & \multirow{12}{*}{$M_{\rm tot}$} & {$\alpha$}  & \multirow{2}{*}{U(-2,8)} \\
& \multirow{13}{*}{PowerlawPeak} & Powerlaw index for the total mass distribution &\\
\cline{3-4}
&& {$m_{\rm min}$} & \multirow{2}{*}{U(4,20)} \\
&& minimum total mass & \\
\cline{3-4}
&& {$m_{\rm max}$} & \multirow{2}{*}{U(80,200)} \\
&& maximum total mass & \\
\cline{3-4}
&& {$\delta_{\rm m}$} & \multirow{2}{*}{U(0,20)} \\
&& the length of mass distribution before it declines & \\
\cline{3-4}
&& {$\mu_{\rm m}$} & \multirow{2}{*}{U(40,120)} \\
&& mean of the Gaussian peak & \\
\cline{3-4}
&& {$\sigma_{\rm m}$} & \multirow{2}{*}{U(2,20)} \\
&& standard deviation the Gaussian peak & \\
\cline{3-4}
&& {$r_{\rm peak}$} & \multirow{2}{*}{U(0,1)} \\
&& fraction of systems in the Gaussian peak & \\
\cline{2-4}
& q & {$\beta$} & \multirow{2}{*}{U(0,8)} \\
& Powerlaw & Powerlaw index for the mass ratio distribution & \\
\cline{1-4}
\multirow{8}{*}{spin} & \multirow{7}{*}{$\chi_{\rm N}$} & {$\mu_{\rm \chi,I}$}  & \multirow{2}{*}{U(0,1)} \\
& \multirow{8}{*}{truncated Gaussians} & mean of the first truncated Gaussian &\\
\cline{3-4}
&& {$\sigma_{\rm \chi,I}$} & \multirow{2}{*}{U(0.001, 0.1)} \\
&& standard deviation of the first truncated Gaussian component & \\
\cline{3-4}
&& {$\mu_{\rm \chi,II}$} & \multirow{2}{*}{U(0, 1)} \\
&& mean of the second truncated Gaussian & \\
\cline{3-4}
&& {$\sigma_{\rm \chi,II}$} & \multirow{2}{*}{U(0.01, 1)} \\
&& standard deviation of the second truncated Gaussian component & \\
\cline{1-4}
\multirow{4}{*}{redshift} & \multirow{3}{*}{$z$} & {$\gamma_{\rm I}$}  & \multirow{2}{*}{U(-4,8)} \\
& \multirow{4}{*}{merger rate $\propto (1+z)^{\gamma_i}$} & Powerlaw index for the merger rate evolution of the first component &\\
\cline{3-4}
&& {$\gamma_{\rm II}$} & \multirow{2}{*}{U(-4, 8)} \\
&& Powerlaw index for the merger rate evolution of the second component & \\
\cline{1-4}
\multirow{8}{*}{$\mathcal{F}_0(M_{\rm tot})$} & \multirow{7}{*}{$-$} & {$f_{\rm L}$}  & \multirow{2}{*}{U(0,1)} \\
& \multirow{8}{*}{Eq.~(\ref{eq:11})} & fraction of the second component at low-mass edge in the local Universe &\\
\cline{3-4}
&& {$f_{\rm H}$} & \multirow{2}{*}{U(0,1)} \\
&& fraction of the second component at high-mass edge in the local Universe & \\
\cline{3-4}
&& {$m_{\rm ref}$} & \multirow{2}{*}{U(10, 150)} \\
&& the location of the reference mass at which $\mathcal{F}_0 = (f_{\rm H}-f_{\rm L})/2$  & \\
\cline{3-4}
&& {$k$} & \multirow{2}{*}{U(0.03, 0.5)} \\
&& rapidness of the evolution & \\
\hline
\hline
\end{tabular}
\tablenotetext{}{{\bf Note.} For the marginal distribution models in Sec.\ref{sec:spinonly}, the hyper-parameters in model M2 have identical priors as the relevant ones in the table, and the prior for $\xi$ is U(0, 1); in model M1, the priors for $\mu_\chi$ and $\sigma_\chi$ are U(0, 1) and U(0.001, 1), respectively.}
\end{table*}

\section{Discussion and Conclusions}\label{sec:summary}

While previous works mainly focused on the $\chi_{\rm eff}$ or individual spins, in this paper we define and study the dimensionless net spin of BBHs. By defining such a quantity, we can study the system's spin not only restricted in the direction aligned with the orbital angular momentum, while avoid the uncertainties and potential degeneracy in the measurement of individual spins. It also allow us to release some assumptions in the modelling (such as whether the spin magnitude or tilted angle for two component black holes follow the same distribution). We find that the over-all distribution of $\chi_{\rm N}$ can be better described by two truncated Gaussian components rather than one component. Assuming these two components represent two sub-population of black holes, their mixing fraction varies with the total mass of the system, and they follows different merger rate evolution across redshift. 

The measured $\chi_{\rm N}$ for GW events and the constraints on their population properties bring new insight into distinguishing the formation channel of BBHs. 
Isolated binary stars are one of the most promising origins of gravitational wave sources. Recent studies suggest that the residual angular momentum in BBHs is mainly provided by the tidal interaction of the first-born black hole (generally considered to be the primary BH in BBH) with the progenitor star of the second-born black hole \citep{2016MNRAS.462..844K,2017ApJ...842..111H,2018A&A...616A..28Q,2020A&A...635A..97B}. As shown in Sec.\ref{sec:chi_N}, on event level, the net angular momentum of some events are more likely to exceed the maximum value that can be provided by the lighter black hole (see also the discussion in \citet{2022ApJ...924..129Q} using the measurements of $\chi_{\rm eff}$ ); on population level, our constraints on the distribution of high $\chi_{\rm N}$ components indicate about half of the black holes in this subclass exceed this limit. The process of binary evolution involves the transfer of mass and angular momentum, and for systems that undergo only stable mass transfer (the SMT channel), a notable fraction of systems may experience mass ratio reversal (MRR), leading to the formation of a spinning primary black hole and thereby allowing the system to have a larger net angular momentum \citep{2022MNRAS.517.2738M}. However, simulations also show that the fraction of systems with non-negligible spin decreases rapidly with the chirp mass \citep{2022MNRAS.516.5737B}, regardless of whether the system undergo MRR. This contradicts our finding that the proportion of systems with large $\chi_{\rm N}$ increases with total mass. Additionally, the trend of the redshift evolution for different BBH total mass is distinct from that found in BBH merger from isolated binary evolution. As shown in \citet{2022ApJ...931...17V}, BBHs with small primary masses (and hence small total mass due to the preference for equal mass in GW observation) merge at higher redshifts than those with high primary masses. 

Several studies suggested isolated Pop III binaries can contribute to heavy BBHs with $\chi_{\rm eff} \gtrsim 0.4$ due to spin up in SMT (see \citet{2024RvMPP...8...13T} and the references therein). While in this work the high $\chi_{\rm N}$ component peaks at $>0.4$, \citet{2020ApJ...894..129S} shown that the $\chi_{\rm eff}$ distribution peaks at much smaller values ($<0.25$) even accounting for the peak may increase with primary mass. This disfavor the case in which the net spin of high-mass systems preferentially aligned with the orbital angular momentum, which is a signature of binary interaction. The pop III or low-metallicity stars, however, could still produce fast spinning individual black holes. The stellar winds in these stars are weaker and thus less efficient to remove the angular momentum stored in the stellar envelope \citep{2022ApJ...932L..19B}. The infall of the envelope can lead to a heavy black hole \citep{2021MNRAS.504..146V,2024MNRAS.529.2980W} with large natal spin. To form the observed BBHs, these remnants should assemble through additional dynamical process. The dynamical process itself can lead to BBHs with high total mass and $\chi_{\rm N}$ through hierarchical merger: by inhering the orbital angular momentum of previous mergers, the second and higher generation BH can significantly increase their spins. Different from binary evolution, this lead to a broadening in $\chi_{\rm eff}$ distribution and a peak $>0.4$ in $\chi_{\rm N}$ distribution (see Fig.\ref{fig:simulation_chis}). These phenomena are revealed by \citet{2022ApJ...932L..19B} and this work. Besides, \citet{2024arXiv240212444Y} pointed out the merger rate for BBHs having primary mass $>40 M_\odot$ in dense star clusters increases much faster than those with small primary masses due to mass segregation in dense star clusters and the early formation of massive BHs, which is consistent with the trend we found. Nevertheless, the (pulsational) pair instability supernovae ((P)PISNe) forbid the formation of BHs with mass ranging from several tens to $\sim 130$ times the solar mass. Though the lower edge of this ``PI mass gap" is still uncertain \citep{2019ApJ...887...53F,2024MNRAS.529.2980W}, the rapidly increase of $\mathcal{F}_0$ within $M_{\rm tot} \sim 80 - 100 M_\odot$ as shown in Fig.\ref{fig:F0} consists with the case in which the gap starts at $\sim 40 - 50 M_\odot$ and the hierarchical merger mainly contributes to the sources with high masses and high $\chi_{\rm N}$. 

In conclusion, the $\chi_{\rm N}$ distribution and it's relation with the total mass and redshift of GW events bring several new challenges to the binary evolution senarios. No matter whether the Pop III/low-metallicity star or hierarchical merger is responsible for the high $\chi_{\rm N}$ component, our results suggest the dynamical process should plays a key role in forming high total mass BBHs. The upcoming O4 data of Advanced LIGO/Virgo/KAGRA will reveal clearer mass dependency and redshift evolution of $\chi_{\rm N}$, and details in the formation of binary black holes could be better studied.

\acknowledgments
 This work was supported by the National Natural Science Foundation of China (NSFC) under grants No. 12203101 and No. 12233011. Tao Zhu is supported by the National Key Research and Development Program of China under Grant No.2020YFC2201503, the National Natural Science Foundation of China under Grants No.12275238, and the Zhejiang Provincial Natural Science Foundation of China under Grants No.LR21A050001.This research has made use of data and software obtained from the Gravitational Wave Open Science Center (https://www.gw-openscience.org), a service of LIGO Laboratory, the LIGO Scientific Collaboration and the Virgo Collaboration. LIGO is funded by the U.S. National Science Foundation. Virgo is funded by the French Centre National de Recherche Scientifique (CNRS), the Italian Istituto Nazionale della Fisica Nucleare (INFN) and the Dutch Nikhef, with contributions by Polish and Hungarian institutes.
 
\vspace{5mm}
\software{Bilby \citep[version 1.1.4, ascl:1901.011, \url{https://git.ligo.org/lscsoft/bilby/}]{2019ascl.soft01011A},
Nessai \citep[version 0.11.0, \url{https://nessai.readthedocs.io/en/latest/}]{2021PhRvD.103j3006W},
PyCBC 
\citep[version 2.2.0, \url{https://github.com/gwastro/pycbc/}]{pycbc2019PASP}
          }

\bibliographystyle{aasjournal}
\bibliography{bibliography}

\begin{thebibliography}{}
\expandafter\ifx\csname natexlab\endcsname\relax\def\natexlab#1{#1}\fi
\providecommand{\url}[1]{\href{#1}{#1}}
\providecommand{\dodoi}[1]{doi:~\href{http://doi.org/#1}{\nolinkurl{#1}}}
\providecommand{\doeprint}[1]{\href{http://ascl.net/#1}{\nolinkurl{http://ascl.net/#1}}}
\providecommand{\doarXiv}[1]{\href{https://arxiv.org/abs/#1}{\nolinkurl{https://arxiv.org/abs/#1}}}

\bibitem[{{Abbott} {et~al.}(2019{\natexlab{a}}){Abbott}, {Abbott}, {Abbott}, {Abraham}, {Acernese}, {Ackley}, {Adams}, {Adhikari}, {Adya}, {Affeldt}, {Agathos}, {Agatsuma}, {Aggarwal}, {Aguiar}, {Aiello}, {Ain}, {Ajith}, {Allen}, {Allocca}, {Aloy}, {Altin}, {Amato}, {Ananyeva}, {Anderson}, {Anderson}, {Angelova}, {Antier}, {Appert}, {Arai}, {Araya}, {Areeda}, {Ar{\`e}ne}, {Arnaud}, {Arun}, {Ascenzi}, {Ashton}, {Aston}, {Astone}, {Aubin}, {Aufmuth}, {AultONeal}, {Austin}, {Avendano}, {Avila-Alvarez}, {Babak}, {Bacon}, {Badaracco}, {Bader}, {Bae}, {Baker}, {Baldaccini}, {Ballardin}, {Ballmer}, {Banagiri}, {Barayoga}, {Barclay}, {Barish}, {Barker}, {Barkett}, {Barnum}, {Barone}, {Barr}, {Barsotti}, {Barsuglia}, {Barta}, {Bartlett}, {Bartos}, {Bassiri}, {Basti}, {Bawaj}, {Bayley}, {Bazzan}, {B{\'e}csy}, {Bejger}, {Belahcene}, {Bell}, {Beniwal}, {Berger}, {Bergmann}, {Bernuzzi}, {Bero}, {Berry}, {Bersanetti}, {Bertolini}, {Betzwieser}, {Bhandare}, {Bidler}, {Bilenko}, {Bilgili}, {Billingsley}, {Birch}, {Birney},
  {Birnholtz}, {Biscans}, {Biscoveanu}, {Bisht}, {Bitossi}, {Bizouard}, {Blackburn}, {Blair}, {Blair}, {Blair}, {Bloemen}, {Bode}, {Boer}, {Boetzel}, {Bogaert}, {Bondu}, {Bonilla}, {Bonnand}, {Booker}, {Boom}, {Booth}, {Bork}, {Boschi}, {Bose}, {Bossie}, {Bossilkov}, {Bosveld}, {Bouffanais}, {Bozzi}, {Bradaschia}, {Brady}, {Bramley}, {Branchesi}, {Brau}, {Briant}, {Briggs}, {Brighenti}, {Brillet}, {Brinkmann}, {Brisson}, {Brockill}, {Brooks}, {Brown}, {Brunett}, {Buikema}, {Bulik}, {Bulten}, {Buonanno}, {Buscicchio}, {Buskulic}, {Buy}, {Byer}, {Cabero}, {Cadonati}, {Cagnoli}, {Cahillane}, {Calder{\'o}n Bustillo}, {Callister}, {Calloni}, {Camp}, {Campbell}, {Canepa}, {Cannon}, {Cao}, {Cao}, {Capocasa}, {Carbognani}, {Caride}, {Carney}, {Carullo}, {Casanueva Diaz}, {Casentini}, {Caudill}, {Cavagli{\`a}}, {Cavalier}, {Cavalieri}, {Cella}, {Cerd{\'a}-Dur{\'a}n}, {Cerretani}, {Cesarini}, {Chaibi}, {Chakravarti}, {Chamberlin}, {Chan}, {Chao}, {Charlton}, {Chase}, {Chassande-Mottin}, {Chatterjee}, {Chaturvedi},
  {Chatziioannou}, {Cheeseboro}, {Chen}, {Chen}, {Chen}, {Cheng}, {Cheong}, {Chia}, {Chincarini}, {Chiummo}, {Cho}, {Cho}, {Cho}, {Christensen}, {Chu}, {Chua}, {Chung}, {Chung}, {Ciani}, {Ciobanu}, {Ciolfi}, {Cipriano}, {Cirone}, {Clara}, {Clark}, {Clearwater}, {Cleva}, {Cocchieri}, {Coccia}, {Cohadon}, {Cohen}, {Colgan}, {Colleoni}, {Collette}, {Collins}, {Cominsky}, {Constancio}, {Conti}, {Cooper}, {Corban}, {Corbitt}, {Cordero-Carri{\'o}n}, {Corley}, {Cornish}, {Corsi}, {Cortese}, {Costa}, {Cotesta}, {Coughlin}, {Coughlin}, {Coulon}, {Countryman}, {Couvares}, {Covas}, {Cowan}, {Coward}, {Cowart}, {Coyne}, {Coyne}, {Creighton}, {Creighton}, {Cripe}, {Croquette}, {Crowder}, {Cullen}, {Cumming}, {Cunningham}, {Cuoco}, {Dal Canton}, {D{\'a}lya}, {Danilishin}, {D'Antonio}, {Danzmann}, {Dasgupta}, {Da Silva Costa}, {Datrier}, {Dattilo}, {Dave}, {Davier}, {Davis}, {Daw}, {DeBra}, {Deenadayalan}, {Degallaix}, {De Laurentis}, {Del{\'e}glise}, {Del Pozzo}, {DeMarchi}, {Demos}, {Dent}, {De Pietri}, {Derby}, {De
  Rosa}, {De Rossi}, {DeSalvo}, {de Varona}, {Dhurandhar}, {D{\'\i}az}, {Dietrich}, {Di Fiore}, {Di Giovanni}, {Di Girolamo}, {Di Lieto}, {Ding}, {Di Pace}, {Di Palma}, {Di Renzo}, {Dmitriev}, {Doctor}, {Donovan}, {Dooley}, {Doravari}, {Dorrington}, {Downes}, {Drago}, {Driggers}, {Du}, {Ducoin}, {Dupej}, {Dwyer}, {Easter}, {Edo}, {Edwards}, {Effler}, {Ehrens}, {Eichholz}, {Eikenberry}, {Eisenmann}, {Eisenstein}, {Essick}, {Estelles}, {Estevez}, {Etienne}, {Etzel}, {Evans}, {Evans}, {Fafone}, {Fair}, {Fairhurst}, {Fan}, {Farinon}, {Farr}, {Farr}, {Fauchon-Jones}, {Favata}, {Fays}, {Fazio}, {Fee}, {Feicht}, {Fejer}, {Feng}, {Fernandez-Galiana}, {Ferrante}, {Ferreira}, {Ferreira}, {Ferrini}, {Fidecaro}, {Fiori}, {Fiorucci}, {Fishbach}, {Fisher}, {Fishner}, {Fitz-Axen}, {Flaminio}, {Fletcher}, {Flynn}, {Fong}, {Font}, {Forsyth}, {Fournier}, {Frasca}, {Frasconi}, {Frei}, {Freise}, {Frey}, {Frey}, {Fritschel}, {Frolov}, {Fulda}, {Fyffe}, {Gabbard}, {Gadre}, {Gaebel}, {Gair}, {Gammaitoni}, {Ganija}, {Gaonkar},
  {Garcia}, {Garc{\'\i}a-Quir{\'o}s}, {Garufi}, {Gateley}, {Gaudio}, {Gaur}, {Gayathri}, {Gemme}, {Genin}, {Gennai}, {George}, {George}, {Gergely}, {Germain}, {Ghonge}, {Ghosh}, {Ghosh}, {Ghosh}, {Giacomazzo}, {Giaime}, {Giardina}, {Giazotto}, {Gill}, {Giordano}, {Glover}, {Godwin}, {Goetz}, {Goetz}, {Goncharov}, {Gonz{\'a}lez}, {Gonzalez Castro}, {Gopakumar}, {Gorodetsky}, {Gossan}, {Gosselin}, {Gouaty}, {Grado}, {Graef}, {Granata}, {Grant}, {Gras}, {Grassia}, {Gray}, {Gray}, {Greco}, {Green}, {Green}, {Gretarsson}, {Groot}, {Grote}, {Grunewald}, {Gruning}, {Guidi}, {Gulati}, {Guo}, {Gupta}, {Gupta}, {Gustafson}, {Gustafson}, {Haegel}, {Halim}, {Hall}, {Hall}, {Hamilton}, {Hammond}, {Haney}, {Hanke}, {Hanks}, {Hanna}, {Hannam}, {Hannuksela}, {Hanson}, {Hardwick}, {Haris}, {Harms}, {Harry}, {Harry}, {Haster}, {Haughian}, {Hayes}, {Healy}, {Heidmann}, {Heintze}, {Heitmann}, {Hello}, {Hemming}, {Hendry}, {Heng}, {Hennig}, {Heptonstall}, {Hernandez Vivanco}, {Heurs}, {Hild}, {Hinderer}, {Hoak}, {Hochheim},
  {Hofman}, {Holgado}, {Holland}, {Holt}, {Holz}, {Hopkins}, {Horst}, {Hough}, {Howell}, {Hoy}, {Hreibi}, {Huerta}, {Huet}, {Hughey}, {Hulko}, {Husa}, {Huttner}, {Huynh-Dinh}, {Idzkowski}, {Iess}, {Ingram}, {Inta}, {Intini}, {Irwin}, {Isa}, {Isac}, {Isi}, {Iyer}, {Izumi}, {Jacqmin}, {Jadhav}, {Jani}, {Janthalur}, {Jaranowski}, {Jenkins}, {Jiang}, {Johnson}, {Jones}, {Jones}, {Jones}, {Jonker}, {Ju}, {Junker}, {Kalaghatgi}, {Kalogera}, {Kamai}, {Kandhasamy}, {Kang}, {Kanner}, {Kapadia}, {Karki}, {Karvinen}, {Kashyap}, {Kasprzack}, {Katsanevas}, {Katsavounidis}, {Katzman}, {Kaufer}, {Kawabe}, {Keerthana}, {K{\'e}f{\'e}lian}, {Keitel}, {Kennedy}, {Key}, {Khalili}, {Khan}, {Khan}, {Khan}, {Khan}, {Khazanov}, {Khursheed}, {Kijbunchoo}, {Kim}, {Kim}, {Kim}, {Kim}, {Kim}, {Kim}, {Kimball}, {King}, {King}, {Kinley-Hanlon}, {Kirchhoff}, {Kissel}, {Kleybolte}, {Klika}, {Klimenko}, {Knowles}, {Koch}, {Koehlenbeck}, {Koekoek}, {Koley}, {Kondrashov}, {Kontos}, {Koper}, {Korobko}, {Korth}, {Kowalska}, {Kozak}, {Kringel},
  {Krishnendu}, {Kr{\'o}lak}, {Kuehn}, {Kumar}, {Kumar}, {Kumar}, {Kumar}, {Kuo}, {Kutynia}, {Kwang}, {Lackey}, {Lai}, {Lam}, {Landry}, {Lane}, {Lang}, {Lange}, {Lantz}, {Lanza}, {Lartaux-Vollard}, {Lasky}, {Laxen}, {Lazzarini}, {Lazzaro}, {Leaci}, {Leavey}, {Lecoeuche}, {Lee}, {Lee}, {Lee}, {Lee}, {Lee}, {Lee}, {Lehmann}, {Lenon}, {Leroy}, {Letendre}, {Levin}, {Li}, {Li}, {Li}, {Li}, {Lin}, {Linde}, {Linker}, {Littenberg}, {Liu}, {Liu}, {Lo}, {Lockerbie}, {London}, {Longo}, {Lorenzini}, {Loriette}, {Lormand}, {Losurdo}, {Lough}, {Lousto}, {Lovelace}, {Lower}, {L{\"u}ck}, {Lumaca}, {Lundgren}, {Lynch}, {Ma}, {Macas}, {Macfoy}, {MacInnis}, {Macleod}, {Macquet}, {Maga{\~n}a-Sandoval}, {Maga{\~n}a Zertuche}, {Magee}, {Majorana}, {Maksimovic}, {Malik}, {Man}, {Mandic}, {Mangano}, {Mansell}, {Manske}, {Mantovani}, {Mapelli}, {Marchesoni}, {Marion}, {M{\'a}rka}, {M{\'a}rka}, {Markakis}, {Markosyan}, {Markowitz}, {Maros}, {Marquina}, {Marsat}, {Martelli}, {Martin}, {Martin}, {Martynov}, {Mason}, {Massera},
  {Masserot}, {Massinger}, {Masso-Reid}, {Mastrogiovanni}, {Matas}, {Matichard}, {Matone}, {Mavalvala}, {Mazumder}, {McCann}, {McCarthy}, {McClelland}, {McCormick}, {McCuller}, {McGuire}, {McIver}, {McManus}, {McRae}, {McWilliams}, {Meacher}, {Meadors}, {Mehmet}, {Mehta}, {Meidam}, {Melatos}, {Mendell}, {Mercer}, {Mereni}, {Merilh}, {Merzougui}, {Meshkov}, {Messenger}, {Messick}, {Metzdorff}, {Meyers}, {Miao}, {Michel}, {Middleton}, {Mikhailov}, {Milano}, {Miller}, {Miller}, {Millhouse}, {Mills}, {Milovich-Goff}, {Minazzoli}, {Minenkov}, {Mishkin}, {Mishra}, {Mistry}, {Mitra}, {Mitrofanov}, {Mitselmakher}, {Mittleman}, {Mo}, {Moffa}, {Mogushi}, {Mohapatra}, {Montani}, {Moore}, {Moraru}, {Moreno}, {Morisaki}, {Mours}, {Mow-Lowry}, {Mukherjee}, {Mukherjee}, {Mukherjee}, {Mukund}, {Mullavey}, {Munch}, {Mu{\~n}iz}, {Muratore}, {Murray}, {Nagar}, {Nardecchia}, {Naticchioni}, {Nayak}, {Neilson}, {Nelemans}, {Nelson}, {Nery}, {Neunzert}, {Ng}, {Ng}, {Nguyen}, {Nichols}, {Nissanke}, {Nocera}, {North}, {Nuttall},
  {Obergaulinger}, {Oberling}, {O'Brien}, {O'Dea}, {Ogin}, {Oh}, {Oh}, {Ohme}, {Ohta}, {Okada}, {Oliver}, {Oppermann}, {Oram}, {O'Reilly}, {Ormiston}, {Ortega}, {O'Shaughnessy}, {Ossokine}, {Ottaway}, {Overmier}, {Owen}, {Pace}, {Pagano}, {Page}, {Pai}, {Pai}, {Palamos}, {Palashov}, {Palomba}, {Pal-Singh}, {Pan}, {Pang}, {Pang}, {Pankow}, {Pannarale}, {Pant}, {Paoletti}, {Paoli}, {Parida}, {Parker}, {Pascucci}, {Pasqualetti}, {Passaquieti}, {Passuello}, {Patil}, {Patricelli}, {Pearlstone}, {Pedersen}, {Pedraza}, {Pedurand}, {Pele}, {Penn}, {Perez}, {Perreca}, {Pfeiffer}, {Phelps}, {Phukon}, {Piccinni}, {Pichot}, {Piergiovanni}, {Pillant}, {Pinard}, {Pirello}, {Pitkin}, {Poggiani}, {Pong}, {Ponrathnam}, {Popolizio}, {Porter}, {Powell}, {Prajapati}, {Prasad}, {Prasai}, {Prasanna}, {Pratten}, {Prestegard}, {Privitera}, {Prodi}, {Prokhorov}, {Puncken}, {Punturo}, {Puppo}, {P{\"u}rrer}, {Qi}, {Quetschke}, {Quinonez}, {Quintero}, {Quitzow-James}, {Raab}, {Radkins}, {Radulescu}, {Raffai}, {Raja}, {Rajan},
  {Rajbhandari}, {Rakhmanov}, {Ramirez}, {Ramos-Buades}, {Rana}, {Rao}, {Rapagnani}, {Raymond}, {Razzano}, {Read}, {Regimbau}, {Rei}, {Reid}, {Reitze}, {Ren}, {Ricci}, {Richardson}, {Richardson}, {Ricker}, {Riles}, {Rizzo}, {Robertson}, {Robie}, {Robinet}, {Rocchi}, {Rolland}, {Rollins}, {Roma}, {Romanelli}, {Romano}, {Romel}, {Romie}, {Rose}, {Rosi{\'n}ska}, {Rosofsky}, {Ross}, {Rowan}, {R{\"u}diger}, {Ruggi}, {Rutins}, {Ryan}, {Sachdev}, {Sadecki}, {Sakellariadou}, {Salconi}, {Saleem}, {Samajdar}, {Sammut}, {Sanchez}, {Sanchez}, {Sanchis-Gual}, {Sandberg}, {Sanders}, {Santiago}, {Sarin}, {Sassolas}, {Sathyaprakash}, {Saulson}, {Sauter}, {Savage}, {Schale}, {Scheel}, {Scheuer}, {Schmidt}, {Schnabel}, {Schofield}, {Sch{\"o}nbeck}, {Schreiber}, {Schulte}, {Schutz}, {Schwalbe}, {Scott}, {Scott}, {Seidel}, {Sellers}, {Sengupta}, {Sennett}, {Sentenac}, {Sequino}, {Sergeev}, {Setyawati}, {Shaddock}, {Shaffer}, {Shahriar}, {Shaner}, {Shao}, {Sharma}, {Shawhan}, {Shen}, {Shink}, {Shoemaker}, {Shoemaker},
  {ShyamSundar}, {Siellez}, {Sieniawska}, {Sigg}, {Silva}, {Singer}, {Singh}, {Singhal}, {Sintes}, {Sitmukhambetov}, {Skliris}, {Slagmolen}, {Slaven-Blair}, {Smith}, {Smith}, {Somala}, {Son}, {Sorazu}, {Sorrentino}, {Souradeep}, {Sowell}, {Spencer}, {Spera}, {Srivastava}, {Srivastava}, {Staats}, {Stachie}, {Standke}, {Steer}, {Steinke}, {Steinlechner}, {Steinlechner}, {Steinmeyer}, {Stevenson}, {Stocks}, {Stone}, {Stops}, {Strain}, {Stratta}, {Strigin}, {Strunk}, {Sturani}, {Stuver}, {Sudhir}, {Summerscales}, {Sun}, {Sunil}, {Suresh}, {Sutton}, {Swinkels}, {Szczepa{\'n}czyk}, {Tacca}, {Tait}, {Talbot}, {Talukder}, {Tanner}, {T{\'a}pai}, {Taracchini}, {Tasson}, {Taylor}, {Thies}, {Thomas}, {Thomas}, {Thondapu}, {Thorne}, {Thrane}, {Tiwari}, {Tiwari}, {Tiwari}, {Toland}, {Tonelli}, {Tornasi}, {Torres-Forn{\'e}}, {Torrie}, {T{\"o}yr{\"a}}, {Travasso}, {Traylor}, {Tringali}, {Trovato}, {Trozzo}, {Trudeau}, {Tsang}, {Tse}, {Tso}, {Tsukada}, {Tsuna}, {Tuyenbayev}, {Ueno}, {Ugolini}, {Unnikrishnan}, {Urban},
  {Usman}, {Vahlbruch}, {Vajente}, {Valdes}, {van Bakel}, {van Beuzekom}, {van den Brand}, {Van Den Broeck}, {Vander-Hyde}, {van der Schaaf}, {van Heijningen}, {van Veggel}, {Vardaro}, {Varma}, {Vass}, {Vas{\'u}th}, {Vecchio}, {Vedovato}, {Veitch}, {Veitch}, {Venkateswara}, {Venugopalan}, {Verkindt}, {Vetrano}, {Vicer{\'e}}, {Viets}, {Vine}, {Vinet}, {Vitale}, {Vo}, {Vocca}, {Vorvick}, {Vyatchanin}, {Wade}, {Wade}, {Wade}, {Walet}, {Walker}, {Wallace}, {Walsh}, {Wang}, {Wang}, {Wang}, {Wang}, {Wang}, {Ward}, {Warden}, {Warner}, {Was}, {Watchi}, {Weaver}, {Wei}, {Weinert}, {Weinstein}, {Weiss}, {Wellmann}, {Wen}, {Wessel}, {We{\ss}els}, {Westhouse}, {Wette}, {Whelan}, {Whiting}, {Whittle}, {Wilken}, {Williams}, {Williamson}, {Willis}, {Willke}, {Wimmer}, {Winkler}, {Wipf}, {Wittel}, {Woan}, {Woehler}, {Wofford}, {Worden}, {Wright}, {Wu}, {Wysocki}, {Xiao}, {Yamamoto}, {Yancey}, {Yang}, {Yap}, {Yazback}, {Yeeles}, {Yu}, {Yu}, {Yuen}, {Yvert}, {Zadro{\.z}ny}, {Zanolin}, {Zelenova}, {Zendri}, {Zevin}, {Zhang},
  {Zhang}, {Zhang}, {Zhao}, {Zhou}, {Zhou}, {Zhu}, {Zimmerman}, {Zlochower}, {Zucker}, {Zweizig}, {LIGO Scientific Collaboration}, \& {Virgo Collaboration}}]{2019ApJ...882L..24A}
{Abbott}, B.~P., {Abbott}, R., {Abbott}, T.~D., {et~al.} 2019{\natexlab{a}}, \apjl, 882, L24, \dodoi{10.3847/2041-8213/ab3800}

\bibitem[{{Abbott} {et~al.}(2019{\natexlab{b}}){Abbott}, {Abbott}, {Abbott}, {Abraham}, {Acernese}, {Ackley}, {Adams}, {Adhikari}, {Adya}, {Affeldt}, {Agathos}, {Agatsuma}, {Aggarwal}, {Aguiar}, {Aiello}, {Ain}, {Ajith}, {Allen}, {Allocca}, {Aloy}, {Altin}, {Amato}, {Ananyeva}, {Anderson}, {Anderson}, {Angelova}, {Antier}, {Appert}, {Arai}, {Araya}, {Areeda}, {Ar{\`e}ne}, {Arnaud}, {Arun}, {Ascenzi}, {Ashton}, {Aston}, {Astone}, {Aubin}, {Aufmuth}, {AultONeal}, {Austin}, {Avendano}, {Avila-Alvarez}, {Babak}, {Bacon}, {Badaracco}, {Bader}, {Bae}, {Baker}, {Baldaccini}, {Ballardin}, {Ballmer}, {Banagiri}, {Barayoga}, {Barclay}, {Barish}, {Barker}, {Barkett}, {Barnum}, {Barone}, {Barr}, {Barsotti}, {Barsuglia}, {Barta}, {Bartlett}, {Bartos}, {Bassiri}, {Basti}, {Bawaj}, {Bayley}, {Bazzan}, {B{\'e}csy}, {Bejger}, {Belahcene}, {Bell}, {Beniwal}, {Berger}, {Bergmann}, {Bernuzzi}, {Bero}, {Berry}, {Bersanetti}, {Bertolini}, {Betzwieser}, {Bhandare}, {Bidler}, {Bilenko}, {Bilgili}, {Billingsley}, {Birch}, {Birney},
  {Birnholtz}, {Biscans}, {Biscoveanu}, {Bisht}, {Bitossi}, {Bizouard}, {Blackburn}, {Blackman}, {Blair}, {Blair}, {Blair}, {Bloemen}, {Bode}, {Boer}, {Boetzel}, {Bogaert}, {Bondu}, {Bonilla}, {Bonnand}, {Booker}, {Boom}, {Booth}, {Bork}, {Boschi}, {Bose}, {Bossie}, {Bossilkov}, {Bosveld}, {Bouffanais}, {Bozzi}, {Bradaschia}, {Brady}, {Bramley}, {Branchesi}, {Brau}, {Briant}, {Briggs}, {Brighenti}, {Brillet}, {Brinkmann}, {Brisson}, {Brockill}, {Brooks}, {Brown}, {Brunett}, {Buikema}, {Bulik}, {Bulten}, {Buonanno}, {Buskulic}, {Bustamante Rosell}, {Buy}, {Byer}, {Cabero}, {Cadonati}, {Cagnoli}, {Cahillane}, {Calder{\'o}n Bustillo}, {Callister}, {Calloni}, {Camp}, {Campbell}, {Canepa}, {Cannon}, {Cao}, {Cao}, {Capocasa}, {Carbognani}, {Caride}, {Carney}, {Carullo}, {Casanueva Diaz}, {Casentini}, {Caudill}, {Cavagli{\`a}}, {Cavalier}, {Cavalieri}, {Cella}, {Cerd{\'a}-Dur{\'a}n}, {Cerretani}, {Cesarini}, {Chaibi}, {Chakravarti}, {Chamberlin}, {Chan}, {Chao}, {Charlton}, {Chase}, {Chassande-Mottin}, {Chatterjee},
  {Chaturvedi}, {Chatziioannou}, {Cheeseboro}, {Chen}, {Chen}, {Chen}, {Cheng}, {Cheong}, {Chia}, {Chincarini}, {Chiummo}, {Cho}, {Cho}, {Cho}, {Christensen}, {Chu}, {Chua}, {Chung}, {Chung}, {Ciani}, {Ciobanu}, {Ciolfi}, {Cipriano}, {Cirone}, {Clara}, {Clark}, {Clearwater}, {Cleva}, {Cocchieri}, {Coccia}, {Cohadon}, {Cohen}, {Colgan}, {Colleoni}, {Collette}, {Collins}, {Cominsky}, {Constancio}, {Conti}, {Cooper}, {Corban}, {Corbitt}, {Cordero-Carri{\'o}n}, {Corley}, {Cornish}, {Corsi}, {Cortese}, {Costa}, {Cotesta}, {Coughlin}, {Coughlin}, {Coulon}, {Countryman}, {Couvares}, {Covas}, {Cowan}, {Coward}, {Cowart}, {Coyne}, {Coyne}, {Creighton}, {Creighton}, {Cripe}, {Croquette}, {Crowder}, {Cullen}, {Cumming}, {Cunningham}, {Cuoco}, {Canton}, {D{\'a}lya}, {Danilishin}, {D'Antonio}, {Danzmann}, {Dasgupta}, {Da Silva Costa}, {Datrier}, {Dattilo}, {Dave}, {Davier}, {Davis}, {Daw}, {DeBra}, {Deenadayalan}, {Degallaix}, {De Laurentis}, {Del{\'e}glise}, {Del Pozzo}, {DeMarchi}, {Demos}, {Dent}, {De Pietri}, {Derby},
  {De Rosa}, {De Rossi}, {DeSalvo}, {de Varona}, {Dhurandhar}, {D{\'\i}az}, {Dietrich}, {Di Fiore}, {Di Giovanni}, {Di Girolamo}, {Di Lieto}, {Ding}, {Di Pace}, {Di Palma}, {Di Renzo}, {Dmitriev}, {Doctor}, {Donovan}, {Dooley}, {Doravari}, {Dorrington}, {Downes}, {Drago}, {Driggers}, {Du}, {Ducoin}, {Dupej}, {Dwyer}, {Easter}, {Edo}, {Edwards}, {Effler}, {Ehrens}, {Eichholz}, {Eikenberry}, {Eisenmann}, {Eisenstein}, {Essick}, {Estelles}, {Estevez}, {Etienne}, {Etzel}, {Evans}, {Evans}, {Fafone}, {Fair}, {Fairhurst}, {Fan}, {Farinon}, {Farr}, {Farr}, {Fauchon-Jones}, {Favata}, {Fays}, {Fazio}, {Fee}, {Feicht}, {Fejer}, {Feng}, {Fernandez-Galiana}, {Ferrante}, {Ferreira}, {Ferreira}, {Ferrini}, {Fidecaro}, {Fiori}, {Fiorucci}, {Fishbach}, {Fisher}, {Fishner}, {Fitz-Axen}, {Flaminio}, {Fletcher}, {Flynn}, {Fong}, {Font}, {Forsyth}, {Fournier}, {Frasca}, {Frasconi}, {Frei}, {Freise}, {Frey}, {Frey}, {Fritschel}, {Frolov}, {Fulda}, {Fyffe}, {Gabbard}, {Gadre}, {Gaebel}, {Gair}, {Gammaitoni}, {Ganija}, {Gaonkar},
  {Garcia}, {Garc{\'\i}a-Quir{\'o}s}, {Garufi}, {Gateley}, {Gaudio}, {Gaur}, {Gayathri}, {Gemme}, {Genin}, {Gennai}, {George}, {George}, {Gergely}, {Germain}, {Ghonge}, {Ghosh}, {Ghosh}, {Ghosh}, {Giacomazzo}, {Giaime}, {Giardina}, {Giazotto}, {Gill}, {Giordano}, {Glover}, {Godwin}, {Goetz}, {Goetz}, {Goncharov}, {Gonz{\'a}lez}, {Gonzalez Castro}, {Gopakumar}, {Gorodetsky}, {Gossan}, {Gosselin}, {Gouaty}, {Grado}, {Graef}, {Granata}, {Grant}, {Gras}, {Grassia}, {Gray}, {Gray}, {Greco}, {Green}, {Green}, {Gretarsson}, {Groot}, {Grote}, {Grunewald}, {Gruning}, {Guidi}, {Gulati}, {Guo}, {Gupta}, {Gupta}, {Gustafson}, {Gustafson}, {Haegel}, {Halim}, {Hall}, {Hall}, {Hamilton}, {Hammond}, {Haney}, {Hanke}, {Hanks}, {Hanna}, {Hannam}, {Hannuksela}, {Hanson}, {Hardwick}, {Haris}, {Harms}, {Harry}, {Harry}, {Haster}, {Haughian}, {Hayes}, {Healy}, {Heidmann}, {Heintze}, {Heitmann}, {Hello}, {Hemming}, {Hendry}, {Heng}, {Hennig}, {Heptonstall}, {Hernandez Vivanco}, {Heurs}, {Hild}, {Hinderer}, {Hoak}, {Hochheim},
  {Hofman}, {Holgado}, {Holland}, {Holt}, {Holz}, {Hopkins}, {Horst}, {Hough}, {Howell}, {Hoy}, {Hreibi}, {Huang}, {Huerta}, {Huet}, {Hughey}, {Hulko}, {Husa}, {Huttner}, {Huynh-Dinh}, {Idzkowski}, {Iess}, {Ingram}, {Inta}, {Intini}, {Irwin}, {Isa}, {Isac}, {Isi}, {Iyer}, {Izumi}, {Jacqmin}, {Jadhav}, {Jani}, {Janthalur}, {Jaranowski}, {Jenkins}, {Jiang}, {Johnson}, {Johnson-McDaniel}, {Jones}, {Jones}, {Jones}, {Jonker}, {Ju}, {Junker}, {Kalaghatgi}, {Kalogera}, {Kamai}, {Kandhasamy}, {Kang}, {Kanner}, {Kapadia}, {Karki}, {Karvinen}, {Kashyap}, {Kasprzack}, {Katsanevas}, {Katsavounidis}, {Katzman}, {Kaufer}, {Kawabe}, {Keerthana}, {K{\'e}f{\'e}lian}, {Keitel}, {Kennedy}, {Key}, {Khalili}, {Khan}, {Khan}, {Khan}, {Khan}, {Khazanov}, {Khursheed}, {Kijbunchoo}, {Kim}, {Kim}, {Kim}, {Kim}, {Kim}, {Kim}, {Kimball}, {King}, {King}, {Kinley-Hanlon}, {Kirchhoff}, {Kissel}, {Kleybolte}, {Klika}, {Klimenko}, {Knowles}, {Koch}, {Koehlenbeck}, {Koekoek}, {Koley}, {Kondrashov}, {Kontos}, {Koper}, {Korobko}, {Korth},
  {Kowalska}, {Kozak}, {Kringel}, {Krishnendu}, {Kr{\'o}lak}, {Kuehn}, {Kumar}, {Kumar}, {Kumar}, {Kumar}, {Kuo}, {Kutynia}, {Kwang}, {Lackey}, {Lai}, {Lam}, {Landry}, {Lane}, {Lang}, {Lange}, {Lantz}, {Lanza}, {Lartaux-Vollard}, {Lasky}, {Laxen}, {Lazzarini}, {Lazzaro}, {Leaci}, {Leavey}, {Lecoeuche}, {Lee}, {Lee}, {Lee}, {Lee}, {Lee}, {Lee}, {Lehmann}, {Lenon}, {Leroy}, {Letendre}, {Levin}, {Li}, {Li}, {Li}, {Li}, {Lin}, {Linde}, {Linker}, {Littenberg}, {Liu}, {Liu}, {Lo}, {Lockerbie}, {London}, {Longo}, {Lorenzini}, {Loriette}, {Lormand}, {Losurdo}, {Lough}, {Lousto}, {Lovelace}, {Lower}, {L{\"u}ck}, {Lumaca}, {Lundgren}, {Lynch}, {Ma}, {Macas}, {Macfoy}, {MacInnis}, {Macleod}, {Macquet}, {Maga{\~n}a-Sandoval}, {Maga{\~n}a Zertuche}, {Magee}, {Majorana}, {Maksimovic}, {Malik}, {Man}, {Mandic}, {Mangano}, {Mansell}, {Manske}, {Mantovani}, {Marchesoni}, {Marion}, {M{\'a}rka}, {M{\'a}rka}, {Markakis}, {Markosyan}, {Markowitz}, {Maros}, {Marquina}, {Marsat}, {Martelli}, {Martin}, {Martin}, {Martynov}, {Mason},
  {Massera}, {Masserot}, {Massinger}, {Masso-Reid}, {Mastrogiovanni}, {Matas}, {Matichard}, {Matone}, {Mavalvala}, {Mazumder}, {McCann}, {McCarthy}, {McClelland}, {McCormick}, {McCuller}, {McGuire}, {McIver}, {McManus}, {McRae}, {McWilliams}, {Meacher}, {Meadors}, {Mehmet}, {Mehta}, {Meidam}, {Melatos}, {Mendell}, {Mercer}, {Mereni}, {Merilh}, {Merzougui}, {Meshkov}, {Messenger}, {Messick}, {Metzdorff}, {Meyers}, {Miao}, {Michel}, {Middleton}, {Mikhailov}, {Milano}, {Miller}, {Miller}, {Millhouse}, {Mills}, {Milovich-Goff}, {Minazzoli}, {Minenkov}, {Mishkin}, {Mishra}, {Mistry}, {Mitra}, {Mitrofanov}, {Mitselmakher}, {Mittleman}, {Mo}, {Moffa}, {Mogushi}, {Mohapatra}, {Montani}, {Moore}, {Moraru}, {Moreno}, {Morisaki}, {Mours}, {Mow-Lowry}, {Mukherjee}, {Mukherjee}, {Mukherjee}, {Mukund}, {Mullavey}, {Munch}, {Mu{\~n}iz}, {Muratore}, {Murray}, {Nagar}, {Nardecchia}, {Naticchioni}, {Nayak}, {Neilson}, {Nelemans}, {Nelson}, {Nery}, {Neunzert}, {Ng}, {Ng}, {Nguyen}, {Nichols}, {Nielsen}, {Nissanke}, {Nitz},
  {Nocera}, {North}, {Nuttall}, {Obergaulinger}, {Oberling}, {O'Brien}, {O'Dea}, {Ogin}, {Oh}, {Oh}, {Ohme}, {Ohta}, {Okada}, {Oliver}, {Oppermann}, {Oram}, {O'Reilly}, {Ormiston}, {Ortega}, {O'Shaughnessy}, {Ossokine}, {Ottaway}, {Overmier}, {Owen}, {Pace}, {Pagano}, {Page}, {Pai}, {Pai}, {Palamos}, {Palashov}, {Palomba}, {Pal-Singh}, {Pan}, {Pang}, {Pang}, {Pankow}, {Pannarale}, {Pant}, {Paoletti}, {Paoli}, {Papa}, {Parida}, {Parker}, {Pascucci}, {Pasqualetti}, {Passaquieti}, {Passuello}, {Patil}, {Patricelli}, {Pearlstone}, {Pedersen}, {Pedraza}, {Pedurand}, {Pele}, {Penn}, {Perego}, {Perez}, {Perreca}, {Pfeiffer}, {Phelps}, {Phukon}, {Piccinni}, {Pichot}, {Piergiovanni}, {Pillant}, {Pinard}, {Pirello}, {Pitkin}, {Poggiani}, {Pong}, {Ponrathnam}, {Popolizio}, {Porter}, {Powell}, {Prajapati}, {Prasad}, {Prasai}, {Prasanna}, {Pratten}, {Prestegard}, {Privitera}, {Prodi}, {Prokhorov}, {Puncken}, {Punturo}, {Puppo}, {P{\"u}rrer}, {Qi}, {Quetschke}, {Quinonez}, {Quintero}, {Quitzow-James}, {Raab}, {Radkins},
  {Radulescu}, {Raffai}, {Raja}, {Rajan}, {Rajbhandari}, {Rakhmanov}, {Ramirez}, {Ramos-Buades}, {Rana}, {Rao}, {Rapagnani}, {Raymond}, {Razzano}, {Read}, {Regimbau}, {Rei}, {Reid}, {Reitze}, {Ren}, {Ricci}, {Richardson}, {Richardson}, {Ricker}, {Riemenschneider}, {Riles}, {Rizzo}, {Robertson}, {Robie}, {Robinet}, {Rocchi}, {Rolland}, {Rollins}, {Roma}, {Romanelli}, {Romano}, {Romel}, {Romie}, {Rose}, {Rosi{\'n}ska}, {Rosofsky}, {Ross}, {Rowan}, {R{\"u}diger}, {Ruggi}, {Rutins}, {Ryan}, {Sachdev}, {Sadecki}, {Sakellariadou}, {Salafia}, {Salconi}, {Saleem}, {Salemi}, {Samajdar}, {Sammut}, {Sanchez}, {Sanchez}, {Sanchis-Gual}, {Sandberg}, {Sanders}, {Santiago}, {Sarin}, {Sassolas}, {Sathyaprakash}, {Saulson}, {Sauter}, {Savage}, {Schale}, {Scheel}, {Scheuer}, {Schmidt}, {Schnabel}, {Schofield}, {Sch{\"o}nbeck}, {Schreiber}, {Schulte}, {Schutz}, {Schwalbe}, {Scott}, {Scott}, {Seidel}, {Sellers}, {Sengupta}, {Sennett}, {Sentenac}, {Sequino}, {Sergeev}, {Setyawati}, {Shaddock}, {Shaffer}, {Shahriar}, {Shaner},
  {Shao}, {Sharma}, {Shawhan}, {Shen}, {Shink}, {Shoemaker}, {Shoemaker}, {ShyamSundar}, {Siellez}, {Sieniawska}, {Sigg}, {Silva}, {Singer}, {Singh}, {Singhal}, {Sintes}, {Sitmukhambetov}, {Skliris}, {Slagmolen}, {Slaven-Blair}, {Smith}, {Smith}, {Somala}, {Son}, {Sorazu}, {Sorrentino}, {Souradeep}, {Sowell}, {Spencer}, {Srivastava}, {Srivastava}, {Staats}, {Stachie}, {Standke}, {Steer}, {Steinke}, {Steinlechner}, {Steinlechner}, {Steinmeyer}, {Stevenson}, {Stocks}, {Stone}, {Stops}, {Strain}, {Stratta}, {Strigin}, {Strunk}, {Sturani}, {Stuver}, {Sudhir}, {Summerscales}, {Sun}, {Sunil}, {Suresh}, {Sutton}, {Swinkels}, {Szczepa{\'n}czyk}, {Tacca}, {Tait}, {Talbot}, {Talukder}, {Tanner}, {T{\'a}pai}, {Taracchini}, {Tasson}, {Taylor}, {Thies}, {Thomas}, {Thomas}, {Thondapu}, {Thorne}, {Thrane}, {Tiwari}, {Tiwari}, {Tiwari}, {Toland}, {Tonelli}, {Tornasi}, {Torres-Forn{\'e}}, {Torrie}, {T{\"o}yr{\"a}}, {Travasso}, {Traylor}, {Tringali}, {Trovato}, {Trozzo}, {Trudeau}, {Tsang}, {Tse}, {Tso}, {Tsukada}, {Tsuna},
  {Tuyenbayev}, {Ueno}, {Ugolini}, {Unnikrishnan}, {Urban}, {Usman}, {Vahlbruch}, {Vajente}, {Valdes}, {van Bakel}, {van Beuzekom}, {van den Brand}, {Van Den Broeck}, {Vander-Hyde}, {van Heijningen}, {van der Schaaf}, {van Veggel}, {Vardaro}, {Varma}, {Vass}, {Vas{\'u}th}, {Vecchio}, {Vedovato}, {Veitch}, {Veitch}, {Venkateswara}, {Venugopalan}, {Verkindt}, {Vetrano}, {Vicer{\'e}}, {Viets}, {Vine}, {Vinet}, {Vitale}, {Vo}, {Vocca}, {Vorvick}, {Vyatchanin}, {Wade}, {Wade}, {Wade}, {Walet}, {Walker}, {Wallace}, {Walsh}, {Wang}, {Wang}, {Wang}, {Wang}, {Wang}, {Ward}, {Warden}, {Warner}, {Was}, {Watchi}, {Weaver}, {Wei}, {Weinert}, {Weinstein}, {Weiss}, {Wellmann}, {Wen}, {Wessel}, {We{\ss}els}, {Westhouse}, {Wette}, {Whelan}, {White}, {Whiting}, {Whittle}, {Wilken}, {Williams}, {Williamson}, {Willis}, {Willke}, {Wimmer}, {Winkler}, {Wipf}, {Wittel}, {Woan}, {Woehler}, {Wofford}, {Worden}, {Wright}, {Wu}, {Wysocki}, {Xiao}, {Yamamoto}, {Yancey}, {Yang}, {Yap}, {Yazback}, {Yeeles}, {Yu}, {Yu}, {Yuen}, {Yvert},
  {Zadro{\.Z}ny}, {Zanolin}, {Zappa}, {Zelenova}, {Zendri}, {Zevin}, {Zhang}, {Zhang}, {Zhang}, {Zhao}, {Zhou}, {Zhou}, {Zhu}, {Zimmerman}, {Zlochower}, {Zucker}, {Zweizig}, {LIGO Scientific Collaboration}, \& {Virgo Collaboration}}]{2019PhRvX...9c1040A}
---. 2019{\natexlab{b}}, Physical Review X, 9, 031040, \dodoi{10.1103/PhysRevX.9.031040}

\bibitem[{{Abbott} {et~al.}(2020){Abbott}, {Abbott}, {Abbott}, {Abraham}, {Acernese}, {Ackley}, {Adams}, {Adya}, {Affeldt}, {Agathos}, {Agatsuma}, {Aggarwal}, {Aguiar}, {Aiello}, {Ain}, {Ajith}, {Akutsu}, {Allen}, {Allocca}, {Aloy}, {Altin}, {Amato}, {Ananyeva}, {Anderson}, {Anderson}, {Ando}, {Angelova}, {Antier}, {Appert}, {Arai}, {Arai}, {Arai}, {Araki}, {Araya}, {Araya}, {Areeda}, {Ar{\`e}ne}, {Aritomi}, {Arnaud}, {Arun}, {Ascenzi}, {Ashton}, {Aso}, {Aston}, {Astone}, {Aubin}, {Aufmuth}, {Aultoneal}, {Austin}, {Avendano}, {Avila-Alvarez}, {Babak}, {Bacon}, {Badaracco}, {Bader}, {Bae}, {Bae}, {Baiotti}, {Bajpai}, {Baker}, {Baldaccini}, {Ballardin}, {Ballmer}, {Banagiri}, {Barayoga}, {Barclay}, {Barish}, {Barker}, {Barkett}, {Barnum}, {Barone}, {Barr}, {Barsotti}, {Barsuglia}, {Barta}, {Bartlett}, {Barton}, {Bartos}, {Bassiri}, {Basti}, {Bawaj}, {Bayley}, {Bazzan}, {B{\'e}csy}, {Bejger}, {Belahcene}, {Bell}, {Beniwal}, {Berger}, {Bergmann}, {Bernuzzi}, {Bero}, {Berry}, {Bersanetti}, {Bertolini},
  {Betzwieser}, {Bhandare}, {Bidler}, {Bilenko}, {Bilgili}, {Billingsley}, {Birch}, {Birney}, {Birnholtz}, {Biscans}, {Biscoveanu}, {Bisht}, {Bitossi}, {Bizouard}, {Blackburn}, {Blair}, {Blair}, {Blair}, {Bloemen}, {Bode}, {Boer}, {Boetzel}, {Bogaert}, {Bondu}, {Bonilla}, {Bonnand}, {Booker}, {Boom}, {Booth}, {Bork}, {Boschi}, {Bose}, {Bossie}, {Bossilkov}, {Bosveld}, {Bouffanais}, {Bozzi}, {Bradaschia}, {Brady}, {Bramley}, {Branchesi}, {Brau}, {Briant}, {Briggs}, {Brighenti}, {Brillet}, {Brinkmann}, {Brisson}, {Brockill}, {Brooks}, {Brown}, {Brown}, {Brunett}, {Buikema}, {Bulik}, {Bulten}, {Buonanno}, {Buskulic}, {Buy}, {Byer}, {Cabero}, {Cadonati}, {Cagnoli}, {Cahillane}, {Bustillo}, {Callister}, {Calloni}, {Camp}, {Campbell}, {Canepa}, {Cannon}, {Cannon}, {Cao}, {Cao}, {Capocasa}, {Carbognani}, {Caride}, {Carney}, {Carullo}, {Casanueva Diaz}, {Casentini}, {Caudill}, {Cavagli{\`a}}, {Cavalier}, {Cavalieri}, {Cella}, {Cerd{\'a}-Dur{\'a}n}, {Cerretani}, {Cesarini}, {Chaibi}, {Chakravarti}, {Chamberlin},
  {Chan}, {Chan}, {Chao}, {Charlton}, {Chase}, {Chassande-Mottin}, {Chatterjee}, {Chaturvedi}, {Chatziioannou}, {Cheeseboro}, {Chen}, {Chen}, {Chen}, {Chen}, {Chen}, {Chen}, {Cheng}, {Cheong}, {Chia}, {Chincarini}, {Chiummo}, {Cho}, {Cho}, {Cho}, {Christensen}, {Chu}, {Chu}, {Chu}, {Chua}, {Chung}, {Chung}, {Ciani}, {Ciobanu}, {Ciolfi}, {Cipriano}, {Cirone}, {Clara}, {Clark}, {Clearwater}, {Cleva}, {Cocchieri}, {Coccia}, {Cohadon}, {Cohen}, {Colgan}, {Colleoni}, {Collette}, {Collins}, {Cominsky}, {Constancio}, {Conti}, {Cooper}, {Corban}, {Corbitt}, {Cordero-Carri{\'o}n}, {Corley}, {Cornish}, {Corsi}, {Cortese}, {Costa}, {Cotesta}, {Coughlin}, {Coughlin}, {Coulon}, {Countryman}, {Couvares}, {Covas}, {Cowan}, {Coward}, {Cowart}, {Coyne}, {Coyne}, {Creighton}, {Creighton}, {Cripe}, {Croquette}, {Crowder}, {Cullen}, {Cumming}, {Cunningham}, {Cuoco}, {Dal Canton}, {D{\'a}lya}, {Danilishin}, {D'Antonio}, {Danzmann}, {Dasgupta}, {da Silva Costa}, {Datrier}, {Dattilo}, {Dave}, {Davier}, {Davis}, {Daw}, {Debra},
  {Deenadayalan}, {Degallaix}, {de Laurentis}, {Del{\'e}glise}, {Pozzo}, {Demarchi}, {Demos}, {Dent}, {de Pietri}, {Derby}, {De Rosa}, {de Rossi}, {Desalvo}, {de Varona}, {Dhurandhar}, {D{\'\i}az}, {Dietrich}, {di Fiore}, {di Giovanni}, {di Girolamo}, {di Lieto}, {Ding}, {di Pace}, {di Palma}, {di Renzo}, {Dmitriev}, {Doctor}, {Doi}, {Donovan}, {Dooley}, {Doravari}, {Dorrington}, {Downes}, {Drago}, {Driggers}, {Du}, {Ducoin}, {Dupej}, {Dwyer}, {Easter}, {Edo}, {Edwards}, {Effler}, {Eguchi}, {Ehrens}, {Eichholz}, {Eikenberry}, {Eisenmann}, {Eisenstein}, {Enomoto}, {Essick}, {Estelles}, {Estevez}, {Etienne}, {Etzel}, {Evans}, {Evans}, {Fafone}, {Fair}, {Fairhurst}, {Fan}, {Farinon}, {Farr}, {Farr}, {Fauchon-Jones}, {Favata}, {Fays}, {Fazio}, {Fee}, {Feicht}, {Fejer}, {Feng}, {Fernandez-Galiana}, {Ferrante}, {Ferreira}, {Ferreira}, {Ferrini}, {Fidecaro}, {Fiori}, {Fiorucci}, {Fishbach}, {Fisher}, {Fishner}, {Fitz-Axen}, {Flaminio}, {Fletcher}, {Flynn}, {Fong}, {Font}, {Forsyth}, {Fournier}, {Frasca}, {Frasconi},
  {Frei}, {Freise}, {Frey}, {Frey}, {Fritschel}, {Frolov}, {Fujii}, {Fukunaga}, {Fukushima}, {Fulda}, {Fyffe}, {Gabbard}, {Gadre}, {Gaebel}, {Gair}, {Gammaitoni}, {Ganija}, {Gaonkar}, {Garcia}, {Garc{\'\i}a-Quir{\'o}s}, {Garufi}, {Gateley}, {Gaudio}, {Gaur}, {Gayathri}, {Ge}, {Gemme}, {Genin}, {Gennai}, {George}, {George}, {Gergely}, {Germain}, {Ghonge}, {Ghosh}, {Ghosh}, {Ghosh}, {Giacomazzo}, {Giaime}, {Giardina}, {Giazotto}, {Gill}, {Giordano}, {Glover}, {Godwin}, {Goetz}, {Goetz}, {Goncharov}, {Gonz{\'a}lez}, {Gonzalez Castro}, {Gopakumar}, {Gorodetsky}, {Gossan}, {Gosselin}, {Gouaty}, {Grado}, {Graef}, {Granata}, {Grant}, {Gras}, {Grassia}, {Gray}, {Gray}, {Greco}, {Green}, {Green}, {Gretarsson}, {Groot}, {Grote}, {Grunewald}, {Gruning}, {Guidi}, {Gulati}, {Guo}, {Gupta}, {Gupta}, {Gustafson}, {Gustafson}, {Haegel}, {Hagiwara}, {Haino}, {Halim}, {Hall}, {Hall}, {Hamilton}, {Hammond}, {Haney}, {Hanke}, {Hanks}, {Hanna}, {Hannam}, {Hannuksela}, {Hanson}, {Hardwick}, {Haris}, {Harms}, {Harry}, {Harry},
  {Hasegawa}, {Haster}, {Haughian}, {Hayakawa}, {Hayama}, {Hayes}, {Healy}, {Heidmann}, {Heintze}, {Heitmann}, {Hello}, {Hemming}, {Hendry}, {Heng}, {Hennig}, {Heptonstall}, {Heurs}, {Hild}, {Himemoto}, {Hinderer}, {Hiranuma}, {Hirata}, {Hirose}, {Hoak}, {Hochheim}, {Hofman}, {Holgado}, {Holland}, {Holt}, {Holz}, {Hong}, {Hopkins}, {Horst}, {Hough}, {Howell}, {Hoy}, {Hreibi}, {Hsieh}, {Huang}, {Huang}, {Huang}, {Huerta}, {Huet}, {Hughey}, {Hulko}, {Husa}, {Huttner}, {Huynh-Dinh}, {Idzkowski}, {Iess}, {Ikenoue}, {Imam}, {Inayoshi}, {Ingram}, {Inoue}, {Inta}, {Intini}, {Ioka}, {Irwin}, {Isa}, {Isac}, {Isi}, {Itoh}, {Iyer}, {Izumi}, {Jacqmin}, {Jadhav}, {Jani}, {Janthalur}, {Jaranowski}, {Jenkins}, {Jiang}, {Johnson}, {Jones}, {Jones}, {Jones}, {Jonker}, {Ju}, {Jung}, {Jung}, {Junker}, {Kajita}, {Kalaghatgi}, {Kalogera}, {Kamai}, {Kamiizumi}, {Kanda}, {Kandhasamy}, {Kang}, {Kanner}, {Kapadia}, {Karki}, {Karvinen}, {Kashyap}, {Kasprzack}, {Katsanevas}, {Katsavounidis}, {Katzman}, {Kaufer}, {Kawabe}, {Kawaguchi},
  {Kawai}, {Kawasaki}, {Keerthana}, {K{\'e}f{\'e}lian}, {Keitel}, {Kennedy}, {Key}, {Khalili}, {Khan}, {Khan}, {Khan}, {Khan}, {Khazanov}, {Khursheed}, {Kijbunchoo}, {Kim}, {Kim}, {Kim}, {Kim}, {Kim}, {Kim}, {Kim}, {Kim}, {Kimball}, {Kimura}, {King}, {King}, {Kinley-Hanlon}, {Kirchhoff}, {Kissel}, {Kita}, {Kitazawa}, {Kleybolte}, {Klika}, {Klimenko}, {Knowles}, {Knyazev}, {Koch}, {Koehlenbeck}, {Koekoek}, {Kojima}, {Kokeyama}, {Koley}, {Komori}, {Kondrashov}, {Kong}, {Kontos}, {Koper}, {Korobko}, {Korth}, {Kotake}, {Kowalska}, {Kozak}, {Kozakai}, {Kozu}, {Kringel}, {Krishnendu}, {Kr{\'o}lak}, {Kuehn}, {Kumar}, {Kumar}, {Kumar}, {Kumar}, {Kumar}, {Kume}, {Kuo}, {Kuo}, {Kuo}, {Kuroyanagi}, {Kusayanagi}, {Kutynia}, {Kwak}, {Kwang}, {Lackey}, {Lai}, {Lam}, {Landry}, {Lane}, {Lang}, {Lange}, {Lantz}, {Lanza}, {Lartaux-Vollard}, {Lasky}, {Laxen}, {Lazzarini}, {Lazzaro}, {Leaci}, {Leavey}, {Lecoeuche}, {Lee}, {Lee}, {Lee}, {Lee}, {Lee}, {Lee}, {Lee}, {Lehmann}, {Lenon}, {Leonardi}, {Leroy}, {Letendre}, {Levin},
  {Li}, {Li}, {Li}, {Li}, {Lin}, {Lin}, {Lin}, {Lin}, {Linde}, {Linker}, {Littenberg}, {Liu}, {Liu}, {Liu}, {Lo}, {Lockerbie}, {London}, {Longo}, {Lorenzini}, {Loriette}, {Lormand}, {Losurdo}, {Lough}, {Lousto}, {Lovelace}, {Lower}, {L{\"u}ck}, {Lumaca}, {Lundgren}, {Luo}, {Lynch}, {Ma}, {Macas}, {Macfoy}, {Macinnis}, {MacLeod}, {Macquet}, {Maga{\~n}a-Sandoval}, {Zertuche}, {Magee}, {Majorana}, {Maksimovic}, {Malik}, {Man}, {Mandic}, {Mangano}, {Mansell}, {Manske}, {Mantovani}, {Marchesoni}, {Marchio}, {Marion}, {M{\'a}rka}, {M{\'a}rka}, {Markakis}, {Markosyan}, {Markowitz}, {Maros}, {Marquina}, {Marsat}, {Martelli}, {Martin}, {Martin}, {Martynov}, {Mason}, {Massera}, {Masserot}, {Massinger}, {Masso-Reid}, {Mastrogiovanni}, {Matas}, {Matichard}, {Matone}, {Mavalvala}, {Mazumder}, {McCann}, {McCarthy}, {McClelland}, {McCormick}, {McCuller}, {McGuire}, {McIver}, {McManus}, {McRae}, {McWilliams}, {Meacher}, {Meadors}, {Mehmet}, {Mehta}, {Meidam}, {Melatos}, {Mendell}, {Mercer}, {Mereni}, {Merilh}, {Merzougui},
  {Meshkov}, {Messenger}, {Messick}, {Metzdorff}, {Meyers}, {Miao}, {Michel}, {Michimura}, {Middleton}, {Mikhailov}, {Milano}, {Miller}, {Miller}, {Millhouse}, {Mills}, {Milovich-Goff}, {Minazzoli}, {Minenkov}, {Mio}, {Mishkin}, {Mishra}, {Mistry}, {Mitra}, {Mitrofanov}, {Mitselmakher}, {Mittleman}, {Miyakawa}, {Miyamoto}, {Miyazaki}, {Miyo}, {Miyoki}, {Mo}, {Moffa}, {Mogushi}, {Mohapatra}, {Montani}, {Moore}, {Moraru}, {Moreno}, {Morisaki}, {Moriwaki}, {Mours}, {Mow-Lowry}, {Mukherjee}, {Mukherjee}, {Mukherjee}, {Mukund}, {Mullavey}, {Munch}, {Mu{\~n}iz}, {Muratore}, {Murray}, {Nagano}, {Nagano}, {Nagar}, {Nakamura}, {Nakano}, {Nakano}, {Nakashima}, {Nardecchia}, {Narikawa}, {Naticchioni}, {Nayak}, {Negishi}, {Neilson}, {Nelemans}, {Nelson}, {Nery}, {Neunzert}, {Ng}, {Ng}, {Nguyen}, {Ni}, {Nichols}, {Nishizawa}, {Nissanke}, {Nocera}, {North}, {Nuttall}, {Obergaulinger}, {Oberling}, {O'Brien}, {Obuchi}, {O'Dea}, {Ogaki}, {Ogin}, {Oh}, {Oh}, {Ohashi}, {Ohishi}, {Ohkawa}, {Ohme}, {Ohta}, {Okada}, {Okutomi},
  {Oliver}, {Oohara}, {Ooi}, {Oppermann}, {Oram}, {O'Reilly}, {Ormiston}, {Ortega}, {O'Shaughnessy}, {Oshino}, {Ossokine}, {Ottaway}, {Overmier}, {Owen}, {Pace}, {Pagano}, {Page}, {Pai}, {Pai}, {Palamos}, {Palashov}, {Palomba}, {Pal-Singh}, {Pan}, {Pan}, {Pang}, {Pang}, {Pang}, {Pankow}, {Pannarale}, {Pant}, {Paoletti}, {Paoli}, {Papa}, {Parida}, {Park}, {Parker}, {Pascucci}, {Pasqualetti}, {Passaquieti}, {Passuello}, {Patil}, {Patricelli}, {Pearlstone}, {Pedersen}, {Pedraza}, {Pedurand}, {Pele}, {Arellano}, {Penn}, {Perez}, {Perreca}, {Pfeiffer}, {Phelps}, {Phukon}, {Piccinni}, {Pichot}, {Piergiovanni}, {Pillant}, {Pinard}, {Pinto}, {Pirello}, {Pitkin}, {Poggiani}, {Pong}, {Ponrathnam}, {Popolizio}, {Porter}, {Powell}, {Prajapati}, {Prasad}, {Prasai}, {Prasanna}, {Pratten}, {Prestegard}, {Privitera}, {Prodi}, {Prokhorov}, {Puncken}, {Punturo}, {Puppo}, {P{\"u}rrer}, {Qi}, {Quetschke}, {Quinonez}, {Quintero}, {Quitzow-James}, {Raab}, {Radkins}, {Radulescu}, {Raffai}, {Raja}, {Rajan}, {Rajbhandari},
  {Rakhmanov}, {Ramirez}, {Ramos-Buades}, {Rana}, {Rao}, {Rapagnani}, {Raymond}, {Razzano}, {Read}, {Regimbau}, {Rei}, {Reid}, {Reitze}, {Ren}, {Ricci}, {Richardson}, {Richardson}, {Ricker}, {Riles}, {Rizzo}, {Robertson}, {Robie}, {Robinet}, {Rocchi}, {Rolland}, {Rollins}, {Roma}, {Romanelli}, {Romano}, {Romel}, {Romie}, {Rose}, {Rosi{\'n}ska}, {Rosofsky}, {Ross}, {Rowan}, {R{\"u}diger}, {Ruggi}, {Rutins}, {Ryan}, {Sachdev}, {Sadecki}, {Sago}, {Saito}, {Saito}, {Sakai}, {Sakai}, {Sakamoto}, {Sakellariadou}, {Sakuno}, {Salconi}, {Saleem}, {Samajdar}, {Sammut}, {Sanchez}, {Sanchez}, {Sanchis-Gual}, {Sandberg}, {Sanders}, {Santiago}, {Sarin}, {Sassolas}, {Sathyaprakash}, {Sato}, {Sato}, {Sauter}, {Savage}, {Sawada}, {Schale}, {Scheel}, {Scheuer}, {Schmidt}, {Schnabel}, {Schofield}, {Sch{\"o}nbeck}, {Schreiber}, {Schulte}, {Schutz}, {Schwalbe}, {Scott}, {Scott}, {Seidel}, {Sekiguchi}, {Sekiguchi}, {Sellers}, {Sengupta}, {Sennett}, {Sentenac}, {Sequino}, {Sergeev}, {Setyawati}, {Shaddock}, {Shaffer}, {Shahriar},
  {Shaner}, {Shao}, {Sharma}, {Shawhan}, {Shen}, {Shibagaki}, {Shimizu}, {Shimoda}, {Shimode}, {Shink}, {Shinkai}, {Shishido}, {Shoda}, {Shoemaker}, {Shoemaker}, {Shyamsundar}, {Siellez}, {Sieniawska}, {Sigg}, {Silva}, {Singer}, {Singh}, {Singhal}, {Sintes}, {Sitmukhambetov}, {Skliris}, {Slagmolen}, {Slaven-Blair}, {Smith}, {Smith}, {Somala}, {Somiya}, {Son}, {Sorazu}, {Sorrentino}, {Sotani}, {Souradeep}, {Sowell}, {Spencer}, {Srivastava}, {Srivastava}, {Staats}, {Stachie}, {Standke}, {Steer}, {Steinke}, {Steinlechner}, {Steinlechner}, {Steinmeyer}, {Stevenson}, {Stocks}, {Stone}, {Stops}, {Strain}, {Stratta}, {Strigin}, {Strunk}, {Sturani}, {Stuver}, {Sudhir}, {Sugimoto}, {Summerscales}, {Sun}, {Sunil}, {Suresh}, {Sutton}, {Suzuki}, {Suzuki}, {Swinkels}, {Szczepa{\'n}czyk}, {Tacca}, {Tagoshi}, {Tait}, {Takahashi}, {Takahashi}, {Takamori}, {Takano}, {Takeda}, {Takeda}, {Talbot}, {Talukder}, {Tanaka}, {Tanaka}, {Tanaka}, {Tanaka}, {Tanaka}, {Tanioka}, {Tanner}, {T{\'a}pai}, {Tapia San Martin}, {Taracchini},
  {Tasson}, {Taylor}, {Telada}, {Thies}, {Thomas}, {Thomas}, {Thondapu}, {Thorne}, {Thrane}, {Tiwari}, {Tiwari}, {Tiwari}, {Toland}, {Tomaru}, {Tomigami}, {Tomura}, {Tonelli}, {Tornasi}, {Torres-Forn{\'e}}, {Torrie}, {T{\"o}yr{\"a}}, {Travasso}, {Traylor}, {Tringali}, {Trovato}, {Trozzo}, {Trudeau}, {Tsang}, {Tsang}, {Tse}, {Tso}, {Tsubono}, {Tsuchida}, {Tsukada}, {Tsuna}, {Tsuzuki}, {Tuyenbayev}, {Uchikata}, {Uchiyama}, {Ueda}, {Uehara}, {Ueno}, {Ueshima}, {Ugolini}, {Unnikrishnan}, {Uraguchi}, {Urban}, {Ushiba}, {Usman}, {Vahlbruch}, {Vajente}, {Valdes}, {van Bakel}, {van Beuzekom}, {van den Brand}, {van den Broeck}, {Vander-Hyde}, {van der Schaaf}, {van Heijningen}, {van Putten}, {van Veggel}, {Vardaro}, {Varma}, {Vass}, {Vas{\'u}th}, {Vecchio}, {Vedovato}, {Veitch}, {Veitch}, {Venkateswara}, {Venugopalan}, {Verkindt}, {Vetrano}, {Vicer{\'e}}, {Viets}, {Vine}, {Vinet}, {Vitale}, {Vivanco}, {Vo}, {Vocca}, {Vorvick}, {Vyatchanin}, {Wade}, {Wade}, {Wade}, {Walet}, {Walker}, {Wallace}, {Walsh}, {Wang}, {Wang},
  {Wang}, {Wang}, {Wang}, {Wang}, {Ward}, {Warden}, {Warner}, {Was}, {Watchi}, {Weaver}, {Wei}, {Weinert}, {Weinstein}, {Weiss}, {Wellmann}, {Wen}, {Wessel}, {We{\ss}els}, {Westhouse}, {Wette}, {Whelan}, {Whiting}, {Whittle}, {Wilken}, {Williams}, {Williamson}, {Willis}, {Willke}, {Wimmer}, {Winkler}, {Wipf}, {Wittel}, {Woan}, {Woehler}, {Wofford}, {Worden}, {Wright}, {Wu}, {Wu}, {Wu}, {Wu}, {Wysocki}, {Xiao}, {Xu}, {Yamada}, {Yamamoto}, {Yamamoto}, {Yamamoto}, {Yamamoto}, {Yancey}, {Yang}, {Yap}, {Yazback}, {Yeeles}, {Yokogawa}, {Yokoyama}, {Yokozawa}, {Yoshioka}, {Yu}, {Yu}, {Yuen}, {Yuzurihara}, {Yvert}, {Zadro{\.z}ny}, {Zanolin}, {Zeidler}, {Zelenova}, {Zendri}, {Zevin}, {Zhang}, {Zhang}, {Zhang}, {Zhao}, {Zhao}, {Zhou}, {Zhou}, {Zhu}, {Zhu}, {Zimmerman}, {Zucker}, {Zweizig}, {Kagra Collaboration}, \& {VIRGO Collaboration}}]{2020LRR....23....3A}
---. 2020, Living Reviews in Relativity, 23, 3, \dodoi{10.1007/s41114-020-00026-9}

\bibitem[{{Abbott} {et~al.}(2021{\natexlab{a}}){Abbott}, {Abbott}, {Abraham}, {Acernese}, {Ackley}, {Adams}, {Adams}, {Adhikari}, {Adya}, {Affeldt}, {Agathos}, {Agatsuma}, {Aggarwal}, {Aguiar}, {Aiello}, {Ain}, {Ajith}, {Akcay}, {Allen}, {Allocca}, {Altin}, {Amato}, {Anand}, {Ananyeva}, {Anderson}, {Anderson}, {Angelova}, {Ansoldi}, {Antelis}, {Antier}, {Appert}, {Arai}, {Araya}, {Areeda}, {Ar{\`e}ne}, {Arnaud}, {Aronson}, {Arun}, {Asali}, {Ascenzi}, {Ashton}, {Aston}, {Astone}, {Aubin}, {Aufmuth}, {AultONeal}, {Austin}, {Avendano}, {Babak}, {Badaracco}, {Bader}, {Bae}, {Baer}, {Bagnasco}, {Baird}, {Ball}, {Ballardin}, {Ballmer}, {Bals}, {Balsamo}, {Baltus}, {Banagiri}, {Bankar}, {Bankar}, {Barayoga}, {Barbieri}, {Barish}, {Barker}, {Barneo}, {Barnum}, {Barone}, {Barr}, {Barsotti}, {Barsuglia}, {Barta}, {Bartlett}, {Bartos}, {Bassiri}, {Basti}, {Bawaj}, {Bayley}, {Bazzan}, {Becher}, {B{\'e}csy}, {Bedakihale}, {Bejger}, {Belahcene}, {Beniwal}, {Benjamin}, {Bennett}, {Bentley}, {Bergamin}, {Berger}, {Bergmann},
  {Bernuzzi}, {Berry}, {Bersanetti}, {Bertolini}, {Betzwieser}, {Bhandare}, {Bhandari}, {Bhattacharjee}, {Bidler}, {Bilenko}, {Billingsley}, {Birney}, {Birnholtz}, {Biscans}, {Bischi}, {Biscoveanu}, {Bisht}, {Bitossi}, {Bizouard}, {Blackburn}, {Blackman}, {Blair}, {Blair}, {Blair}, {Blanch}, {Bobba}, {Bode}, {Boer}, {Boetzel}, {Bogaert}, {Boldrini}, {Bondu}, {Bonilla}, {Bonnand}, {Booker}, {Boom}, {Bork}, {Boschi}, {Bose}, {Bossilkov}, {Boudart}, {Bouffanais}, {Bozzi}, {Bradaschia}, {Brady}, {Bramley}, {Branchesi}, {Brau}, {Breschi}, {Briant}, {Briggs}, {Brighenti}, {Brillet}, {Brinkmann}, {Brockill}, {Brooks}, {Brooks}, {Brown}, {Brunett}, {Bruno}, {Bruntz}, {Buikema}, {Bulik}, {Bulten}, {Buonanno}, {Buscicchio}, {Buskulic}, {Byer}, {Cabero}, {Cadonati}, {Caesar}, {Cagnoli}, {Cahillane}, {Calder{\'o}n Bustillo}, {Callaghan}, {Callister}, {Calloni}, {Camp}, {Canepa}, {Cannon}, {Cao}, {Cao}, {Carapella}, {Carbognani}, {Carney}, {Carpinelli}, {Carullo}, {Carver}, {Casanueva Diaz}, {Casentini}, {Caudill},
  {Cavagli{\`a}}, {Cavalier}, {Cavalieri}, {Cella}, {Cerd{\'a}-Dur{\'a}n}, {Cesarini}, {Chaibi}, {Chakravarti}, {Chan}, {Chan}, {Chandra}, {Chanial}, {Chao}, {Charlton}, {Chase}, {Chassande-Mottin}, {Chatterjee}, {Chattopadhyay}, {Chaturvedi}, {Chatziioannou}, {Chen}, {Chen}, {Chen}, {Chen}, {Cheng}, {Cheong}, {Chia}, {Chiadini}, {Chierici}, {Chincarini}, {Chiummo}, {Cho}, {Cho}, {Cho}, {Choate}, {Christensen}, {Chu}, {Chua}, {Chung}, {Chung}, {Ciani}, {Ciecielag}, {Cie{\'s}lar}, {Cifaldi}, {Ciobanu}, {Ciolfi}, {Cipriano}, {Cirone}, {Clara}, {Clark}, {Clark}, {Clarke}, {Clearwater}, {Clesse}, {Cleva}, {Coccia}, {Cohadon}, {Cohen}, {Colleoni}, {Collette}, {Collins}, {Colpi}, {Constancio}, {Conti}, {Cooper}, {Corban}, {Corbitt}, {Cordero-Carri{\'o}n}, {Corezzi}, {Corley}, {Cornish}, {Corre}, {Corsi}, {Cortese}, {Costa}, {Cotesta}, {Coughlin}, {Coughlin}, {Coulon}, {Countryman}, {Cousins}, {Couvares}, {Covas}, {Coward}, {Cowart}, {Coyne}, {Coyne}, {Creighton}, {Creighton}, {Croquette}, {Crowder}, {Cudell},
  {Cullen}, {Cumming}, {Cummings}, {Cunningham}, {Cuoco}, {Cury{\l}o}, {Canton}, {D{\'a}lya}, {Dana}, {DaneshgaranBajastani}, {D'Angelo}, {Danila}, {Danilishin}, {D'Antonio}, {Danzmann}, {Darsow-Fromm}, {Dasgupta}, {Datrier}, {Dattilo}, {Dave}, {Davier}, {Davies}, {Davis}, {Daw}, {Dean}, {DeBra}, {Deenadayalan}, {Degallaix}, {De Laurentis}, {Del{\'e}glise}, {Del Favero}, {De Lillo}, {De Lillo}, {Del Pozzo}, {DeMarchi}, {De Matteis}, {D'Emilio}, {Demos}, {Denker}, {Dent}, {Depasse}, {De Pietri}, {De Rosa}, {De Rossi}, {DeSalvo}, {de Varona}, {Dhurandhar}, {D{\'\i}az}, {Diaz-Ortiz}, {Didio}, {Dietrich}, {Di Fiore}, {DiFronzo}, {Di Giorgio}, {Di Giovanni}, {Di Giovanni}, {Di Girolamo}, {Di Lieto}, {Ding}, {Di Pace}, {Di Palma}, {Di Renzo}, {Divakarla}, {Dmitriev}, {Doctor}, {D'Onofrio}, {Donovan}, {Dooley}, {Doravari}, {Dorrington}, {Downes}, {Drago}, {Driggers}, {Du}, {Ducoin}, {Dupej}, {Durante}, {D'Urso}, {Duverne}, {Dwyer}, {Easter}, {Eddolls}, {Edelman}, {Edo}, {Edy}, {Effler}, {Eichholz}, {Eikenberry},
  {Eisenmann}, {Eisenstein}, {Ejlli}, {Errico}, {Essick}, {Estell{\'e}s}, {Estevez}, {Etienne}, {Etzel}, {Evans}, {Evans}, {Ewing}, {Fafone}, {Fair}, {Fairhurst}, {Fan}, {Farah}, {Farinon}, {Farr}, {Farr}, {Fauchon-Jones}, {Favata}, {Fays}, {Fazio}, {Feicht}, {Fejer}, {Feng}, {Fenyvesi}, {Ferguson}, {Fernandez-Galiana}, {Ferrante}, {Ferreira}, {Fidecaro}, {Figura}, {Fiori}, {Fiorucci}, {Fishbach}, {Fisher}, {Fishner}, {Fittipaldi}, {Fitz-Axen}, {Fiumara}, {Flaminio}, {Floden}, {Flynn}, {Fong}, {Font}, {Forsyth}, {Fournier}, {Frasca}, {Frasconi}, {Frei}, {Freise}, {Frey}, {Frey}, {Fritschel}, {Frolov}, {Fronz{\'e}}, {Fulda}, {Fyffe}, {Gabbard}, {Gadre}, {Gaebel}, {Gair}, {Gais}, {Galaudage}, {Gamba}, {Ganapathy}, {Ganguly}, {Gaonkar}, {Garaventa}, {Garc{\'\i}a-Quir{\'o}s}, {Garufi}, {Gateley}, {Gaudio}, {Gayathri}, {Gemme}, {Gennai}, {George}, {George}, {George}, {Gergely}, {Ghonge}, {Ghosh}, {Ghosh}, {Ghosh}, {Giacomazzo}, {Giacoppo}, {Giaime}, {Giardina}, {Gibson}, {Gier}, {Gill}, {Giri}, {Glanzer},
  {Gleckl}, {Godwin}, {Goetz}, {Goetz}, {Gohlke}, {Goncharov}, {Gonz{\'a}lez}, {Gopakumar}, {Gossan}, {Gosselin}, {Gouaty}, {Grace}, {Grado}, {Granata}, {Granata}, {Grant}, {Gras}, {Grassia}, {Gray}, {Gray}, {Greco}, {Green}, {Green}, {Gretarsson}, {Griggs}, {Grignani}, {Grimaldi}, {Grimes}, {Grimm}, {Grote}, {Grunewald}, {Gruning}, {Guerrero}, {Guidi}, {Guimaraes}, {Guix{\'e}}, {Gulati}, {Guo}, {Gupta}, {Gupta}, {Gupta}, {Gustafson}, {Gustafson}, {Guzman}, {Haegel}, {Halim}, {Hall}, {Hamilton}, {Hammond}, {Haney}, {Hanke}, {Hanks}, {Hanna}, {Hannam}, {Hannuksela}, {Hannuksela}, {Hansen}, {Hansen}, {Hanson}, {Harder}, {Hardwick}, {Haris}, {Harms}, {Harry}, {Harry}, {Hartwig}, {Hasskew}, {Haster}, {Haughian}, {Hayes}, {Healy}, {Heidmann}, {Heintze}, {Heinze}, {Heinzel}, {Heitmann}, {Hellman}, {Hello}, {Helmling-Cornell}, {Hemming}, {Hendry}, {Heng}, {Hennes}, {Hennig}, {Hennig}, {Hernandez Vivanco}, {Heurs}, {Hild}, {Hill}, {Hines}, {Hochheim}, {Hofgard}, {Hofman}, {Hohmann}, {Holgado}, {Holland}, {Hollows},
  {Holmes}, {Holt}, {Holz}, {Hopkins}, {Horst}, {Hough}, {Howell}, {Hoy}, {Hoyland}, {Huang}, {H{\"u}bner}, {Huddart}, {Huerta}, {Hughey}, {Hui}, {Husa}, {Huttner}, {Hutzler}, {Huxford}, {Huynh-Dinh}, {Idzkowski}, {Iess}, {Imperato}, {Inchauspe}, {Ingram}, {Intini}, {Isi}, {Iyer}, {JaberianHamedan}, {Jacqmin}, {Jadhav}, {Jadhav}, {James}, {Jani}, {Janssens}, {Janthalur}, {Jaranowski}, {Jariwala}, {Jaume}, {Jenkins}, {Jeunon}, {Jiang}, {Johns}, {Johnson-McDaniel}, {Jones}, {Jones}, {Jones}, {Jones}, {Jones}, {Jonker}, {Ju}, {Junker}, {Kalaghatgi}, {Kalogera}, {Kamai}, {Kandhasamy}, {Kang}, {Kanner}, {Kapadia}, {Kapasi}, {Karathanasis}, {Karki}, {Kashyap}, {Kasprzack}, {Kastaun}, {Katsanevas}, {Katsavounidis}, {Katzman}, {Kawabe}, {K{\'e}f{\'e}lian}, {Keitel}, {Key}, {Khadka}, {Khalili}, {Khan}, {Khan}, {Khazanov}, {Khetan}, {Khursheed}, {Kijbunchoo}, {Kim}, {Kim}, {Kim}, {Kim}, {Kim}, {Kim}, {Kimball}, {King}, {Kinley-Hanlon}, {Kirchhoff}, {Kissel}, {Kleybolte}, {Klimenko}, {Knowles}, {Knyazev}, {Koch},
  {Koehlenbeck}, {Koekoek}, {Koley}, {Kolstein}, {Komori}, {Kondrashov}, {Kontos}, {Koper}, {Korobko}, {Korth}, {Kovalam}, {Kozak}, {Kr{\"a}mer}, {Kringel}, {Krishnendu}, {Kr{\'o}lak}, {Kuehn}, {Kumar}, {Kumar}, {Kumar}, {Kumar}, {Kuns}, {Kwang}, {Lackey}, {Laghi}, {Lalande}, {Lam}, {Lamberts}, {Landry}, {Lane}, {Lang}, {Lange}, {Lantz}, {Lanza}, {La Rosa}, {Lartaux-Vollard}, {Lasky}, {Laxen}, {Lazzarini}, {Lazzaro}, {Leaci}, {Leavey}, {Lecoeuche}, {Lee}, {Lee}, {Lee}, {Lee}, {Lehmann}, {Leon}, {Leroy}, {Letendre}, {Levin}, {Li}, {Li}, {Li}, {Li}, {Li}, {Linde}, {Linker}, {Linley}, {Littenberg}, {Liu}, {Liu}, {Llorens-Monteagudo}, {Lo}, {Lockwood}, {London}, {Longo}, {Lorenzini}, {Loriette}, {Lormand}, {Losurdo}, {Lough}, {Lousto}, {Lovelace}, {L{\"u}ck}, {Lumaca}, {Lundgren}, {Ma}, {Macas}, {MacInnis}, {Macleod}, {MacMillan}, {Macquet}, {Maga{\~n}a Hernandez}, {Maga{\~n}a-Sandoval}, {Magazz{\`u}}, {Magee}, {Majorana}, {Maksimovic}, {Maliakal}, {Malik}, {Man}, {Mandic}, {Mangano}, {Mansell}, {Manske},
  {Mantovani}, {Mapelli}, {Marchesoni}, {Marion}, {M{\'a}rka}, {M{\'a}rka}, {Markakis}, {Markosyan}, {Markowitz}, {Maros}, {Marquina}, {Marsat}, {Martelli}, {Martin}, {Martin}, {Martinez}, {Martinez}, {Martynov}, {Masalehdan}, {Mason}, {Massera}, {Masserot}, {Massinger}, {Masso-Reid}, {Mastrogiovanni}, {Matas}, {Mateu-Lucena}, {Matichard}, {Matiushechkina}, {Mavalvala}, {Maynard}, {McCann}, {McCarthy}, {McClelland}, {McCormick}, {McCuller}, {McGuire}, {McIsaac}, {McIver}, {McManus}, {McRae}, {McWilliams}, {Meacher}, {Meadors}, {Mehmet}, {Mehta}, {Melatos}, {Melchor}, {Mendell}, {Menendez-Vazquez}, {Mercer}, {Mereni}, {Merfeld}, {Merilh}, {Merritt}, {Merzougui}, {Meshkov}, {Messenger}, {Messick}, {Metzdorff}, {Meyers}, {Meylahn}, {Mhaske}, {Miani}, {Miao}, {Michaloliakos}, {Michel}, {Middleton}, {Milano}, {Miller}, {Millhouse}, {Mills}, {Milotti}, {Milovich-Goff}, {Minazzoli}, {Minenkov}, {Mir}, {Mishkin}, {Mishra}, {Mistry}, {Mitra}, {Mitrofanov}, {Mitselmakher}, {Mittleman}, {Mo}, {Mogushi}, {Mohapatra},
  {Mohite}, {Molina}, {Molina-Ruiz}, {Mondin}, {Montani}, {Moore}, {Moraru}, {Morawski}, {Moreno}, {Morisaki}, {Mours}, {Mow-Lowry}, {Mozzon}, {Muciaccia}, {Mukherjee}, {Mukherjee}, {Mukherjee}, {Mukherjee}, {Mukund}, {Mullavey}, {Munch}, {Mu{\~n}iz}, {Murray}, {Nadji}, {Nagar}, {Nardecchia}, {Naticchioni}, {Nayak}, {Neil}, {Neilson}, {Nelemans}, {Nelson}, {Nery}, {Neunzert}, {Nitz}, {Ng}, {Ng}, {Nguyen}, {Nguyen}, {Nguyen}, {Nichols}, {Nissanke}, {Nocera}, {Noh}, {North}, {Nothard}, {Nuttall}, {Oberling}, {O'Brien}, {O'Dell}, {Oganesyan}, {Ogin}, {Oh}, {Oh}, {Ohme}, {Ohta}, {Okada}, {Olivetto}, {Oppermann}, {Oram}, {O'Reilly}, {Ormiston}, {Ortega}, {O'Shaughnessy}, {Ossokine}, {Osthelder}, {Ottaway}, {Overmier}, {Owen}, {Pace}, {Pagano}, {Page}, {Pagliaroli}, {Pai}, {Pai}, {Palamos}, {Palashov}, {Palomba}, {Pan}, {Panda}, {Pang}, {Pankow}, {Pannarale}, {Pant}, {Paoletti}, {Paoli}, {Paolone}, {Parker}, {Pascucci}, {Pasqualetti}, {Passaquieti}, {Passuello}, {Patel}, {Patricelli}, {Payne}, {Pechsiri},
  {Pedraza}, {Pegoraro}, {Pele}, {Penn}, {Perego}, {Perez}, {P{\'e}rigois}, {Perreca}, {Perri{\`e}s}, {Petermann}, {Petterson}, {Pfeiffer}, {Pham}, {Phukon}, {Piccinni}, {Pichot}, {Piendibene}, {Piergiovanni}, {Pierini}, {Pierro}, {Pillant}, {Pilo}, {Pinard}, {Pinto}, {Piotrzkowski}, {Pirello}, {Pitkin}, {Placidi}, {Plastino}, {Pluchar}, {Poggiani}, {Polini}, {Pong}, {Ponrathnam}, {Popolizio}, {Porter}, {Poverman}, {Powell}, {Pracchia}, {Prajapati}, {Prasai}, {Prasanna}, {Pratten}, {Prestegard}, {Principe}, {Prodi}, {Prokhorov}, {Prosposito}, {Prudenzi}, {Puecher}, {Punturo}, {Puosi}, {Puppo}, {P{\"u}rrer}, {Qi}, {Quetschke}, {Quinonez}, {Quitzow-James}, {Raab}, {Raaijmakers}, {Radkins}, {Radulesco}, {Raffai}, {Rafferty}, {Rail}, {Raja}, {Rajan}, {Rajbhandari}, {Rakhmanov}, {Ramirez}, {Ramirez}, {Ramos-Buades}, {Rana}, {Rao}, {Rapagnani}, {Rapol}, {Ratto}, {Raymond}, {Razzano}, {Read}, {Regimbau}, {Rei}, {Reid}, {Reitze}, {Rettegno}, {Ricci}, {Richardson}, {Richardson}, {Richardson}, {Ricker},
  {Riemenschneider}, {Riles}, {Rizzo}, {Robertson}, {Robinet}, {Rocchi}, {Rocha}, {Rodriguez}, {Rodriguez-Soto}, {Rolland}, {Rollins}, {Roma}, {Romanelli}, {Romano}, {Romel}, {Romero}, {Romero-Shaw}, {Romie}, {Ronchini}, {Rose}, {Rose}, {Rose}, {Rosell}, {Rosi{\'n}ska}, {Rosofsky}, {Ross}, {Rowan}, {Rowlinson}, {Roy}, {Roy}, {Ruggi}, {Ryan}, {Sachdev}, {Sadecki}, {Sadiq}, {Sakellariadou}, {Salafia}, {Salconi}, {Saleem}, {Samajdar}, {Sanchez}, {Sanchez}, {Sanchez}, {Sanchis-Gual}, {Sanders}, {Sandles}, {Santiago}, {Santos}, {Saravanan}, {Sarin}, {Sassolas}, {Sathyaprakash}, {Sauter}, {Savage}, {Savant}, {Sawant}, {Sayah}, {Schaetzl}, {Schale}, {Scheel}, {Scheuer}, {Schindler-Tyka}, {Schmidt}, {Schnabel}, {Schofield}, {Sch{\"o}nbeck}, {Schreiber}, {Schulte}, {Schutz}, {Schwarm}, {Schwartz}, {Scott}, {Scott}, {Seglar-Arroyo}, {Seidel}, {Sellers}, {Sengupta}, {Sennett}, {Sentenac}, {Sequino}, {Sergeev}, {Setyawati}, {Shaffer}, {Shahriar}, {Sharifi}, {Sharma}, {Sharma}, {Shawhan}, {Shen}, {Shikauchi}, {Shink},
  {Shoemaker}, {Shoemaker}, {Shukla}, {ShyamSundar}, {Sieniawska}, {Sigg}, {Singer}, {Singh}, {Singh}, {Singha}, {Singhal}, {Sintes}, {Sipala}, {Skliris}, {Slagmolen}, {Slaven-Blair}, {Smetana}, {Smith}, {Smith}, {Somala}, {Son}, {Soni}, {Soni}, {Sorazu}, {Sordini}, {Sorrentino}, {Sorrentino}, {Soulard}, {Souradeep}, {Sowell}, {Spencer}, {Spera}, {Srivastava}, {Srivastava}, {Staats}, {Stachie}, {Steer}, {Steinhoff}, {Steinke}, {Steinlechner}, {Steinlechner}, {Steinmeyer}, {Stevenson}, {Stolle-McAllister}, {Stops}, {Stover}, {Strain}, {Stratta}, {Strunk}, {Sturani}, {Stuver}, {S{\"u}dbeck}, {Sudhagar}, {Sudhir}, {Suh}, {Summerscales}, {Sun}, {Sun}, {Sunil}, {Sur}, {Suresh}, {Sutton}, {Swinkels}, {Szczepa{\'n}czyk}, {Tacca}, {Tait}, {Talbot}, {Tanasijczuk}, {Tanner}, {Tao}, {Tapia}, {Tapia San Martin}, {Tasson}, {Taylor}, {Tenorio}, {Terkowski}, {Thirugnanasambandam}, {Thomas}, {Thomas}, {Thomas}, {Thompson}, {Thondapu}, {Thorne}, {Thrane}, {Tiwari}, {Tiwari}, {Tiwari}, {Toland}, {Tolley}, {Tonelli}, {Tornasi},
  {Torres-Forn{\'e}}, {Torrie}, {e Melo}, {T{\"o}yr{\"a}}, {Tran}, {Trapananti}, {Travasso}, {Traylor}, {Tringali}, {Tripathee}, {Trovato}, {Trudeau}, {Tsai}, {Tsang}, {Tse}, {Tso}, {Tsukada}, {Tsuna}, {Tsutsui}, {Turconi}, {Ubhi}, {Udall}, {Ueno}, {Ugolini}, {Unnikrishnan}, {Urban}, {Usman}, {Utina}, {Vahlbruch}, {Vajente}, {Vajpeyi}, {Valdes}, {Valentini}, {Valsan}, {van Bakel}, {van Beuzekom}, {van den Brand}, {Van Den Broeck}, {Vander-Hyde}, {van der Schaaf}, {van Heijningen}, {Vardaro}, {Vargas}, {Varma}, {Vass}, {Vas{\'u}th}, {Vecchio}, {Vedovato}, {Veitch}, {Veitch}, {Venkateswara}, {Venneberg}, {Venugopalan}, {Verkindt}, {Verma}, {Veske}, {Vetrano}, {Vicer{\'e}}, {Viets}, {Vijaykumar}, {Villa-Ortega}, {Vinet}, {Vitale}, {Vo}, {Vocca}, {Vorvick}, {Vyatchanin}, {Wade}, {Wade}, {Wade}, {Walet}, {Walker}, {Wallace}, {Wallace}, {Walsh}, {Wang}, {Wang}, {Wang}, {Wang}, {Ward}, {Warner}, {Was}, {Washington}, {Watchi}, {Weaver}, {Wei}, {Weinert}, {Weinstein}, {Weiss}, {Wellmann}, {Wen}, {We{\ss}els},
  {Westhouse}, {Wette}, {Whelan}, {White}, {White}, {Whiting}, {Whittle}, {Wilken}, {Williams}, {Williams}, {Williamson}, {Willis}, {Willke}, {Wilson}, {Wimmer}, {Winkler}, {Wipf}, {Woan}, {Woehler}, {Wofford}, {Wong}, {Wrangel}, {Wright}, {Wu}, {Wysocki}, {Xiao}, {Yamamoto}, {Yang}, {Yang}, {Yang}, {Yap}, {Yeeles}, {Yoon}, {Yu}, {Yu}, {Yuen}, {Zadro{\.Z}ny}, {Zanolin}, {Zelenova}, {Zendri}, {Zevin}, {Zhang}, {Zhang}, {Zhang}, {Zhang}, {Zhao}, {Zhao}, {Zheng}, {Zhou}, {Zhou}, {Zhu}, {Zimmerman}, {Zlochower}, {Zucker}, {Zweizig}, {LIGO Scientific Collaboration}, \& {Virgo Collaboration}}]{2021PhRvX..11b1053A}
{Abbott}, R., {Abbott}, T.~D., {Abraham}, S., {et~al.} 2021{\natexlab{a}}, Physical Review X, 11, 021053, \dodoi{10.1103/PhysRevX.11.021053}

\bibitem[{{Abbott} {et~al.}(2021{\natexlab{b}}){Abbott}, {Abbott}, {Abraham}, {Acernese}, {Ackley}, {Adams}, {Adams}, {Adhikari}, {et~al.}}]{2021ApJ...913L...7A}
---. 2021{\natexlab{b}}, \apjl, 913, L7, \dodoi{10.3847/2041-8213/abe949}

\bibitem[{{Abbott} {et~al.}(2023{\natexlab{a}}){Abbott}, {Abbott}, {Acernese}, {Ackley}, {Adams}, {Adhikari}, {Adhikari}, {Adya}, {Affeldt}, {Agarwal}, {Agathos}, {Agatsuma}, {Aggarwal}, {Aguiar}, {Aiello}, {Ain}, {Ajith}, {Akcay}, {Akutsu}, {Albanesi}, {Allocca}, {Altin}, {Amato}, {Anand}, {Anand}, {Ananyeva}, {Anderson}, {Anderson}, {Ando}, {Andrade}, {Andres}, {Andri{\'c}}, {Angelova}, {Ansoldi}, {Antelis}, {Antier}, {Appert}, {Arai}, {Arai}, {Arai}, {Araki}, {Araya}, {Araya}, {Areeda}, {Ar{\`e}ne}, {Aritomi}, {Arnaud}, {Arogeti}, {Aronson}, {Arun}, {Asada}, {Asali}, {Ashton}, {Aso}, {Assiduo}, {Aston}, {Astone}, {Aubin}, {Austin}, {Babak}, {Badaracco}, {Bader}, {Badger}, {Bae}, {Bae}, {Baer}, {Bagnasco}, {Bai}, {Baiotti}, {Baird}, {Bajpai}, {Ball}, {Ballardin}, {Ballmer}, {Balsamo}, {Baltus}, {Banagiri}, {Bankar}, {Barayoga}, {Barbieri}, {Barish}, {Barker}, {Barneo}, {Barone}, {Barr}, {Barsotti}, {Barsuglia}, {Barta}, {Bartlett}, {Barton}, {Bartos}, {Bassiri}, {Basti}, {Bawaj}, {Bayley}, {Baylor},
  {Bazzan}, {B{\'e}csy}, {Bedakihale}, {Bejger}, {Belahcene}, {Benedetto}, {Beniwal}, {Bennett}, {Bentley}, {Benyaala}, {Bergamin}, {Berger}, {Bernuzzi}, {Berry}, {Bersanetti}, {Bertolini}, {Betzwieser}, {Beveridge}, {Bhandare}, {Bhardwaj}, {Bhattacharjee}, {Bhaumik}, {Bilenko}, {Billingsley}, {Bini}, {Birney}, {Birnholtz}, {Biscans}, {Bischi}, {Biscoveanu}, {Bisht}, {Biswas}, {Bitossi}, {Bizouard}, {Blackburn}, {Blair}, {Blair}, {Blair}, {Bobba}, {Bode}, {Boer}, {Bogaert}, {Boldrini}, {Bonavena}, {Bondu}, {Bonilla}, {Bonnand}, {Booker}, {Boom}, {Bork}, {Boschi}, {Bose}, {Bose}, {Bossilkov}, {Boudart}, {Bouffanais}, {Bozzi}, {Bradaschia}, {Brady}, {Bramley}, {Branch}, {Branchesi}, {Brandt}, {Brau}, {Breschi}, {Briant}, {Briggs}, {Brillet}, {Brinkmann}, {Brockill}, {Brooks}, {Brooks}, {Brown}, {Brunett}, {Bruno}, {Bruntz}, {Bryant}, {Bulik}, {Bulten}, {Buonanno}, {Buscicchio}, {Buskulic}, {Buy}, {Byer}, {Davies}, {Cadonati}, {Cagnoli}, {Cahillane}, {Bustillo}, {Callaghan}, {Callister}, {Calloni}, {Cameron},
  {Camp}, {Canepa}, {Canevarolo}, {Cannavacciuolo}, {Cannon}, {Cao}, {Cao}, {Capocasa}, {Capote}, {Carapella}, {Carbognani}, {Carlin}, {Carney}, {Carpinelli}, {Carrillo}, {Carullo}, {Carver}, {Diaz}, {Casentini}, {Castaldi}, {Caudill}, {Cavagli{\`a}}, {Cavalier}, {Cavalieri}, {Ceasar}, {Cella}, {Cerd{\'a}-Dur{\'a}n}, {Cesarini}, {Chaibi}, {Chakravarti}, {Subrahmanya}, {Champion}, {Chan}, {Chan}, {Chan}, {Chan}, {Chan}, {Chandra}, {Chanial}, {Chao}, {Chapman-Bird}, {Charlton}, {Chase}, {Chassande-Mottin}, {Chatterjee}, {Chatterjee}, {Chatterjee}, {Chaturvedi}, {Chaty}, {Chatziioannou}, {Chen}, {Chen}, {Chen}, {Chen}, {Chen}, {Chen}, {Chen}, {Chen}, {Cheng}, {Cheong}, {Cheung}, {Chia}, {Chiadini}, {Chiang}, {Chiarini}, {Chierici}, {Chincarini}, {Chiofalo}, {Chiummo}, {Cho}, {Cho}, {Choudhary}, {Choudhary}, {Christensen}, {Chu}, {Chu}, {Chu}, {Chua}, {Chung}, {Ciani}, {Ciecielag}, {Cie{\'s}lar}, {Cifaldi}, {Ciobanu}, {Ciolfi}, {Cipriano}, {Cirone}, {Clara}, {Clark}, {Clark}, {Clarke}, {Clearwater}, {Clesse},
  {Cleva}, {Coccia}, {Codazzo}, {Cohadon}, {Cohen}, {Cohen}, {Colleoni}, {Collette}, {Colombo}, {Colpi}, {Compton}, {Constancio}, {Conti}, {Cooper}, {Corban}, {Corbitt}, {Cordero-Carri{\'o}n}, {Corezzi}, {Corley}, {Cornish}, {Corre}, {Corsi}, {Cortese}, {Costa}, {Cotesta}, {Coughlin}, {Coulon}, {Countryman}, {Cousins}, {Couvares}, {Coward}, {Cowart}, {Coyne}, {Coyne}, {Creighton}, {Creighton}, {Criswell}, {Croquette}, {Crowder}, {Cudell}, {Cullen}, {Cumming}, {Cummings}, {Cunningham}, {Cuoco}, {Cury{\l}o}, {Dabadie}, {Canton}, {Dall'Osso}, {D{\'a}lya}, {Dana}, {Daneshgaranbajastani}, {D'Angelo}, {Danila}, {Danilishin}, {D'Antonio}, {Danzmann}, {Darsow-Fromm}, {Dasgupta}, {Datrier}, {Dattilo}, {Dave}, {Davier}, {Davis}, {Davis}, {Daw}, {de Alarc{\'o}n}, {Dean}, {Debra}, {Deenadayalan}, {Degallaix}, {de Laurentis}, {Del{\'e}glise}, {Del Favero}, {de Lillo}, {de Lillo}, {Del Pozzo}, {Demarchi}, {de Matteis}, {D'Emilio}, {Demos}, {Dent}, {Depasse}, {de Pietri}, {De Rosa}, {de Rossi}, {Desalvo}, {de Simone},
  {Dhurandhar}, {D{\'\i}az}, {Diaz-Ortiz}, {Didio}, {Dietrich}, {di Fiore}, {di Fronzo}, {di Giorgio}, {di Giovanni}, {di Giovanni}, {di Girolamo}, {di Lieto}, {Ding}, {di Pace}, {di Palma}, {di Renzo}, {Divakarla}, {Dmitriev}, {Doctor}, {D'Onofrio}, {Donovan}, {Dooley}, {Doravari}, {Dorrington}, {Drago}, {Driggers}, {Drori}, {Ducoin}, {Dupej}, {Durante}, {D'Urso}, {Duverne}, {Dwyer}, {Eassa}, {Easter}, {Ebersold}, {Eckhardt}, {Eddolls}, {Edelman}, {Edo}, {Edy}, {Effler}, {Eguchi}, {Eichholz}, {Eikenberry}, {Eisenmann}, {Eisenstein}, {Ejlli}, {Engelby}, {Enomoto}, {Errico}, {Essick}, {Estell{\'e}s}, {Estevez}, {Etienne}, {Etzel}, {Evans}, {Evans}, {Ewing}, {Fafone}, {Fair}, {Fairhurst}, {Farah}, {Farinon}, {Farr}, {Farr}, {Farrow}, {Fauchon-Jones}, {Favaro}, {Favata}, {Fays}, {Fazio}, {Feicht}, {Fejer}, {Fenyvesi}, {Ferguson}, {Fernandez-Galiana}, {Ferrante}, {Ferreira}, {Fidecaro}, {Figura}, {Fiori}, {Fishbach}, {Fisher}, {Fittipaldi}, {Fiumara}, {Flaminio}, {Floden}, {Fong}, {Font}, {Fornal}, {Forsyth},
  {Franke}, {Frasca}, {Frasconi}, {Frederick}, {Freed}, {Frei}, {Freise}, {Frey}, {Fritschel}, {Frolov}, {Fronz{\'e}}, {Fujii}, {Fujikawa}, {Fukunaga}, {Fukushima}, {Fulda}, {Fyffe}, {Gabbard}, {Gabella}, {Gadre}, {Gair}, {Gais}, {Galaudage}, {Gamba}, {Ganapathy}, {Ganguly}, {Gao}, {Gaonkar}, {Garaventa}, {Garc{\'\i}a}, {Garc{\'\i}a-N{\'u}{\~n}ez}, {Garc{\'\i}a-Quir{\'o}s}, {Garufi}, {Gateley}, {Gaudio}, {Gayathri}, {Ge}, {Gemme}, {Gennai}, {George}, {George}, {Gerberding}, {Gergely}, {Gewecke}, {Ghonge}, {Ghosh}, {Ghosh}, {Ghosh}, {Ghosh}, {Giacomazzo}, {Giacoppo}, {Giaime}, {Giardina}, {Gibson}, {Gier}, {Giesler}, {Giri}, {Gissi}, {Glanzer}, {Gleckl}, {Godwin}, {Goetz}, {Goetz}, {Gohlke}, {Golomb}, {Goncharov}, {Gonz{\'a}lez}, {Gopakumar}, {Gosselin}, {Gouaty}, {Gould}, {Grace}, {Grado}, {Granata}, {Granata}, {Grant}, {Gras}, {Grassia}, {Gray}, {Gray}, {Greco}, {Green}, {Green}, {Gretarsson}, {Gretarsson}, {Griffith}, {Griffiths}, {Griggs}, {Grignani}, {Grimaldi}, {Grimm}, {Grote}, {Grunewald}, {Gruning},
  {Guerra}, {Guidi}, {Guimaraes}, {Guix{\'e}}, {Gulati}, {Guo}, {Guo}, {Gupta}, {Gupta}, {Gupta}, {Gustafson}, {Gustafson}, {Guzman}, {Ha}, {Haegel}, {Hagiwara}, {Haino}, {Halim}, {Hall}, {Hamilton}, {Hammond}, {Han}, {Haney}, {Hanks}, {Hanna}, {Hannam}, {Hannuksela}, {Hansen}, {Hansen}, {Hanson}, {Harder}, {Hardwick}, {Haris}, {Harms}, {Harry}, {Harry}, {Hartwig}, {Hasegawa}, {Haskell}, {Hasskew}, {Haster}, {Hattori}, {Haughian}, {Hayakawa}, {Hayama}, {Hayes}, {Healy}, {Heidmann}, {Heidt}, {Heintze}, {Heinze}, {Heinzel}, {Heitmann}, {Hellman}, {Hello}, {Helmling-Cornell}, {Hemming}, {Hendry}, {Heng}, {Hennes}, {Hennig}, {Hennig}, {Hernandez}, {Hernandez Vivanco}, {Heurs}, {Hild}, {Hill}, {Himemoto}, {Hines}, {Hiranuma}, {Hirata}, {Hirose}, {Hochheim}, {Hofman}, {Hohmann}, {Holcomb}, {Holland}, {Holley-Bockelmann}, {Hollows}, {Holmes}, {Holt}, {Holz}, {Hong}, {Hopkins}, {Hough}, {Hourihane}, {Howell}, {Hoy}, {Hoyland}, {Hreibi}, {Hsieh}, {Hsu}, {Huang}, {Huang}, {Huang}, {Huang}, {Huang}, {Huang},
  {H{\"u}bner}, {Huddart}, {Hughey}, {Hui}, {Hui}, {Husa}, {Huttner}, {Huxford}, {Huynh-Dinh}, {Ide}, {Idzkowski}, {Iess}, {Ikenoue}, {Imam}, {Inayoshi}, {Ingram}, {Inoue}, {Ioka}, {Isi}, {Isleif}, {Ito}, {Itoh}, {Iyer}, {Izumi}, {Jaberianhamedan}, {Jacqmin}, {Jadhav}, {Jadhav}, {James}, {Jan}, {Jani}, {Janquart}, {Janssens}, {Janthalur}, {Jaranowski}, {Jariwala}, {Jaume}, {Jenkins}, {Jenner}, {Jeon}, {Jeunon}, {Jia}, {Jin}, {Johns}, {Johnson-McDaniel}, {Jones}, {Jones}, {Jones}, {Jones}, {Jones}, {Jonker}, {Ju}, {Jung}, {Jung}, {Junker}, {Juste}, {Kaihotsu}, {Kajita}, {Kakizaki}, {Kalaghatgi}, {Kalogera}, {Kamai}, {Kamiizumi}, {Kanda}, {Kandhasamy}, {Kang}, {Kanner}, {Kao}, {Kapadia}, {Kapasi}, {Karat}, {Karathanasis}, {Karki}, {Kashyap}, {Kasprzack}, {Kastaun}, {Katsanevas}, {Katsavounidis}, {Katzman}, {Kaur}, {Kawabe}, {Kawaguchi}, {Kawai}, {Kawasaki}, {K{\'e}f{\'e}lian}, {Keitel}, {Key}, {Khadka}, {Khalili}, {Khan}, {Khazanov}, {Khetan}, {Khursheed}, {Kijbunchoo}, {Kim}, {Kim}, {Kim}, {Kim}, {Kim}, {Kim},
  {Kimball}, {Kimura}, {Kinley-Hanlon}, {Kirchhoff}, {Kissel}, {Kita}, {Kitazawa}, {Kleybolte}, {Klimenko}, {Knee}, {Knowles}, {Knyazev}, {Koch}, {Koekoek}, {Kojima}, {Kokeyama}, {Koley}, {Kolitsidou}, {Kolstein}, {Komori}, {Kondrashov}, {Kong}, {Kontos}, {Koper}, {Korobko}, {Kotake}, {Kovalam}, {Kozak}, {Kozakai}, {Kozu}, {Kringel}, {Krishnendu}, {Kr{\'o}lak}, {Kuehn}, {Kuei}, {Kuijer}, {Kulkarni}, {Kumar}, {Kumar}, {Kumar}, {Kumar}, {Kume}, {Kuns}, {Kuo}, {Kuo}, {Kuromiya}, {Kuroyanagi}, {Kusayanagi}, {Kuwahara}, {Kwak}, {Lagabbe}, {Laghi}, {Lalande}, {Lam}, {Lamberts}, {Landry}, {Lane}, {Lang}, {Lange}, {Lantz}, {La Rosa}, {Lartaux-Vollard}, {Lasky}, {Laxen}, {Lazzarini}, {Lazzaro}, {Leaci}, {Leavey}, {Lecoeuche}, {Lee}, {Lee}, {Lee}, {Lee}, {Lee}, {Lee}, {Lehmann}, {Lema{\^\i}tre}, {Leonardi}, {Leroy}, {Letendre}, {Levesque}, {Levin}, {Leviton}, {Leyde}, {Li}, {Li}, {Li}, {Li}, {Li}, {Li}, {Lin}, {Lin}, {Lin}, {Lin}, {Lin}, {Linde}, {Linker}, {Linley}, {Littenberg}, {Liu}, {Liu}, {Liu}, {Liu}, {Llamas},
  {Llorens-Monteagudo}, {Lo}, {Lockwood}, {Loh}, {London}, {Longo}, {Lopez}, {Portilla}, {Lorenzini}, {Loriette}, {Lormand}, {Losurdo}, {Lott}, {Lough}, {Lousto}, {Lovelace}, {Lucaccioni}, {L{\"u}ck}, {Lumaca}, {Lundgren}, {Luo}, {Lynam}, {Macas}, {Macinnis}, {MacLeod}, {MacMillan}, {Macquet}, {Hernandez}, {Magazz{\`u}}, {Magee}, {Maggiore}, {Magnozzi}, {Mahesh}, {Majorana}, {Makarem}, {Maksimovic}, {Maliakal}, {Malik}, {Man}, {Mandic}, {Mangano}, {Mango}, {Mansell}, {Manske}, {Mantovani}, {Mapelli}, {Marchesoni}, {Marchio}, {Marion}, {Mark}, {M{\'a}rka}, {M{\'a}rka}, {Markakis}, {Markosyan}, {Markowitz}, {Maros}, {Marquina}, {Marsat}, {Martelli}, {Martin}, {Martin}, {Martinez}, {Martinez}, {Martinez}, {Martinovic}, {Martynov}, {Marx}, {Masalehdan}, {Mason}, {Massera}, {Masserot}, {Massinger}, {Masso-Reid}, {Mastrogiovanni}, {Matas}, {Mateu-Lucena}, {Matichard}, {Matiushechkina}, {Mavalvala}, {McCann}, {McCarthy}, {McClelland}, {McClincy}, {McCormick}, {McCuller}, {McGhee}, {McGuire}, {McIsaac}, {McIver},
  {McRae}, {McWilliams}, {Meacher}, {Mehmet}, {Mehta}, {Meijer}, {Melatos}, {Melchor}, {Mendell}, {Menendez-Vazquez}, {Menoni}, {Mercer}, {Mereni}, {Merfeld}, {Merilh}, {Merritt}, {Merzougui}, {Meshkov}, {Messenger}, {Messick}, {Meyers}, {Meylahn}, {Mhaske}, {Miani}, {Miao}, {Michaloliakos}, {Michel}, {Michimura}, {Middleton}, {Milano}, {Miller}, {Miller}, {Miller}, {Millhouse}, {Mills}, {Milotti}, {Minazzoli}, {Minenkov}, {Mio}, {Mir}, {Miravet-Ten{\'e}s}, {Mishra}, {Mishra}, {Mistry}, {Mitra}, {Mitrofanov}, {Mitselmakher}, {Mittleman}, {Miyakawa}, {Miyamoto}, {Miyazaki}, {Miyo}, {Miyoki}, {Mo}, {Modafferi}, {Moguel}, {Mogushi}, {Mohapatra}, {Mohite}, {Molina}, {Molina-Ruiz}, {Mondin}, {Montani}, {Moore}, {Moraru}, {Morawski}, {More}, {Moreno}, {Moreno}, {Mori}, {Morisaki}, {Moriwaki}, {Morr{\'a}s}, {Mours}, {Mow-Lowry}, {Mozzon}, {Muciaccia}, {Mukherjee}, {Mukherjee}, {Mukherjee}, {Mukherjee}, {Mukherjee}, {Mukund}, {Mullavey}, {Munch}, {Mu{\~n}iz}, {Murray}, {Musenich}, {Muusse}, {Nadji}, {Nagano},
  {Nagano}, {Nagar}, {Nakamura}, {Nakano}, {Nakano}, {Nakashima}, {Nakayama}, {Napolano}, {Nardecchia}, {Narikawa}, {Naticchioni}, {Nayak}, {Nayak}, {Negishi}, {Neil}, {Neilson}, {Nelemans}, {Nelson}, {Nery}, {Neubauer}, {Neunzert}, {Ng}, {Ng}, {Nguyen}, {Nguyen}, {Nguyen}, {Quynh}, {Ni}, {Nichols}, {Nishizawa}, {Nissanke}, {Nitoglia}, {Nocera}, {Norman}, {North}, {Nozaki}, {Siles}, {Nuttall}, {Oberling}, {O'Brien}, {Obuchi}, {O'Dell}, {Oelker}, {Ogaki}, {Oganesyan}, {Oh}, {Oh}, {Oh}, {Ohashi}, {Ohishi}, {Ohkawa}, {Ohme}, {Ohta}, {Okada}, {Okutani}, {Okutomi}, {Olivetto}, {Oohara}, {Ooi}, {Oram}, {O'Reilly}, {Ormiston}, {Ormsby}, {Ortega}, {O'Shaughnessy}, {O'Shea}, {Oshino}, {Ossokine}, {Osthelder}, {Otabe}, {Ottaway}, {Overmier}, {Pace}, {Pagano}, {Page}, {Pagliaroli}, {Pai}, {Pai}, {Palamos}, {Palashov}, {Palomba}, {Pan}, {Pan}, {Panda}, {Pang}, {Pang}, {Pankow}, {Pannarale}, {Pant}, {Panther}, {Paoletti}, {Paoli}, {Paolone}, {Parisi}, {Park}, {Park}, {Parker}, {Pascucci}, {Pasqualetti}, {Passaquieti},
  {Passuello}, {Patel}, {Pathak}, {Patricelli}, {Patron}, {Paul}, {Payne}, {Pedraza}, {Pegoraro}, {Pele}, {Arellano}, {Penn}, {Perego}, {Pereira}, {Pereira}, {Perez}, {P{\'e}rigois}, {Perkins}, {Perreca}, {Perri{\`e}s}, {Petermann}, {Petterson}, {Pfeiffer}, {Pham}, {Phukon}, {Piccinni}, {Pichot}, {Piendibene}, {Piergiovanni}, {Pierini}, {Pierro}, {Pillant}, {Pillas}, {Pilo}, {Pinard}, {Pinto}, {Pinto}, {Piotrzkowski}, {Piotrzkowski}, {Pirello}, {Pitkin}, {Placidi}, {Planas}, {Plastino}, {Pluchar}, {Poggiani}, {Polini}, {Pong}, {Ponrathnam}, {Popolizio}, {Porter}, {Poulton}, {Powell}, {Pracchia}, {Pradier}, {Prajapati}, {Prasai}, {Prasanna}, {Pratten}, {Principe}, {Prodi}, {Prokhorov}, {Prosposito}, {Prudenzi}, {Puecher}, {Punturo}, {Puosi}, {Puppo}, {P{\"u}rrer}, {Qi}, {Quetschke}, {Quitzow-James}, {Qutob}, {Raab}, {Raaijmakers}, {Radkins}, {Radulesco}, {Raffai}, {Rail}, {Raja}, {Rajan}, {Ramirez}, {Ramirez}, {Ramos-Buades}, {Rana}, {Rapagnani}, {Rapol}, {Ray}, {Raymond}, {Raza}, {Razzano}, {Read}, {Rees},
  {Regimbau}, {Rei}, {Reid}, {Reid}, {Reitze}, {Relton}, {Renzini}, {Rettegno}, {Reza}, {Rezac}, {Ricci}, {Richards}, {Richardson}, {Richardson}, {Riemenschneider}, {Riles}, {Rinaldi}, {Rink}, {Rizzo}, {Robertson}, {Robie}, {Robinet}, {Rocchi}, {Rodriguez}, {Rolland}, {Rollins}, {Romanelli}, {Romano}, {Romel}, {Romero-Rodr{\'\i}guez}, {Romero-Shaw}, {Romie}, {Ronchini}, {Rosa}, {Rose}, {Rosi{\'n}ska}, {Ross}, {Rowan}, {Rowlinson}, {Roy}, {Roy}, {Roy}, {Rozza}, {Ruggi}, {Ruiz-Rocha}, {Ryan}, {Sachdev}, {Sadecki}, {Sadiq}, {Sago}, {Saito}, {Saito}, {Sakai}, {Sakai}, {Sakellariadou}, {Sakuno}, {Salafia}, {Salconi}, {Saleem}, {Salemi}, {Samajdar}, {Sanchez}, {Sanchez}, {Sanchez}, {Sanchis-Gual}, {Sanders}, {Sanuy}, {Saravanan}, {Sarin}, {Sassolas}, {Satari}, {Sathyaprakash}, {Sato}, {Sato}, {Sauter}, {Savage}, {Sawada}, {Sawant}, {Sawant}, {Sayah}, {Schaetzl}, {Scheel}, {Scheuer}, {Schiworski}, {Schmidt}, {Schmidt}, {Schnabel}, {Schneewind}, {Schofield}, {Sch{\"o}nbeck}, {Schulte}, {Schutz}, {Schwartz}, {Scott},
  {Scott}, {Seglar-Arroyo}, {Sekiguchi}, {Sekiguchi}, {Sellers}, {Sengupta}, {Sentenac}, {Seo}, {Sequino}, {Sergeev}, {Setyawati}, {Shaffer}, {Shahriar}, {Shams}, {Shao}, {Sharma}, {Sharma}, {Shawhan}, {Shcheblanov}, {Shibagaki}, {Shikauchi}, {Shimizu}, {Shimoda}, {Shimode}, {Shinkai}, {Shishido}, {Shoda}, {Shoemaker}, {Shoemaker}, {Shyamsundar}, {Sieniawska}, {Sigg}, {Singer}, {Singh}, {Singh}, {Singha}, {Sintes}, {Sipala}, {Skliris}, {Slagmolen}, {Slaven-Blair}, {Smetana}, {Smith}, {Smith}, {Soldateschi}, {Somala}, {Somiya}, {Son}, {Soni}, {Soni}, {Sordini}, {Sorrentino}, {Sorrentino}, {Sotani}, {Soulard}, {Souradeep}, {Sowell}, {Spagnuolo}, {Spencer}, {Spera}, {Srinivasan}, {Srivastava}, {Srivastava}, {Staats}, {Stachie}, {Steer}, {Steinhoff}, {Steinlechner}, {Steinlechner}, {Stevenson}, {Stops}, {Stover}, {Strain}, {Strang}, {Stratta}, {Strunk}, {Sturani}, {Stuver}, {Sudhagar}, {Sudhir}, {Sugimoto}, {Suh}, {Sullivan}, {Sullivan}, {Summerscales}, {Sun}, {Sun}, {Sunil}, {Sur}, {Suresh}, {Sutton}, {Suzuki},
  {Suzuki}, {Swinkels}, {Szczepa{\'n}czyk}, {Szewczyk}, {Tacca}, {Tagoshi}, {Tait}, {Takahashi}, {Takahashi}, {Takamori}, {Takano}, {Takeda}, {Takeda}, {Talbot}, {Talbot}, {Tanaka}, {Tanaka}, {Tanaka}, {Tanaka}, {Tanaka}, {Tanasijczuk}, {Tanioka}, {Tanner}, {Tao}, {Tao}, {Mart{\'\i}n}, {Taranto}, {Tasson}, {Telada}, {Tenorio}, {Terhune}, {Terkowski}, {Thirugnanasambandam}, {Thomas}, {Thomas}, {Thomas}, {Thompson}, {Thondapu}, {Thorne}, {Thrane}, {Tiwari}, {Tiwari}, {Tiwari}, {Toivonen}, {Toland}, {Tolley}, {Tomaru}, {Tomigami}, {Tomura}, {Tonelli}, {Torres-Forn{\'e}}, {Torrie}, {E Melo}, {T{\"o}yr{\"a}}, {Trapananti}, {Travasso}, {Traylor}, {Trevor}, {Tringali}, {Tripathee}, {Troiano}, {Trovato}, {Trozzo}, {Trudeau}, {Tsai}, {Tsai}, {Tsang}, {Tsang}, {Tsao}, {Tse}, {Tso}, {Tsubono}, {Tsuchida}, {Tsukada}, {Tsuna}, {Tsutsui}, {Tsuzuki}, {Turbang}, {Turconi}, {Tuyenbayev}, {Ubhi}, {Uchikata}, {Uchiyama}, {Udall}, {Ueda}, {Uehara}, {Ueno}, {Ueshima}, {Unnikrishnan}, {Uraguchi}, {Urban}, {Ushiba}, {Utina},
  {Vahlbruch}, {Vajente}, {Vajpeyi}, {Valdes}, {Valentini}, {Valsan}, {van Bakel}, {van Beuzekom}, {van den Brand}, {van den Broeck}, {Vander-Hyde}, {van der Schaaf}, {van Heijningen}, {Vanosky}, {van Putten}, {van Remortel}, {Vardaro}, {Vargas}, {Varma}, {Vas{\'u}th}, {Vecchio}, {Vedovato}, {Veitch}, {Veitch}, {Venneberg}, {Venugopalan}, {Verkindt}, {Verma}, {Verma}, {Veske}, {Vetrano}, {Vicer{\'e}}, {Vidyant}, {Viets}, {Vijaykumar}, {Villa-Ortega}, {Vinet}, {Virtuoso}, {Vitale}, {Vo}, {Vocca}, {von Reis}, {von Wrangel}, {Vorvick}, {Vyatchanin}, {Wade}, {Wade}, {Wagner}, {Walet}, {Walker}, {Wallace}, {Wallace}, {Walsh}, {Wang}, {Wang}, {Wang}, {Ward}, {Warner}, {Was}, {Washimi}, {Washington}, {Watchi}, {Weaver}, {Webster}, {Weinert}, {Weinstein}, {Weiss}, {Weller}, {Weller}, {Wellmann}, {Wen}, {We{\ss}els}, {Wette}, {Whelan}, {White}, {Whiting}, {Whittle}, {Wilken}, {Williams}, {Williams}, {Williams}, {Williamson}, {Willis}, {Willke}, {Wilson}, {Winkler}, {Wipf}, {Wlodarczyk}, {Woan}, {Woehler}, {Wofford},
  {Wong}, {Wu}, {Wu}, {Wu}, {Wu}, {Wysocki}, {Xiao}, {Xu}, {Yamada}, {Yamamoto}, {Yamamoto}, {Yamamoto}, {Yamamoto}, {Yamashita}, {Yamazaki}, {Yang}, {Yang}, {Yang}, {Yang}, {Yang}, {Yap}, {Yeeles}, {Yelikar}, {Ying}, {Yokogawa}, {Yokoyama}, {Yokozawa}, {Yoo}, {Yoshioka}, {Yu}, {Yu}, {Yuzurihara}, {Zadro{\.z}ny}, {Zanolin}, {Zeidler}, {Zelenova}, {Zendri}, {Zevin}, {Zhan}, {Zhang}, {Zhang}, {Zhang}, {Zhang}, {Zhang}, {Zhao}, {Zhao}, {Zhao}, {Zhao}, {Zheng}, {Zhou}, {Zhou}, {Zhu}, {Zhu}, {Zimmerman}, {Zlochower}, {Zucker}, {Zweizig}, {Ligo Scientific Collaboration}, \& {Kagra Collaboration}}]{2023PhRvX..13d1039A}
{Abbott}, R., {Abbott}, T.~D., {Acernese}, F., {et~al.} 2023{\natexlab{a}}, Physical Review X, 13, 041039, \dodoi{10.1103/PhysRevX.13.041039}

\bibitem[{{Abbott} {et~al.}(2023{\natexlab{b}}){Abbott}, {Abbott}, {Acernese}, {Ackley}, {Adams}, {Adhikari}, {Adhikari}, {Adya}, {Affeldt}, {Agarwal}, {Agathos}, {Agatsuma}, {Aggarwal}, {Aguiar}, {Aiello}, {Ain}, {Ajith}, {Akutsu}, {de Alarc{\'o}n}, {Akcay}, {Albanesi}, {Allocca}, {Altin}, {Amato}, {Anand}, {Anand}, {Ananyeva}, {Anderson}, {Anderson}, {Ando}, {Andrade}, {Andres}, {Andri{\'c}}, {Angelova}, {Ansoldi}, {Antelis}, {Antier}, {Antonini}, {Appert}, {Arai}, {Arai}, {Arai}, {Araki}, {Araya}, {Araya}, {Areeda}, {Ar{\`e}ne}, {Aritomi}, {Arnaud}, {Arogeti}, {Aronson}, {Arun}, {Asada}, {Asali}, {Ashton}, {Aso}, {Assiduo}, {Aston}, {Astone}, {Aubin}, {Austin}, {Babak}, {Badaracco}, {Bader}, {Badger}, {Bae}, {Bae}, {Baer}, {Bagnasco}, {Bai}, {Baiotti}, {Baird}, {Bajpai}, {Ball}, {Ballardin}, {Ballmer}, {Balsamo}, {Baltus}, {Banagiri}, {Bankar}, {Barayoga}, {Barbieri}, {Barish}, {Barker}, {Barneo}, {Barone}, {Barr}, {Barsotti}, {Barsuglia}, {Barta}, {Bartlett}, {Barton}, {Bartos}, {Bassiri}, {Basti},
  {Bawaj}, {Bayley}, {Baylor}, {Bazzan}, {B{\'e}csy}, {Bedakihale}, {Bejger}, {Belahcene}, {Benedetto}, {Beniwal}, {Bennett}, {Bentley}, {Benyaala}, {Bergamin}, {Berger}, {Bernuzzi}, {Berry}, {Bersanetti}, {Bertolini}, {Betzwieser}, {Beveridge}, {Bhandare}, {Bhardwaj}, {Bhattacharjee}, {Bhaumik}, {Bilenko}, {Billingsley}, {Bini}, {Birney}, {Birnholtz}, {Biscans}, {Bischi}, {Biscoveanu}, {Bisht}, {Biswas}, {Bitossi}, {Bizouard}, {Blackburn}, {Blair}, {Blair}, {Blair}, {Bobba}, {Bode}, {Boer}, {Bogaert}, {Boldrini}, {Bonavena}, {Bondu}, {Bonilla}, {Bonnand}, {Booker}, {Boom}, {Bork}, {Boschi}, {Bose}, {Bose}, {Bossilkov}, {Boudart}, {Bouffanais}, {Bozzi}, {Bradaschia}, {Brady}, {Bramley}, {Branch}, {Branchesi}, {Brandt}, {Brau}, {Breschi}, {Briant}, {Briggs}, {Brillet}, {Brinkmann}, {Brockill}, {Brooks}, {Brooks}, {Brown}, {Brunett}, {Bruno}, {Bruntz}, {Bryant}, {Bulik}, {Bulten}, {Buonanno}, {Buscicchio}, {Buskulic}, {Buy}, {Byer}, {Cadonati}, {Cagnoli}, {Cahillane}, {Bustillo}, {Callaghan}, {Callister},
  {Calloni}, {Cameron}, {Camp}, {Canepa}, {Canevarolo}, {Cannavacciuolo}, {Cannon}, {Cao}, {Cao}, {Capocasa}, {Capote}, {Carapella}, {Carbognani}, {Carlin}, {Carney}, {Carpinelli}, {Carrillo}, {Carullo}, {Carver}, {Diaz}, {Casentini}, {Castaldi}, {Caudill}, {Cavagli{\`a}}, {Cavalier}, {Cavalieri}, {Ceasar}, {Cella}, {Cerd{\'a}-Dur{\'a}n}, {Cesarini}, {Chaibi}, {Chakravarti}, {Subrahmanya}, {Champion}, {Chan}, {Chan}, {Chan}, {Chan}, {Chan}, {Chandra}, {Chanial}, {Chao}, {Chapman-Bird}, {Charlton}, {Chase}, {Chassande-Mottin}, {Chatterjee}, {Chatterjee}, {Chatterjee}, {Chaturvedi}, {Chaty}, {Chatziioannou}, {Chen}, {Chen}, {Chen}, {Chen}, {Chen}, {Chen}, {Chen}, {Chen}, {Cheng}, {Cheong}, {Cheung}, {Chia}, {Chiadini}, {Chiang}, {Chiarini}, {Chierici}, {Chincarini}, {Chiofalo}, {Chiummo}, {Cho}, {Cho}, {Choudhary}, {Choudhary}, {Christensen}, {Chu}, {Chu}, {Chu}, {Chua}, {Chung}, {Ciani}, {Ciecielag}, {Cie{\'s}lar}, {Cifaldi}, {Ciobanu}, {Ciolfi}, {Cipriano}, {Cirone}, {Clara}, {Clark}, {Clark}, {Clarke},
  {Clearwater}, {Clesse}, {Cleva}, {Coccia}, {Codazzo}, {Cohadon}, {Cohen}, {Cohen}, {Colleoni}, {Collette}, {Colombo}, {Colpi}, {Compton}, {Constancio}, {Conti}, {Cooper}, {Corban}, {Corbitt}, {Cordero-Carri{\'o}n}, {Corezzi}, {Corley}, {Cornish}, {Corre}, {Corsi}, {Cortese}, {Costa}, {Cotesta}, {Coughlin}, {Coulon}, {Countryman}, {Cousins}, {Couvares}, {Coward}, {Cowart}, {Coyne}, {Coyne}, {Creighton}, {Creighton}, {Criswell}, {Croquette}, {Crowder}, {Cudell}, {Cullen}, {Cumming}, {Cummings}, {Cunningham}, {Cuoco}, {Cury{\l}o}, {Dabadie}, {Canton}, {Dall'Osso}, {D{\'a}lya}, {Dana}, {Daneshgaranbajastani}, {D'Angelo}, {Danila}, {Danilishin}, {D'Antonio}, {Danzmann}, {Darsow-Fromm}, {Dasgupta}, {Datrier}, {Datta}, {Dattilo}, {Dave}, {Davier}, {Davies}, {Davis}, {Davis}, {Daw}, {Dean}, {Debra}, {Deenadayalan}, {Degallaix}, {de Laurentis}, {Del{\'e}glise}, {Del Favero}, {de Lillo}, {de Lillo}, {Del Pozzo}, {Demarchi}, {de Matteis}, {D'Emilio}, {Demos}, {Dent}, {Depasse}, {de Pietri}, {De Rosa}, {de Rossi},
  {Desalvo}, {de Simone}, {Dhurandhar}, {D{\'\i}az}, {Diaz-Ortiz}, {Didio}, {Dietrich}, {di Fiore}, {di Fronzo}, {di Giorgio}, {di Giovanni}, {di Giovanni}, {di Girolamo}, {di Lieto}, {Ding}, {di Pace}, {di Palma}, {di Renzo}, {Divakarla}, {Dmitriev}, {Doctor}, {D'Onofrio}, {Donovan}, {Dooley}, {Doravari}, {Dorrington}, {Drago}, {Driggers}, {Drori}, {Ducoin}, {Dupej}, {Durante}, {D'Urso}, {Duverne}, {Dwyer}, {Eassa}, {Easter}, {Ebersold}, {Eckhardt}, {Eddolls}, {Edelman}, {Edo}, {Edy}, {Effler}, {Eguchi}, {Eichholz}, {Eikenberry}, {Eisenmann}, {Eisenstein}, {Ejlli}, {Engelby}, {Enomoto}, {Errico}, {Essick}, {Estell{\'e}s}, {Estevez}, {Etienne}, {Etzel}, {Evans}, {Evans}, {Ewing}, {Fafone}, {Fair}, {Fairhurst}, {Farah}, {Farinon}, {Farr}, {Farr}, {Farrow}, {Fauchon-Jones}, {Favaro}, {Favata}, {Fays}, {Fazio}, {Feicht}, {Fejer}, {Fenyvesi}, {Ferguson}, {Fernandez-Galiana}, {Ferrante}, {Ferreira}, {Fidecaro}, {Figura}, {Fiori}, {Fishbach}, {Fisher}, {Fittipaldi}, {Fiumara}, {Flaminio}, {Floden}, {Fong}, {Font},
  {Fornal}, {Forsyth}, {Franke}, {Frasca}, {Frasconi}, {Frederick}, {Freed}, {Frei}, {Freise}, {Frey}, {Fritschel}, {Frolov}, {Fronz{\'e}}, {Fujii}, {Fujikawa}, {Fukunaga}, {Fukushima}, {Fulda}, {Fyffe}, {Gabbard}, {Gadre}, {Gair}, {Gais}, {Galaudage}, {Gamba}, {Ganapathy}, {Ganguly}, {Gao}, {Gaonkar}, {Garaventa}, {Garc{\'\i}a}, {Garc{\'\i}a-N{\'u}{\~n}ez}, {Garc{\'\i}a-Quir{\'o}s}, {Garufi}, {Gateley}, {Gaudio}, {Gayathri}, {Ge}, {Gemme}, {Gennai}, {George}, {George}, {Gerberding}, {Gergely}, {Gewecke}, {Ghonge}, {Ghosh}, {Ghosh}, {Ghosh}, {Ghosh}, {Giacomazzo}, {Giacoppo}, {Giaime}, {Giardina}, {Gibson}, {Gier}, {Giesler}, {Giri}, {Gissi}, {Glanzer}, {Gleckl}, {Godwin}, {Golomb}, {Goetz}, {Goetz}, {Gohlke}, {Goncharov}, {Gonz{\'a}lez}, {Gopakumar}, {Gosselin}, {Gouaty}, {Gould}, {Grace}, {Grado}, {Granata}, {Granata}, {Grant}, {Gras}, {Grassia}, {Gray}, {Gray}, {Greco}, {Green}, {Green}, {Gretarsson}, {Gretarsson}, {Griffith}, {Griffiths}, {Griggs}, {Grignani}, {Grimaldi}, {Grimm}, {Grote}, {Grunewald},
  {Gruning}, {Guerra}, {Guidi}, {Guimaraes}, {Guix{\'e}}, {Gulati}, {Guo}, {Guo}, {Gupta}, {Gupta}, {Gupta}, {Gustafson}, {Gustafson}, {Guzman}, {Ha}, {Haegel}, {Hagiwara}, {Haino}, {Halim}, {Hall}, {Hamilton}, {Hammond}, {Han}, {Haney}, {Hanks}, {Hanna}, {Hannam}, {Hannuksela}, {Hansen}, {Hansen}, {Hanson}, {Harder}, {Hardwick}, {Haris}, {Harms}, {Harry}, {Harry}, {Hartwig}, {Hasegawa}, {Haskell}, {Hasskew}, {Haster}, {Hattori}, {Haughian}, {Hayakawa}, {Hayama}, {Hayes}, {Healy}, {Heidmann}, {Heidt}, {Heintze}, {Heinze}, {Heinzel}, {Heitmann}, {Hellman}, {Hello}, {Helmling-Cornell}, {Hemming}, {Hendry}, {Heng}, {Hennes}, {Hennig}, {Hennig}, {Hernandez}, {Vivanco}, {Heurs}, {Hild}, {Hill}, {Himemoto}, {Hines}, {Hiranuma}, {Hirata}, {Hirose}, {Hochheim}, {Hofman}, {Hohmann}, {Holcomb}, {Holland}, {Hollows}, {Holmes}, {Holt}, {Holz}, {Hong}, {Hopkins}, {Hough}, {Hourihane}, {Howell}, {Hoy}, {Hoyland}, {Hreibi}, {Hsieh}, {Hsu}, {Huang}, {Huang}, {Huang}, {Huang}, {Huang}, {Huang}, {H{\"u}bner}, {Huddart},
  {Hughey}, {Hui}, {Hui}, {Husa}, {Huttner}, {Huxford}, {Huynh-Dinh}, {Ide}, {Idzkowski}, {Iess}, {Ikenoue}, {Imam}, {Inayoshi}, {Ingram}, {Inoue}, {Ioka}, {Isi}, {Isleif}, {Ito}, {Itoh}, {Iyer}, {Izumi}, {Jaberianhamedan}, {Jacqmin}, {Jadhav}, {Jadhav}, {James}, {Jan}, {Jani}, {Janquart}, {Janssens}, {Janthalur}, {Jaranowski}, {Jariwala}, {Jaume}, {Jenkins}, {Jenner}, {Jeon}, {Jeunon}, {Jia}, {Jin}, {Johns}, {Jones}, {Jones}, {Jones}, {Jones}, {Jones}, {Jonker}, {Ju}, {Jung}, {Jung}, {Junker}, {Juste}, {Kaihotsu}, {Kajita}, {Kakizaki}, {Kalaghatgi}, {Kalogera}, {Kamai}, {Kamiizumi}, {Kanda}, {Kandhasamy}, {Kang}, {Kanner}, {Kao}, {Kapadia}, {Kapasi}, {Karat}, {Karathanasis}, {Karki}, {Kashyap}, {Kasprzack}, {Kastaun}, {Katsanevas}, {Katsavounidis}, {Katzman}, {Kaur}, {Kawabe}, {Kawaguchi}, {Kawai}, {Kawasaki}, {K{\'e}f{\'e}lian}, {Keitel}, {Key}, {Khadka}, {Khalili}, {Khan}, {Khazanov}, {Khetan}, {Khursheed}, {Kijbunchoo}, {Kim}, {Kim}, {Kim}, {Kim}, {Kim}, {Kim}, {Kimball}, {Kimura}, {Kinley-Hanlon},
  {Kirchhoff}, {Kissel}, {Kita}, {Kitazawa}, {Kleybolte}, {Klimenko}, {Knee}, {Knowles}, {Knyazev}, {Koch}, {Koekoek}, {Kojima}, {Kokeyama}, {Koley}, {Kolitsidou}, {Kolstein}, {Komori}, {Kondrashov}, {Kong}, {Kontos}, {Koper}, {Korobko}, {Kotake}, {Kovalam}, {Kozak}, {Kozakai}, {Kozu}, {Kringel}, {Krishnendu}, {Kr{\'o}lak}, {Kuehn}, {Kuei}, {Kuijer}, {Kulkarni}, {Kumar}, {Kumar}, {Kumar}, {Kumar}, {Kume}, {Kuns}, {Kuo}, {Kuo}, {Kuromiya}, {Kuroyanagi}, {Kusayanagi}, {Kuwahara}, {Kwak}, {Lagabbe}, {Laghi}, {Lalande}, {Lam}, {Lamberts}, {Landry}, {Landry}, {Lane}, {Lang}, {Lange}, {Lantz}, {La Rosa}, {Lartaux-Vollard}, {Lasky}, {Laxen}, {Lazzarini}, {Lazzaro}, {Leaci}, {Leavey}, {Lecoeuche}, {Lee}, {Lee}, {Lee}, {Lee}, {Lee}, {Lee}, {Lehmann}, {Lema{\^\i}tre}, {Leonardi}, {Leroy}, {Letendre}, {Levesque}, {Levin}, {Leviton}, {Leyde}, {Li}, {Li}, {Li}, {Li}, {Li}, {Li}, {Lin}, {Lin}, {Lin}, {Lin}, {Lin}, {Linde}, {Linker}, {Linley}, {Littenberg}, {Liu}, {Liu}, {Liu}, {Liu}, {Llamas}, {Llorens-Monteagudo}, {Lo},
  {Lockwood}, {Loh}, {London}, {Longo}, {Lopez}, {Portilla}, {Lorenzini}, {Loriette}, {Lormand}, {Losurdo}, {Lott}, {Lough}, {Lousto}, {Lovelace}, {Lucaccioni}, {L{\"u}ck}, {Lumaca}, {Lundgren}, {Luo}, {Lynam}, {Macas}, {Macinnis}, {MacLeod}, {MacMillan}, {Macquet}, {Hernandez}, {Magazz{\`u}}, {Magee}, {Maggiore}, {Magnozzi}, {Mahesh}, {Majorana}, {Makarem}, {Maksimovic}, {Maliakal}, {Malik}, {Man}, {Mandic}, {Mangano}, {Mango}, {Mansell}, {Manske}, {Mantovani}, {Mapelli}, {Marchesoni}, {Marchio}, {Marion}, {Mark}, {M{\'a}rka}, {M{\'a}rka}, {Markakis}, {Markosyan}, {Markowitz}, {Maros}, {Marquina}, {Marsat}, {Martelli}, {Martin}, {Martin}, {Martinez}, {Martinez}, {Martinez}, {Martinovic}, {Martynov}, {Marx}, {Masalehdan}, {Mason}, {Massera}, {Masserot}, {Massinger}, {Masso-Reid}, {Mastrogiovanni}, {Matas}, {Mateu-Lucena}, {Matichard}, {Matiushechkina}, {Mavalvala}, {McCann}, {McCarthy}, {McClelland}, {McClincy}, {McCormick}, {McCuller}, {McGhee}, {McGuire}, {McIsaac}, {McIver}, {McRae}, {McWilliams},
  {Meacher}, {Mehmet}, {Mehta}, {Meijer}, {Melatos}, {Melchor}, {Mendell}, {Menendez-Vazquez}, {Menoni}, {Mercer}, {Mereni}, {Merfeld}, {Merilh}, {Merritt}, {Merzougui}, {Meshkov}, {Messenger}, {Messick}, {Meyers}, {Meylahn}, {Mhaske}, {Miani}, {Miao}, {Michaloliakos}, {Michel}, {Michimura}, {Middleton}, {Milano}, {Miller}, {Miller}, {Miller}, {Miller}, {Millhouse}, {Mills}, {Milotti}, {Minazzoli}, {Minenkov}, {Mio}, {Mir}, {Miravet-Ten{\'e}s}, {Mishra}, {Mishra}, {Mistry}, {Mitra}, {Mitrofanov}, {Mitselmakher}, {Mittleman}, {Miyakawa}, {Miyamoto}, {Miyazaki}, {Miyo}, {Miyoki}, {Mo}, {Modafferi}, {Moguel}, {Mogushi}, {Mohapatra}, {Mohite}, {Molina}, {Molina-Ruiz}, {Mondin}, {Montani}, {Moore}, {Moraru}, {Morawski}, {More}, {Moreno}, {Moreno}, {Mori}, {Morisaki}, {Moriwaki}, {Morr{\'a}s}, {Mours}, {Mow-Lowry}, {Mozzon}, {Muciaccia}, {Mukherjee}, {Mukherjee}, {Mukherjee}, {Mukherjee}, {Mukherjee}, {Mukund}, {Mullavey}, {Munch}, {Mu{\~n}iz}, {Murray}, {Musenich}, {Muusse}, {Nadji}, {Nagano}, {Nagano}, {Nagar},
  {Nakamura}, {Nakano}, {Nakano}, {Nakashima}, {Nakayama}, {Napolano}, {Nardecchia}, {Narikawa}, {Naticchioni}, {Nayak}, {Nayak}, {Negishi}, {Neil}, {Neilson}, {Nelemans}, {Nelson}, {Nery}, {Neubauer}, {Neunzert}, {Ng}, {Ng}, {Nguyen}, {Nguyen}, {Nguyen}, {Quynh}, {Ni}, {Nichols}, {Nishizawa}, {Nissanke}, {Nitoglia}, {Nocera}, {Norman}, {North}, {Nozaki}, {Siles}, {Nuttall}, {Oberling}, {O'Brien}, {Obuchi}, {O'Dell}, {Oelker}, {Ogaki}, {Oganesyan}, {Oh}, {Oh}, {Oh}, {Ohashi}, {Ohishi}, {Ohkawa}, {Ohme}, {Ohta}, {Okada}, {Okutani}, {Okutomi}, {Olivetto}, {Oohara}, {Ooi}, {Oram}, {O'Reilly}, {Ormiston}, {Ormsby}, {Ortega}, {O'Shaughnessy}, {O'Shea}, {Oshino}, {Ossokine}, {Osthelder}, {Otabe}, {Ottaway}, {Overmier}, {Pace}, {Pagano}, {Page}, {Pagliaroli}, {Pai}, {Pai}, {Palamos}, {Palashov}, {Palomba}, {Pan}, {Pan}, {Panda}, {Pang}, {Pang}, {Pankow}, {Pannarale}, {Pant}, {Panther}, {Paoletti}, {Paoli}, {Paolone}, {Parisi}, {Park}, {Park}, {Parker}, {Pascucci}, {Pasqualetti}, {Passaquieti}, {Passuello}, {Patel},
  {Pathak}, {Patricelli}, {Patron}, {Paul}, {Payne}, {Pedraza}, {Pegoraro}, {Pele}, {Arellano}, {Penn}, {Perego}, {Pereira}, {Pereira}, {Perez}, {P{\'e}rigois}, {Perkins}, {Perreca}, {Perri{\`e}s}, {Petermann}, {Petterson}, {Pfeiffer}, {Pham}, {Phukon}, {Piccinni}, {Pichot}, {Piendibene}, {Piergiovanni}, {Pierini}, {Pierro}, {Pillant}, {Pillas}, {Pilo}, {Pinard}, {Pinto}, {Pinto}, {Piotrzkowski}, {Piotrzkowski}, {Pirello}, {Pitkin}, {Placidi}, {Planas}, {Plastino}, {Pluchar}, {Poggiani}, {Polini}, {Pong}, {Ponrathnam}, {Popolizio}, {Porter}, {Poulton}, {Powell}, {Pracchia}, {Pradier}, {Prajapati}, {Prasai}, {Prasanna}, {Pratten}, {Principe}, {Prodi}, {Prokhorov}, {Prosposito}, {Prudenzi}, {Puecher}, {Punturo}, {Puosi}, {Puppo}, {P{\"u}rrer}, {Qi}, {Quetschke}, {Quitzow-James}, {Raab}, {Raaijmakers}, {Radkins}, {Radulesco}, {Raffai}, {Rail}, {Raja}, {Rajan}, {Ramirez}, {Ramirez}, {Ramos-Buades}, {Rana}, {Rapagnani}, {Rapol}, {Ray}, {Raymond}, {Raza}, {Razzano}, {Read}, {Rees}, {Regimbau}, {Rei}, {Reid},
  {Reid}, {Reitze}, {Relton}, {Renzini}, {Rettegno}, {Reza}, {Rezac}, {Ricci}, {Richards}, {Richardson}, {Richardson}, {Riemenschneider}, {Riles}, {Rinaldi}, {Rink}, {Rizzo}, {Robertson}, {Robie}, {Robinet}, {Rocchi}, {Rodriguez}, {Rolland}, {Rollins}, {Romanelli}, {Romano}, {Romel}, {Romero-Rodr{\'\i}guez}, {Romero-Shaw}, {Romie}, {Ronchini}, {Rosa}, {Rose}, {Rosi{\'n}ska}, {Ross}, {Rowan}, {Rowlinson}, {Roy}, {Roy}, {Roy}, {Rozza}, {Ruggi}, {Ryan}, {Sachdev}, {Sadecki}, {Sadiq}, {Sago}, {Saito}, {Saito}, {Sakai}, {Sakai}, {Sakellariadou}, {Sakuno}, {Salafia}, {Salconi}, {Saleem}, {Salemi}, {Samajdar}, {Sanchez}, {Sanchez}, {Sanchez}, {Sanchis-Gual}, {Sanders}, {Sanuy}, {Saravanan}, {Sarin}, {Sassolas}, {Satari}, {Sathyaprakash}, {Sato}, {Sato}, {Sauter}, {Savage}, {Sawada}, {Sawant}, {Sawant}, {Sayah}, {Schaetzl}, {Scheel}, {Scheuer}, {Schiworski}, {Schmidt}, {Schmidt}, {Schnabel}, {Schneewind}, {Schofield}, {Sch{\"o}nbeck}, {Schulte}, {Schutz}, {Schwartz}, {Scott}, {Scott}, {Seglar-Arroyo}, {Sekiguchi},
  {Sekiguchi}, {Sellers}, {Sengupta}, {Sentenac}, {Seo}, {Sequino}, {Sergeev}, {Setyawati}, {Shaffer}, {Shahriar}, {Shams}, {Shao}, {Sharma}, {Sharma}, {Shawhan}, {Shcheblanov}, {Shibagaki}, {Shikauchi}, {Shimizu}, {Shimoda}, {Shimode}, {Shinkai}, {Shishido}, {Shoda}, {Shoemaker}, {Shoemaker}, {Shyamsundar}, {Sieniawska}, {Sigg}, {Singer}, {Singh}, {Singh}, {Singha}, {Sintes}, {Sipala}, {Skliris}, {Slagmolen}, {Slaven-Blair}, {Smetana}, {Smith}, {Smith}, {Soldateschi}, {Somala}, {Somiya}, {Son}, {Soni}, {Soni}, {Sordini}, {Sorrentino}, {Sorrentino}, {Sotani}, {Soulard}, {Souradeep}, {Sowell}, {Spagnuolo}, {Spencer}, {Spera}, {Srinivasan}, {Srivastava}, {Srivastava}, {Staats}, {Stachie}, {Steer}, {Steinhoff}, {Steinlechner}, {Steinlechner}, {Stevenson}, {Stops}, {Stover}, {Strain}, {Strang}, {Stratta}, {Strunk}, {Sturani}, {Stuver}, {Sudhagar}, {Sudhir}, {Sugimoto}, {Suh}, {Sullivan}, {Summerscales}, {Sun}, {Sun}, {Sunil}, {Sur}, {Suresh}, {Sutton}, {Suzuki}, {Suzuki}, {Swinkels}, {Szczepa{\'n}czyk},
  {Szewczyk}, {Tacca}, {Tagoshi}, {Tait}, {Takahashi}, {Takahashi}, {Takamori}, {Takano}, {Takeda}, {Takeda}, {Talbot}, {Talbot}, {Tanaka}, {Tanaka}, {Tanaka}, {Tanaka}, {Tanaka}, {Tanasijczuk}, {Tanioka}, {Tanner}, {Tao}, {Tao}, {Mart{\'\i}n}, {Taranto}, {Tasson}, {Telada}, {Tenorio}, {Terhune}, {Terkowski}, {Thirugnanasambandam}, {Thomas}, {Thomas}, {Thomas}, {Thompson}, {Thondapu}, {Thorne}, {Thrane}, {Tiwari}, {Tiwari}, {Tiwari}, {Toivonen}, {Toland}, {Tolley}, {Tomaru}, {Tomigami}, {Tomura}, {Tonelli}, {Torres-Forn{\'e}}, {Torrie}, {E Melo}, {T{\"o}yr{\"a}}, {Trapananti}, {Travasso}, {Traylor}, {Trevor}, {Tringali}, {Tripathee}, {Troiano}, {Trovato}, {Trozzo}, {Trudeau}, {Tsai}, {Tsai}, {Tsang}, {Tsang}, {Tsao}, {Tse}, {Tso}, {Tsubono}, {Tsuchida}, {Tsukada}, {Tsuna}, {Tsutsui}, {Tsuzuki}, {Turbang}, {Turconi}, {Tuyenbayev}, {Ubhi}, {Uchikata}, {Uchiyama}, {Udall}, {Ueda}, {Uehara}, {Ueno}, {Ueshima}, {Unnikrishnan}, {Uraguchi}, {Urban}, {Ushiba}, {Utina}, {Vahlbruch}, {Vajente}, {Vajpeyi}, {Valdes},
  {Valentini}, {Valsan}, {van Bakel}, {van Beuzekom}, {van den Brand}, {van den Broeck}, {Vander-Hyde}, {van der Schaaf}, {van Heijningen}, {Vanosky}, {van Putten}, {van Remortel}, {Vardaro}, {Vargas}, {Varma}, {Vas{\'u}th}, {Vecchio}, {Vedovato}, {Veitch}, {Veitch}, {Venneberg}, {Venugopalan}, {Verkindt}, {Verma}, {Verma}, {Veske}, {Vetrano}, {Vicer{\'e}}, {Vidyant}, {Viets}, {Vijaykumar}, {Villa-Ortega}, {Vinet}, {Virtuoso}, {Vitale}, {Vo}, {Vocca}, {von Reis}, {von Wrangel}, {Vorvick}, {Vyatchanin}, {Wade}, {Wade}, {Wagner}, {Walet}, {Walker}, {Wallace}, {Wallace}, {Walsh}, {Wang}, {Wang}, {Wang}, {Ward}, {Warner}, {Was}, {Washimi}, {Washington}, {Watchi}, {Weaver}, {Webster}, {Weinert}, {Weinstein}, {Weiss}, {Weller}, {Wellmann}, {Wen}, {We{\ss}els}, {Wette}, {Whelan}, {White}, {Whiting}, {Whittle}, {Wilken}, {Williams}, {Williams}, {Williamson}, {Willis}, {Willke}, {Wilson}, {Winkler}, {Wipf}, {Wlodarczyk}, {Woan}, {Woehler}, {Wofford}, {Wong}, {Wu}, {Wu}, {Wu}, {Wu}, {Wysocki}, {Xiao}, {Xu}, {Yamada},
  {Yamamoto}, {Yamamoto}, {Yamamoto}, {Yamamoto}, {Yamashita}, {Yamazaki}, {Yang}, {Yang}, {Yang}, {Yang}, {Yang}, {Yap}, {Yeeles}, {Yelikar}, {Ying}, {Yokogawa}, {Yokoyama}, {Yokozawa}, {Yoo}, {Yoshioka}, {Yu}, {Yu}, {Yuzurihara}, {Zadro{\.z}ny}, {Zanolin}, {Zeidler}, {Zelenova}, {Zendri}, {Zevin}, {Zhan}, {Zhang}, {Zhang}, {Zhang}, {Zhang}, {Zhang}, {Zhao}, {Zhao}, {Zhao}, {Zhao}, {Zheng}, {Zhou}, {Zhou}, {Zhu}, {Zhu}, {Zimmerman}, {Zlochower}, {Zucker}, {Zweizig}, {LIGO Scientific Collaboration}, {VIRGO Collaboration}, \& {KAGRA Collaboration}}]{2023PhRvX..13a1048A}
---. 2023{\natexlab{b}}, Physical Review X, 13, 011048, \dodoi{10.1103/PhysRevX.13.011048}

\bibitem[{{Abbott} {et~al.}(2024){Abbott}, {Abbott}, {Acernese}, {Ackley}, {Adams}, {Adhikari}, {Adhikari}, {Adya}, {Affeldt}, {Agarwal}, {Agathos}, {Agatsuma}, {Aggarwal}, {Aguiar}, {Aiello}, {Ain}, {Ajith}, {Albanesi}, {Allocca}, {Altin}, {Amato}, {Anand}, {Anand}, {Ananyeva}, {Anderson}, {Anderson}, {Andrade}, {Andres}, {Andri{\'c}}, {Angelova}, {Ansoldi}, {Antelis}, {Antier}, {Appert}, {Arai}, {Araya}, {Areeda}, {Ar{\`e}ne}, {Arnaud}, {Aronson}, {Arun}, {Asali}, {Ashton}, {Assiduo}, {Aston}, {Astone}, {Aubin}, {Austin}, {Babak}, {Badaracco}, {Bader}, {Badger}, {Bae}, {Baer}, {Bagnasco}, {Bai}, {Baird}, {Ball}, {Ballardin}, {Ballmer}, {Balsamo}, {Baltus}, {Banagiri}, {Bankar}, {Barayoga}, {Barbieri}, {Barish}, {Barker}, {Barneo}, {Barone}, {Barr}, {Barsotti}, {Barsuglia}, {Barta}, {Bartlett}, {Barton}, {Bartos}, {Bassiri}, {Basti}, {Bawaj}, {Bayley}, {Baylor}, {Bazzan}, {B{\'e}csy}, {Bedakihale}, {Bejger}, {Belahcene}, {Benedetto}, {Beniwal}, {Bennett}, {Bentley}, {BenYaala}, {Bergamin}, {Berger},
  {Bernuzzi}, {Berry}, {Bersanetti}, {Bertolini}, {Betzwieser}, {Beveridge}, {Bhandare}, {Bhardwaj}, {Bhattacharjee}, {Bhaumik}, {Bilenko}, {Billingsley}, {Bini}, {Birney}, {Birnholtz}, {Biscans}, {Bischi}, {Biscoveanu}, {Bisht}, {Biswas}, {Bitossi}, {Bizouard}, {Blackburn}, {Blair}, {Blair}, {Blair}, {Bobba}, {Bode}, {Boer}, {Bogaert}, {Boldrini}, {Bonavena}, {Bondu}, {Bonilla}, {Bonnand}, {Booker}, {Boom}, {Bork}, {Boschi}, {Bose}, {Bose}, {Bossilkov}, {Boudart}, {Bouffanais}, {Bozzi}, {Bradaschia}, {Brady}, {Bramley}, {Branch}, {Branchesi}, {Brau}, {Breschi}, {Briant}, {Briggs}, {Brillet}, {Brinkmann}, {Brockill}, {Brooks}, {Brooks}, {Brown}, {Brunett}, {Bruno}, {Bruntz}, {Bryant}, {Bulik}, {Bulten}, {Buonanno}, {Buscicchio}, {Buskulic}, {Buy}, {Byer}, {Cadonati}, {Cagnoli}, {Cahillane}, {Bustillo}, {Callaghan}, {Callister}, {Calloni}, {Cameron}, {Camp}, {Canepa}, {Canevarolo}, {Cannavacciuolo}, {Cannon}, {Cao}, {Capote}, {Carapella}, {Carbognani}, {Carlin}, {Carney}, {Carpinelli}, {Carrillo}, {Carullo},
  {Carver}, {Diaz}, {Casentini}, {Castaldi}, {Caudill}, {Cavagli{\`a}}, {Cavalier}, {Cavalieri}, {Ceasar}, {Cella}, {Cerd{\'a}-Dur{\'a}n}, {Cesarini}, {Chaibi}, {Chakravarti}, {Subrahmanya}, {Champion}, {Chan}, {Chan}, {Chan}, {Chan}, {Chandra}, {Chanial}, {Chao}, {Charlton}, {Chase}, {Chassande-Mottin}, {Chatterjee}, {Chatterjee}, {Chatterjee}, {Chattopadhyay}, {Chaturvedi}, {Chaty}, {Chatziioannou}, {Chen}, {Chen}, {Chen}, {Chen}, {Chen}, {Cheng}, {Cheong}, {Cheung}, {Chia}, {Chiadini}, {Chiarini}, {Chierici}, {Chincarini}, {Chiofalo}, {Chiummo}, {Cho}, {Cho}, {Choudhary}, {Choudhary}, {Christensen}, {Chu}, {Chua}, {Chung}, {Ciani}, {Ciecielag}, {Cie{\'s}lar}, {Cifaldi}, {Ciobanu}, {Ciolfi}, {Cipriano}, {Cirone}, {Clara}, {Clark}, {Clark}, {Clarke}, {Clearwater}, {Clesse}, {Cleva}, {Coccia}, {Codazzo}, {Cohadon}, {Cohen}, {Cohen}, {Colleoni}, {Collette}, {Colombo}, {Colpi}, {Compton}, {Constancio}, {Conti}, {Cooper}, {Corban}, {Corbitt}, {Cordero-Carri{\'o}n}, {Corezzi}, {Corley}, {Cornish}, {Corre},
  {Corsi}, {Cortese}, {Costa}, {Cotesta}, {Coughlin}, {Coulon}, {Countryman}, {Cousins}, {Couvares}, {Coward}, {Cowart}, {Coyne}, {Coyne}, {Creighton}, {Creighton}, {Criswell}, {Croquette}, {Crowder}, {Cudell}, {Cullen}, {Cumming}, {Cummings}, {Cunningham}, {Cuoco}, {Cury{\l}o}, {Dabadie}, {Canton}, {Dall'Osso}, {D{\'a}lya}, {Dana}, {DaneshgaranBajastani}, {D'Angelo}, {Danila}, {Danilishin}, {D'Antonio}, {Danzmann}, {Darsow-Fromm}, {Dasgupta}, {Datrier}, {Datta}, {Dattilo}, {Dave}, {Davier}, {Davies}, {Davis}, {Davis}, {Daw}, {Dean}, {DeBra}, {Deenadayalan}, {Degallaix}, {De Laurentis}, {Del{\'e}glise}, {Del Favero}, {De Lillo}, {De Lillo}, {Del Pozzo}, {DeMarchi}, {De Matteis}, {D'Emilio}, {Demos}, {Dent}, {Depasse}, {De Pietri}, {De Rosa}, {De Rossi}, {DeSalvo}, {De Simone}, {Dhurandhar}, {D{\'\i}az}, {Diaz-Ortiz}, {Didio}, {Dietrich}, {Di Fiore}, {Di Fronzo}, {Di Giorgio}, {Di Giovanni}, {Di Giovanni}, {Di Girolamo}, {Di Lieto}, {Ding}, {Di Pace}, {Di Palma}, {Di Renzo}, {Divakarla}, {Divyajyoti},
  {Doctor}, {D'Onofrio}, {Donovan}, {Dooley}, {Doravari}, {Dorrington}, {Drago}, {Driggers}, {Drori}, {Ducoin}, {Dupej}, {Durante}, {D'Urso}, {Duverne}, {Dwyer}, {Eassa}, {Easter}, {Ebersold}, {Eckhardt}, {Eddolls}, {Edelman}, {Edo}, {Edy}, {Effler}, {Eichholz}, {Eikenberry}, {Eisenmann}, {Eisenstein}, {Ejlli}, {Engelby}, {Errico}, {Essick}, {Estell{\'e}s}, {Estevez}, {Etienne}, {Etzel}, {Evans}, {Evans}, {Ewing}, {Fafone}, {Fair}, {Fairhurst}, {Fanning}, {Farah}, {Farinon}, {Farr}, {Farr}, {Farrow}, {Fauchon-Jones}, {Favaro}, {Favata}, {Fays}, {Fazio}, {Feicht}, {Fejer}, {Fenyvesi}, {Ferguson}, {Fernandez-Galiana}, {Ferrante}, {Ferreira}, {Fidecaro}, {Figura}, {Fiori}, {Fishbach}, {Fisher}, {Fittipaldi}, {Fiumara}, {Flaminio}, {Floden}, {Fong}, {Font}, {Fornal}, {Forsyth}, {Franke}, {Frasca}, {Frasconi}, {Frederick}, {Freed}, {Frei}, {Freise}, {Frey}, {Fritschel}, {Frolov}, {Fronz{\'e}}, {Fulda}, {Fyffe}, {Gabbard}, {Gabella}, {Gadre}, {Gair}, {Gais}, {Galaudage}, {Gamba}, {Ganapathy}, {Ganguly}, {Gaonkar},
  {Garaventa}, {Garc{\'\i}a}, {Garc{\'\i}a-N{\'u}{\~n}ez}, {Garc{\'\i}a-Quir{\'o}s}, {Garufi}, {Gateley}, {Gaudio}, {Gayathri}, {Gemme}, {Gennai}, {George}, {George}, {Gerberding}, {Gergely}, {Gewecke}, {Ghonge}, {Ghosh}, {Ghosh}, {Ghosh}, {Ghosh}, {Giacomazzo}, {Giacoppo}, {Giaime}, {Giardina}, {Gibson}, {Gier}, {Giesler}, {Giri}, {Gissi}, {Glanzer}, {Gleckl}, {Godwin}, {Goetz}, {Goetz}, {Gohlke}, {Goncharov}, {Gonz{\'a}lez}, {Gopakumar}, {Gosselin}, {Gouaty}, {Gould}, {Grace}, {Grado}, {Granata}, {Granata}, {Grant}, {Gras}, {Grassia}, {Gray}, {Gray}, {Greco}, {Green}, {Green}, {Gretarsson}, {Gretarsson}, {Griffith}, {Griffiths}, {Griggs}, {Grignani}, {Grimaldi}, {Grimm}, {Grote}, {Grunewald}, {Gruning}, {Guerra}, {Guidi}, {Guimaraes}, {Guix{\'e}}, {Gulati}, {Guo}, {Guo}, {Gupta}, {Gupta}, {Gupta}, {Gustafson}, {Gustafson}, {Guzman}, {Haegel}, {Halim}, {Hall}, {Hamilton}, {Hammond}, {Haney}, {Hanks}, {Hanna}, {Hannam}, {Hannuksela}, {Hansen}, {Hansen}, {Hanson}, {Harder}, {Hardwick}, {Haris}, {Harms},
  {Harry}, {Harry}, {Hartwig}, {Haskell}, {Hasskew}, {Haster}, {Haughian}, {Hayes}, {Healy}, {Heidmann}, {Heidt}, {Heintze}, {Heinze}, {Heinzel}, {Heitmann}, {Hellman}, {Hello}, {Helmling-Cornell}, {Hemming}, {Hendry}, {Heng}, {Hennes}, {Hennig}, {Hennig}, {Hernandez}, {Vivanco}, {Heurs}, {Hild}, {Hill}, {Hines}, {Hochheim}, {Hofman}, {Hohmann}, {Holcomb}, {Holland}, {Holley-Bockelmann}, {Hollows}, {Holmes}, {Holt}, {Holz}, {Hopkins}, {Hough}, {Hourihane}, {Howell}, {Hoy}, {Hoyland}, {Hreibi}, {Hsu}, {Huang}, {H{\"u}bner}, {Huddart}, {Hughey}, {Hui}, {Husa}, {Huttner}, {Huxford}, {Huynh-Dinh}, {Idzkowski}, {Iess}, {Ingram}, {Isi}, {Isleif}, {Iyer}, {JaberianHamedan}, {Jacqmin}, {Jadhav}, {Jadhav}, {James}, {Jan}, {Jani}, {Janquart}, {Janssens}, {Janthalur}, {Jaranowski}, {Jariwala}, {Jaume}, {Jenkins}, {Jenner}, {Jeunon}, {Jia}, {Johns}, {Johnson-McDaniel}, {Jones}, {Jones}, {Jones}, {Jones}, {Jones}, {Jonker}, {Ju}, {Junker}, {Juste}, {Kalaghatgi}, {Kalogera}, {Kamai}, {Kandhasamy}, {Kang}, {Kanner}, {Kao},
  {Kapadia}, {Kapasi}, {Karat}, {Karathanasis}, {Karki}, {Kashyap}, {Kasprzack}, {Kastaun}, {Katsanevas}, {Katsavounidis}, {Katzman}, {Kaur}, {Kawabe}, {K{\'e}f{\'e}lian}, {Keitel}, {Key}, {Khadka}, {Khalili}, {Khan}, {Khazanov}, {Khetan}, {Khursheed}, {Kijbunchoo}, {Kim}, {Kim}, {Kim}, {Kim}, {Kim}, {Kimball}, {Kinley-Hanlon}, {Kirchhoff}, {Kissel}, {Kleybolte}, {Klimenko}, {Knee}, {Knowles}, {Knyazev}, {Koch}, {Koekoek}, {Koley}, {Kolitsidou}, {Kolstein}, {Komori}, {Kondrashov}, {Kontos}, {Koper}, {Korobko}, {Kovalam}, {Kozak}, {Kringel}, {Krishnendu}, {Kr{\'o}lak}, {Kuehn}, {Kuei}, {Kuijer}, {Kulkarni}, {Kumar}, {Kumar}, {Kumar}, {Kumar}, {Kuns}, {Kuwahara}, {Lagabbe}, {Laghi}, {Lalande}, {Lam}, {Lamberts}, {Landry}, {Lane}, {Lang}, {Lange}, {Lantz}, {La Rosa}, {Lartaux-Vollard}, {Lasky}, {Laxen}, {Lazzarini}, {Lazzaro}, {Leaci}, {Leavey}, {Lecoeuche}, {Lee}, {Lee}, {Lee}, {Lee}, {Lehmann}, {Lema{\^\i}tre}, {Leroy}, {Letendre}, {Levesque}, {Levin}, {Leviton}, {Leyde}, {Li}, {Li}, {Li}, {Li}, {Li}, {Linde},
  {Linker}, {Linley}, {Littenberg}, {Liu}, {Liu}, {Liu}, {Llamas}, {Llorens-Monteagudo}, {Lo}, {Lockwood}, {London}, {Longo}, {Lopez}, {Portilla}, {Lorenzini}, {Loriette}, {Lormand}, {Losurdo}, {Lott}, {Lough}, {Lousto}, {Lovelace}, {Lucaccioni}, {L{\"u}ck}, {Lumaca}, {Lundgren}, {Lynam}, {Macas}, {MacInnis}, {Macleod}, {MacMillan}, {Macquet}, {Hernandez}, {Magazz{\`u}}, {Magee}, {Maggiore}, {Magnozzi}, {Mahesh}, {Majorana}, {Makarem}, {Maksimovic}, {Maliakal}, {Malik}, {Man}, {Mandic}, {Mangano}, {Mango}, {Mansell}, {Manske}, {Mantovani}, {Mapelli}, {Marchesoni}, {Marion}, {Mark}, {M{\'a}rka}, {M{\'a}rka}, {Markakis}, {Markosyan}, {Markowitz}, {Maros}, {Marquina}, {Marsat}, {Martelli}, {Martin}, {Martin}, {Martinez}, {Martinez}, {Martinez}, {Martinovic}, {Martynov}, {Marx}, {Masalehdan}, {Mason}, {Massera}, {Masserot}, {Massinger}, {Masso-Reid}, {Mastrogiovanni}, {Matas}, {Mateu-Lucena}, {Matichard}, {Matiushechkina}, {Mavalvala}, {McCann}, {McCarthy}, {McClelland}, {McClincy}, {McCormick}, {McCuller},
  {McGhee}, {McGuire}, {McIsaac}, {McIver}, {McRae}, {McWilliams}, {Meacher}, {Mehmet}, {Mehta}, {Meijer}, {Melatos}, {Melchor}, {Mendell}, {Menendez-Vazquez}, {Menoni}, {Mercer}, {Mereni}, {Merfeld}, {Merilh}, {Merritt}, {Merzougui}, {Meshkov}, {Messenger}, {Messick}, {Meyers}, {Meylahn}, {Mhaske}, {Miani}, {Miao}, {Michaloliakos}, {Michel}, {Middleton}, {Milano}, {Miller}, {Miller}, {Miller}, {Millhouse}, {Mills}, {Milotti}, {Minazzoli}, {Minenkov}, {Mir}, {Miravet-Ten{\'e}s}, {Mishra}, {Mishra}, {Mistry}, {Mitra}, {Mitrofanov}, {Mitselmakher}, {Mittleman}, {Mo}, {Moguel}, {Mogushi}, {Mohapatra}, {Mohite}, {Molina}, {Molina-Ruiz}, {Mondin}, {Montani}, {Moore}, {Moraru}, {Morawski}, {More}, {Moreno}, {Moreno}, {Morisaki}, {Mours}, {Mow-Lowry}, {Mozzon}, {Muciaccia}, {Mukherjee}, {Mukherjee}, {Mukherjee}, {Mukherjee}, {Mukherjee}, {Mukund}, {Mullavey}, {Munch}, {Mu{\~n}iz}, {Murray}, {Musenich}, {Muusse}, {Nadji}, {Nagar}, {Napolano}, {Nardecchia}, {Naticchioni}, {Nayak}, {Nayak}, {Neil}, {Neilson},
  {Nelemans}, {Nelson}, {Nery}, {Neubauer}, {Neunzert}, {Ng}, {Ng}, {Nguyen}, {Nguyen}, {Nguyen}, {Nichols}, {Nissanke}, {Nitoglia}, {Nocera}, {Norman}, {North}, {Nuttall}, {Oberling}, {O'Brien}, {O'Dell}, {Oelker}, {Oganesyan}, {Oh}, {Oh}, {Ohme}, {Ohta}, {Okada}, {Olivetto}, {Oram}, {O'Reilly}, {Ormiston}, {Ormsby}, {Ortega}, {O'Shaughnessy}, {O'Shea}, {Ossokine}, {Osthelder}, {Ottaway}, {Overmier}, {Pace}, {Pagano}, {Page}, {Pagliaroli}, {Pai}, {Pai}, {Palamos}, {Palashov}, {Palomba}, {Pan}, {Panda}, {Pang}, {Pankow}, {Pannarale}, {Pant}, {Panther}, {Paoletti}, {Paoli}, {Paolone}, {Park}, {Parker}, {Pascucci}, {Pasqualetti}, {Passaquieti}, {Passuello}, {Patel}, {Pathak}, {Patricelli}, {Patron}, {Patrone}, {Paul}, {Payne}, {Pedraza}, {Pegoraro}, {Pele}, {Penn}, {Perego}, {Pereira}, {Pereira}, {Perez}, {P{\'e}rigois}, {Perkins}, {Perreca}, {Perri{\`e}s}, {Petermann}, {Petterson}, {Pfeiffer}, {Pham}, {Phukon}, {Piccinni}, {Pichot}, {Piendibene}, {Piergiovanni}, {Pierini}, {Pierro}, {Pillant}, {Pillas},
  {Pilo}, {Pinard}, {Pinto}, {Pinto}, {Piotrzkowski}, {Pirello}, {Pitkin}, {Placidi}, {Planas}, {Plastino}, {Pluchar}, {Poggiani}, {Polini}, {Pong}, {Ponrathnam}, {Popolizio}, {Porter}, {Poulton}, {Powell}, {Pracchia}, {Pradier}, {Prajapati}, {Prasai}, {Prasanna}, {Pratten}, {Principe}, {Prodi}, {Prokhorov}, {Prosposito}, {Prudenzi}, {Puecher}, {Punturo}, {Puosi}, {Puppo}, {P{\"u}rrer}, {Qi}, {Quetschke}, {Quitzow-James}, {Raab}, {Raaijmakers}, {Radkins}, {Radulesco}, {Raffai}, {Rail}, {Raja}, {Rajan}, {Ramirez}, {Ramirez}, {Ramos-Buades}, {Rana}, {Rapagnani}, {Rapol}, {Ray}, {Raymond}, {Raza}, {Razzano}, {Read}, {Rees}, {Regimbau}, {Rei}, {Reid}, {Reid}, {Reitze}, {Relton}, {Renzini}, {Rettegno}, {Reza}, {Rezac}, {Ricci}, {Richards}, {Richardson}, {Richardson}, {Riemenschneider}, {Riles}, {Rinaldi}, {Rink}, {Rizzo}, {Robertson}, {Robie}, {Robinet}, {Rocchi}, {Rodriguez}, {Rolland}, {Rollins}, {Romanelli}, {Romano}, {Romel}, {Romero-Rodr{\'\i}guez}, {Romero-Shaw}, {Romie}, {Ronchini}, {Rosa}, {Rose},
  {Rosell}, {Rosi{\'n}ska}, {Ross}, {Rowan}, {Rowlinson}, {Roy}, {Roy}, {Roy}, {Rozza}, {Ruggi}, {Ruiz-Rocha}, {Ryan}, {Sachdev}, {Sadecki}, {Sadiq}, {Sakellariadou}, {Salafia}, {Salconi}, {Saleem}, {Salemi}, {Samajdar}, {Sanchez}, {Sanchez}, {Sanchez}, {Sanchis-Gual}, {Sanders}, {Sanuy}, {Saravanan}, {Sarin}, {Sassolas}, {Satari}, {Sauter}, {Savage}, {Sawant}, {Sawant}, {Sayah}, {Schaetzl}, {Scheel}, {Scheuer}, {Schiworski}, {Schmidt}, {Schmidt}, {Schnabel}, {Schneewind}, {Schofield}, {Sch{\"o}nbeck}, {Schulte}, {Schutz}, {Schwartz}, {Scott}, {Scott}, {Seglar-Arroyo}, {Sellers}, {Sengupta}, {Sentenac}, {Seo}, {Sequino}, {Sergeev}, {Setyawati}, {Shaffer}, {Shahriar}, {Shams}, {Sharma}, {Sharma}, {Shawhan}, {Shcheblanov}, {Shikauchi}, {Shoemaker}, {Shoemaker}, {ShyamSundar}, {Sieniawska}, {Sigg}, {Singer}, {Singh}, {Singh}, {Singha}, {Sintes}, {Sipala}, {Skliris}, {Slagmolen}, {Slaven-Blair}, {Smetana}, {Smith}, {Smith}, {Soldateschi}, {Somala}, {Son}, {Soni}, {Soni}, {Sordini}, {Sorrentino}, {Sorrentino},
  {Soulard}, {Souradeep}, {Sowell}, {Spagnuolo}, {Spencer}, {Spera}, {Srinivasan}, {Srivastava}, {Srivastava}, {Staats}, {Stachie}, {Steer}, {Steinhoff}, {Steinlechner}, {Steinlechner}, {Stevenson}, {Stops}, {Stover}, {Strain}, {Strang}, {Stratta}, {Strunk}, {Sturani}, {Stuver}, {Sudhagar}, {Sudhir}, {Suh}, {Summerscales}, {Sun}, {Sun}, {Sunil}, {Sur}, {Suresh}, {Sutton}, {Swinkels}, {Szczepa{\'n}czyk}, {Szewczyk}, {Tacca}, {Tait}, {Talbot}, {Talbot}, {Tanasijczuk}, {Tanner}, {Tao}, {Tao}, {Mart{\'\i}n}, {Taranto}, {Tasson}, {Tenorio}, {Terhune}, {Terkowski}, {Thirugnanasambandam}, {Thomas}, {Thomas}, {Thomas}, {Thompson}, {Thondapu}, {Thorne}, {Thrane}, {Tiwari}, {Tiwari}, {Tiwari}, {Toivonen}, {Toland}, {Tolley}, {Tonelli}, {Torres-Forn{\'e}}, {Torrie}, {e Melo}, {T{\"o}yr{\"a}}, {Trapananti}, {Travasso}, {Traylor}, {Trevor}, {Tringali}, {Tripathee}, {Troiano}, {Trovato}, {Trozzo}, {Trudeau}, {Tsai}, {Tsai}, {Tsang}, {Tse}, {Tso}, {Tsukada}, {Tsuna}, {Tsutsui}, {Turbang}, {Turconi}, {Ubhi}, {Udall}, {Ueno},
  {Unnikrishnan}, {Urban}, {Utina}, {Vahlbruch}, {Vajente}, {Vajpeyi}, {Valdes}, {Valentini}, {Valsan}, {van Bakel}, {van Beuzekom}, {van den Brand}, {Van Den Broeck}, {Vander-Hyde}, {van der Schaaf}, {van Heijningen}, {Vanosky}, {van Remortel}, {Vardaro}, {Vargas}, {Varma}, {Vas{\'u}th}, {Vecchio}, {Vedovato}, {Veitch}, {Veitch}, {Venneberg}, {Venugopalan}, {Verkindt}, {Verma}, {Verma}, {Veske}, {Vetrano}, {Vicer{\'e}}, {Vidyant}, {Viets}, {Vijaykumar}, {Villa-Ortega}, {Vinet}, {Virtuoso}, {Vitale}, {Vo}, {Vocca}, {von Reis}, {von Wrangel}, {Vorvick}, {Vyatchanin}, {Wade}, {Wade}, {Wagner}, {Walet}, {Walker}, {Wallace}, {Wallace}, {Walsh}, {Wang}, {Wang}, {Ward}, {Warner}, {Was}, {Washington}, {Watchi}, {Weaver}, {Webster}, {Weinert}, {Weinstein}, {Weiss}, {Weller}, {Weller}, {Wellmann}, {Wen}, {We{\ss}els}, {Wette}, {Whelan}, {White}, {Whiting}, {Whittle}, {Wilken}, {Williams}, {Williams}, {Williamson}, {Willis}, {Willke}, {Wilson}, {Winkler}, {Wipf}, {Wlodarczyk}, {Woan}, {Woehler}, {Wofford}, {Wong},
  {Wu}, {Wysocki}, {Xiao}, {Yamamoto}, {Yang}, {Yang}, {Yang}, {Yang}, {Yap}, {Yeeles}, {Yelikar}, {Ying}, {Yoo}, {Yu}, {Yu}, {Zadro{\.Z}ny}, {Zanolin}, {Zelenova}, {Zendri}, {Zevin}, {Zhang}, {Zhang}, {Zhang}, {Zhang}, {Zhao}, {Zhao}, {Zhao}, {Zhou}, {Zhou}, {Zhu}, {Zimmerman}, {Zlochower}, {Zucker}, {Zweizig}, {LIGO Scientific Collaboration}, \& {the Virgo Collaboration}}]{2024PhRvD.109b2001A}
---. 2024, \prd, 109, 022001, \dodoi{10.1103/PhysRevD.109.022001}

\bibitem[{{Ajith} {et~al.}(2011){Ajith}, {Hannam}, {Husa}, {Chen}, {Br{\"u}gmann}, {Dorband}, {M{\"u}ller}, {Ohme}, {Pollney}, {Reisswig}, {Santamar{\'\i}a}, \& {Seiler}}]{2011PhRvL.106x1101A}
{Ajith}, P., {Hannam}, M., {Husa}, S., {et~al.} 2011, \prl, 106, 241101, \dodoi{10.1103/PhysRevLett.106.241101}

\bibitem[{{Ashton} {et~al.}(2019{\natexlab{a}}){Ashton}, {H{\"u}bner}, {Lasky}, {Talbot}, {Ackley}, {Biscoveanu}, {Chu}, {Divakarla}, {Easter}, {Goncharov}, {Hernandez Vivanco}, {Harms}, {Lower}, {Meadors}, {Melchor}, {Payne}, {Pitkin}, {Powell}, {Sarin}, {Smith}, \& {Thrane}}]{2019ApJS..241...27A}
{Ashton}, G., {H{\"u}bner}, M., {Lasky}, P.~D., {et~al.} 2019{\natexlab{a}}, \apjs, 241, 27, \dodoi{10.3847/1538-4365/ab06fc}

\bibitem[{{Ashton} {et~al.}(2019{\natexlab{b}}){Ashton}, {H{\"u}bner}, {Lasky}, {Talbot}, {Ackley}, {Biscoveanu}, {Chu}, {Divarkala}, {Easter}, {Goncharov}, {Hernandez Vivanco}, {Harms}, {Lower}, {Meadors}, {Melchor}, {Payne}, {Pitkin}, {Powell}, {Sarin}, {Smith}, \& {Thrane}}]{2019ascl.soft01011A}
---. 2019{\natexlab{b}}, {Bilby: Bayesian inference library}, Astrophysics Source Code Library, record ascl:1901.011

\bibitem[{{Bavera} {et~al.}(2020){Bavera}, {Fragos}, {Qin}, {Zapartas}, {Neijssel}, {Mandel}, {Batta}, {Gaebel}, {Kimball}, \& {Stevenson}}]{2020A&A...635A..97B}
{Bavera}, S.~S., {Fragos}, T., {Qin}, Y., {et~al.} 2020, \aap, 635, A97, \dodoi{10.1051/0004-6361/201936204}

\bibitem[{{Biscoveanu} {et~al.}(2022){Biscoveanu}, {Callister}, {Haster}, {Ng}, {Vitale}, \& {Farr}}]{2022ApJ...932L..19B}
{Biscoveanu}, S., {Callister}, T.~A., {Haster}, C.-J., {et~al.} 2022, \apjl, 932, L19, \dodoi{10.3847/2041-8213/ac71a8}

\bibitem[{{Biscoveanu} {et~al.}(2021){Biscoveanu}, {Isi}, {Vitale}, \& {Varma}}]{2021PhRvL.126q1103B}
{Biscoveanu}, S., {Isi}, M., {Vitale}, S., \& {Varma}, V. 2021, \prl, 126, 171103, \dodoi{10.1103/PhysRevLett.126.171103}

\bibitem[{{Biwer} {et~al.}(2019){Biwer}, {Capano}, {De}, {Cabero}, {Brown}, {Nitz}, \& {Raymond}}]{pycbc2019PASP}
{Biwer}, C.~M., {Capano}, C.~D., {De}, S., {et~al.} 2019, \pasp, 131, 024503, \dodoi{10.1088/1538-3873/aaef0b}

\bibitem[{{Broekgaarden} {et~al.}(2022){Broekgaarden}, {Berger}, {Stevenson}, {Justham}, {Mandel}, {Chru{\'s}li{\'n}ska}, {van Son}, {Wagg}, {Vigna-G{\'o}mez}, {de Mink}, {Chattopadhyay}, \& {Neijssel}}]{2022MNRAS.516.5737B}
{Broekgaarden}, F.~S., {Berger}, E., {Stevenson}, S., {et~al.} 2022, \mnras, 516, 5737, \dodoi{10.1093/mnras/stac1677}

\bibitem[{{Callister} {et~al.}(2021){Callister}, {Haster}, {Ng}, {Vitale}, \& {Farr}}]{2021ApJ...922L...5C}
{Callister}, T.~A., {Haster}, C.-J., {Ng}, K. K.~Y., {Vitale}, S., \& {Farr}, W.~M. 2021, \apjl, 922, L5, \dodoi{10.3847/2041-8213/ac2ccc}

\bibitem[{{Callister} {et~al.}(2022){Callister}, {Miller}, {Chatziioannou}, \& {Farr}}]{2022ApJ...937L..13C}
{Callister}, T.~A., {Miller}, S.~J., {Chatziioannou}, K., \& {Farr}, W.~M. 2022, \apjl, 937, L13, \dodoi{10.3847/2041-8213/ac847e}

\bibitem[{{Edelman} {et~al.}(2022){Edelman}, {Doctor}, {Godfrey}, \& {Farr}}]{2022ApJ...924..101E}
{Edelman}, B., {Doctor}, Z., {Godfrey}, J., \& {Farr}, B. 2022, \apj, 924, 101, \dodoi{10.3847/1538-4357/ac3667}

\bibitem[{{Farmer} {et~al.}(2019){Farmer}, {Renzo}, {de Mink}, {Marchant}, \& {Justham}}]{2019ApJ...887...53F}
{Farmer}, R., {Renzo}, M., {de Mink}, S.~E., {Marchant}, P., \& {Justham}, S. 2019, \apj, 887, 53, \dodoi{10.3847/1538-4357/ab518b}

\bibitem[{{Galaudage} {et~al.}(2021){Galaudage}, {Talbot}, {Nagar}, {Jain}, {Thrane}, \& {Mandel}}]{2021ApJ...921L..15G}
{Galaudage}, S., {Talbot}, C., {Nagar}, T., {et~al.} 2021, \apjl, 921, L15, \dodoi{10.3847/2041-8213/ac2f3c}

\bibitem[{{Garc{\'\i}a-Bellido} {et~al.}(2021){Garc{\'\i}a-Bellido}, {Nu{\~n}o Siles}, \& {Ruiz Morales}}]{2021PDU....3100791G}
{Garc{\'\i}a-Bellido}, J., {Nu{\~n}o Siles}, J.~F., \& {Ruiz Morales}, E. 2021, Physics of the Dark Universe, 31, 100791, \dodoi{10.1016/j.dark.2021.100791}

\bibitem[{{Hannam} {et~al.}(2014){Hannam}, {Schmidt}, {Boh{\'e}}, {Haegel}, {Husa}, {Ohme}, {Pratten}, \& {P{\"u}rrer}}]{2014PhRvL.113o1101H}
{Hannam}, M., {Schmidt}, P., {Boh{\'e}}, A., {et~al.} 2014, \prl, 113, 151101, \dodoi{10.1103/PhysRevLett.113.151101}

\bibitem[{{Hotokezaka} \& {Piran}(2017)}]{2017ApJ...842..111H}
{Hotokezaka}, K., \& {Piran}, T. 2017, \apj, 842, 111, \dodoi{10.3847/1538-4357/aa6f61}

\bibitem[{{Kushnir} {et~al.}(2016){Kushnir}, {Zaldarriaga}, {Kollmeier}, \& {Waldman}}]{2016MNRAS.462..844K}
{Kushnir}, D., {Zaldarriaga}, M., {Kollmeier}, J.~A., \& {Waldman}, R. 2016, \mnras, 462, 844, \dodoi{10.1093/mnras/stw1684}

\bibitem[{{Li} {et~al.}(2024){Li}, {Tang}, {Gao}, {Wu}, \& {Wang}}]{2024arXiv240409668L}
{Li}, Y.-J., {Tang}, S.-P., {Gao}, S.-J., {Wu}, D.-C., \& {Wang}, Y.-Z. 2024, arXiv e-prints, arXiv:2404.09668, \dodoi{10.48550/arXiv.2404.09668}

\bibitem[{{Li} {et~al.}(2021){Li}, {Wang}, {Han}, {Tang}, {Yuan}, {Fan}, \& {Wei}}]{2021ApJ...917...33L}
{Li}, Y.-J., {Wang}, Y.-Z., {Han}, M.-Z., {et~al.} 2021, \apj, 917, 33, \dodoi{10.3847/1538-4357/ac0971}

\bibitem[{{Li} {et~al.}(2023){Li}, {Wang}, {Tang}, \& {Fan}}]{2023arXiv230302973L}
{Li}, Y.-J., {Wang}, Y.-Z., {Tang}, S.-P., \& {Fan}, Y.-Z. 2023, arXiv e-prints, arXiv:2303.02973, \dodoi{10.48550/arXiv.2303.02973}

\bibitem[{{Li} {et~al.}(2022){Li}, {Wang}, {Tang}, {Yuan}, {Fan}, \& {Wei}}]{2022ApJ...933L..14L}
{Li}, Y.-J., {Wang}, Y.-Z., {Tang}, S.-P., {et~al.} 2022, \apjl, 933, L14, \dodoi{10.3847/2041-8213/ac78dd}

\bibitem[{{Mandel} \& {Fragos}(2020)}]{2020ApJ...895L..28M}
{Mandel}, I., \& {Fragos}, T. 2020, \apjl, 895, L28, \dodoi{10.3847/2041-8213/ab8e41}

\bibitem[{{Mandel} \& {Smith}(2021)}]{2021ApJ...922L..14M}
{Mandel}, I., \& {Smith}, R. J.~E. 2021, \apjl, 922, L14, \dodoi{10.3847/2041-8213/ac35dd}

\bibitem[{{Miller} {et~al.}(2020){Miller}, {Callister}, \& {Farr}}]{2020ApJ...895..128M}
{Miller}, S., {Callister}, T.~A., \& {Farr}, W.~M. 2020, \apj, 895, 128, \dodoi{10.3847/1538-4357/ab80c0}

\bibitem[{{Mould} {et~al.}(2022){Mould}, {Gerosa}, {Broekgaarden}, \& {Steinle}}]{2022MNRAS.517.2738M}
{Mould}, M., {Gerosa}, D., {Broekgaarden}, F.~S., \& {Steinle}, N. 2022, \mnras, 517, 2738, \dodoi{10.1093/mnras/stac2859}

\bibitem[{{Qin} {et~al.}(2018){Qin}, {Fragos}, {Meynet}, {Andrews}, {S{\o}rensen}, \& {Song}}]{2018A&A...616A..28Q}
{Qin}, Y., {Fragos}, T., {Meynet}, G., {et~al.} 2018, \aap, 616, A28, \dodoi{10.1051/0004-6361/201832839}

\bibitem[{{Qin} {et~al.}(2022{\natexlab{a}}){Qin}, {Wang}, {Wu}, {Meynet}, \& {Song}}]{2022ApJ...924..129Q}
{Qin}, Y., {Wang}, Y.-Z., {Wu}, D.-H., {Meynet}, G., \& {Song}, H. 2022{\natexlab{a}}, \apj, 924, 129, \dodoi{10.3847/1538-4357/ac3982}

\bibitem[{{Qin} {et~al.}(2022{\natexlab{b}}){Qin}, {Wang}, {Bavera}, {Wu}, {Meynet}, {Wang}, {Hu}, {Zhu}, {Wu}, {Shu}, {Peng}, {Song}, \& {Wei}}]{2022ApJ...941..179Q}
{Qin}, Y., {Wang}, Y.-Z., {Bavera}, S.~S., {et~al.} 2022{\natexlab{b}}, \apj, 941, 179, \dodoi{10.3847/1538-4357/aca40c}

\bibitem[{{Romero-Shaw} {et~al.}(2022){Romero-Shaw}, {Thrane}, \& {Lasky}}]{2022PASA...39...25R}
{Romero-Shaw}, I.~M., {Thrane}, E., \& {Lasky}, P.~D. 2022, PASA, 39, e025, \dodoi{10.1017/pasa.2022.24}

\bibitem[{{Roulet} {et~al.}(2021){Roulet}, {Chia}, {Olsen}, {Dai}, {Venumadhav}, {Zackay}, \& {Zaldarriaga}}]{2021PhRvD.104h3010R}
{Roulet}, J., {Chia}, H.~S., {Olsen}, S., {et~al.} 2021, \prd, 104, 083010, \dodoi{10.1103/PhysRevD.104.083010}

\bibitem[{{Roulet} \& {Zaldarriaga}(2019)}]{2019MNRAS.484.4216R}
{Roulet}, J., \& {Zaldarriaga}, M. 2019, \mnras, 484, 4216, \dodoi{10.1093/mnras/stz226}

\bibitem[{{Safarzadeh} {et~al.}(2020){Safarzadeh}, {Farr}, \& {Ramirez-Ruiz}}]{2020ApJ...894..129S}
{Safarzadeh}, M., {Farr}, W.~M., \& {Ramirez-Ruiz}, E. 2020, \apj, 894, 129, \dodoi{10.3847/1538-4357/ab80be}

\bibitem[{{Santamar{\'\i}a} {et~al.}(2010){Santamar{\'\i}a}, {Ohme}, {Ajith}, {Br{\"u}gmann}, {Dorband}, {Hannam}, {Husa}, {M{\"o}sta}, {Pollney}, {Reisswig}, {Robinson}, {Seiler}, \& {Krishnan}}]{2010PhRvD..82f4016S}
{Santamar{\'\i}a}, L., {Ohme}, F., {Ajith}, P., {et~al.} 2010, \prd, 82, 064016, \dodoi{10.1103/PhysRevD.82.064016}

\bibitem[{{Schmidt} {et~al.}(2015){Schmidt}, {Ohme}, \& {Hannam}}]{2015PhRvD..91b4043S}
{Schmidt}, P., {Ohme}, F., \& {Hannam}, M. 2015, \prd, 91, 024043, \dodoi{10.1103/PhysRevD.91.024043}

\bibitem[{{Talbot} \& {Thrane}(2017)}]{2017PhRvD..96b3012T}
{Talbot}, C., \& {Thrane}, E. 2017, \prd, 96, 023012, \dodoi{10.1103/PhysRevD.96.023012}

\bibitem[{{Talbot} \& {Thrane}(2018)}]{2018ApJ...856..173T}
---. 2018, \apj, 856, 173, \dodoi{10.3847/1538-4357/aab34c}

\bibitem[{{Tanikawa}(2024)}]{2024RvMPP...8...13T}
{Tanikawa}, A. 2024, Reviews of Modern Plasma Physics, 8, 13, \dodoi{10.1007/s41614-024-00153-8}

\bibitem[{{Thrane} \& {Talbot}(2019)}]{2019PASA...36...10T}
{Thrane}, E., \& {Talbot}, C. 2019, \pasa, 36, e010, \dodoi{10.1017/pasa.2019.2}

\bibitem[{{Tiwari} \& {Fairhurst}(2021)}]{2021ApJ...913L..19T}
{Tiwari}, V., \& {Fairhurst}, S. 2021, \apjl, 913, L19, \dodoi{10.3847/2041-8213/abfbe7}

\bibitem[{{Tong} {et~al.}(2022){Tong}, {Galaudage}, \& {Thrane}}]{2022PhRvD.106j3019T}
{Tong}, H., {Galaudage}, S., \& {Thrane}, E. 2022, \prd, 106, 103019, \dodoi{10.1103/PhysRevD.106.103019}

\bibitem[{{van Son} {et~al.}(2022){van Son}, {de Mink}, {Callister}, {Justham}, {Renzo}, {Wagg}, {Broekgaarden}, {Kummer}, {Pakmor}, \& {Mandel}}]{2022ApJ...931...17V}
{van Son}, L.~A.~C., {de Mink}, S.~E., {Callister}, T., {et~al.} 2022, \apj, 931, 17, \dodoi{10.3847/1538-4357/ac64a3}

\bibitem[{{Veske} {et~al.}(2021){Veske}, {Bartos}, {M{\'a}rka}, \& {M{\'a}rka}}]{2021ApJ...922..258V}
{Veske}, D., {Bartos}, I., {M{\'a}rka}, Z., \& {M{\'a}rka}, S. 2021, \apj, 922, 258, \dodoi{10.3847/1538-4357/ac27ac}

\bibitem[{{Vink} {et~al.}(2021){Vink}, {Higgins}, {Sander}, \& {Sabhahit}}]{2021MNRAS.504..146V}
{Vink}, J.~S., {Higgins}, E.~R., {Sander}, A. A.~C., \& {Sabhahit}, G.~N. 2021, \mnras, 504, 146, \dodoi{10.1093/mnras/stab842}

\bibitem[{{Vitale} {et~al.}(2022){Vitale}, {Biscoveanu}, \& {Talbot}}]{2022A&A...668L...2V}
{Vitale}, S., {Biscoveanu}, S., \& {Talbot}, C. 2022, \aap, 668, L2, \dodoi{10.1051/0004-6361/202245084}

\bibitem[{{Wang} {et~al.}(2022){Wang}, {Li}, {Vink}, {Fan}, {Tang}, {Qin}, \& {Wei}}]{2022ApJ...941L..39W}
{Wang}, Y.-Z., {Li}, Y.-J., {Vink}, J.~S., {et~al.} 2022, \apjl, 941, L39, \dodoi{10.3847/2041-8213/aca89f}

\bibitem[{{Wang} {et~al.}(2021){Wang}, {Tang}, {Liang}, {Han}, {Li}, {Jin}, {Fan}, \& {Wei}}]{2021ApJ...913...42W}
{Wang}, Y.-Z., {Tang}, S.-P., {Liang}, Y.-F., {et~al.} 2021, \apj, 913, 42, \dodoi{10.3847/1538-4357/abf5df}

\bibitem[{{Williams} {et~al.}(2021){Williams}, {Veitch}, \& {Messenger}}]{2021PhRvD.103j3006W}
{Williams}, M.~J., {Veitch}, J., \& {Messenger}, C. 2021, \prd, 103, 103006, \dodoi{10.1103/PhysRevD.103.103006}

\bibitem[{{Winch} {et~al.}(2024){Winch}, {Vink}, {Higgins}, \& {Sabhahitf}}]{2024MNRAS.529.2980W}
{Winch}, E. R.~J., {Vink}, J.~S., {Higgins}, E.~R., \& {Sabhahitf}, G.~N. 2024, \mnras, 529, 2980, \dodoi{10.1093/mnras/stae393}

\bibitem[{{Wysocki} {et~al.}(2019){Wysocki}, {Lange}, \& {O'Shaughnessy}}]{2019PhRvD.100d3012W}
{Wysocki}, D., {Lange}, J., \& {O'Shaughnessy}, R. 2019, \prd, 100, 043012, \dodoi{10.1103/PhysRevD.100.043012}

\bibitem[{{Ye} \& {Fishbach}(2024)}]{2024arXiv240212444Y}
{Ye}, C.~S., \& {Fishbach}, M. 2024, arXiv e-prints, arXiv:2402.12444, \dodoi{10.48550/arXiv.2402.12444}

\bibitem[{{Zevin} {et~al.}(2021){Zevin}, {Bavera}, {Berry}, {Kalogera}, {Fragos}, {Marchant}, {Rodriguez}, {Antonini}, {Holz}, \& {Pankow}}]{2021ApJ...910..152Z}
{Zevin}, M., {Bavera}, S.~S., {Berry}, C. P.~L., {et~al.} 2021, \apj, 910, 152, \dodoi{10.3847/1538-4357/abe40e}

\end{thebibliography}

\clearpage

\appendix

\counterwithin{figure}{section}
\counterwithin{table}{section}

\section{Supplementary Figures}\label{sec:app_post}

\begin{figure*}
	\centering
\includegraphics[width=1\linewidth]{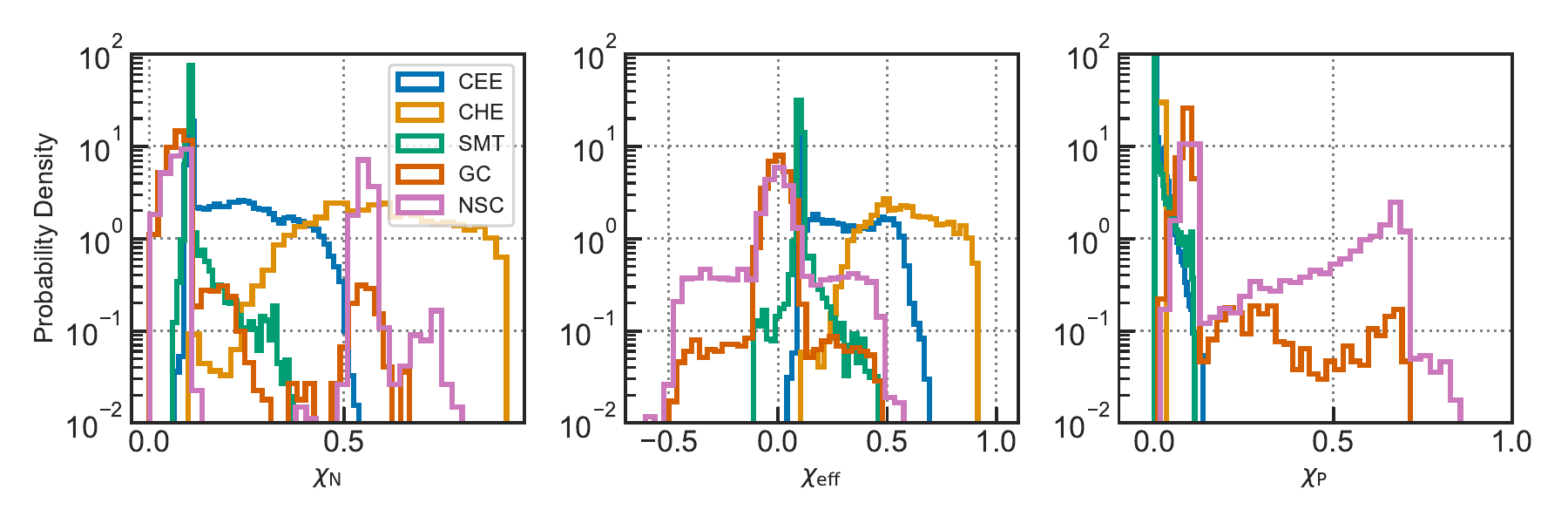} 
\caption{Theoretical prediction of $\chi_{\rm N}$ (left), $\chi_{\rm eff}$ (middle) and $\chi_{\rm p}$ (right) distributions for five different channels: the common envelope evolution (CEE), chemically homogeneous evolution (CHE), stable mass transfer (SMT), the dynamical capture in globular clusters (GC) as weill as in neuclear star clusters (NSC). The simulated data are taken from \citep{2021ApJ...910..152Z}, and we use the subset of data with common envelop efficiency $\alpha=1$ and BH natal spin of $0.1$. 
}
\label{fig:simulation_chis}
\end{figure*}

\begin{figure*}
	\centering  
\includegraphics[width=0.3\linewidth]{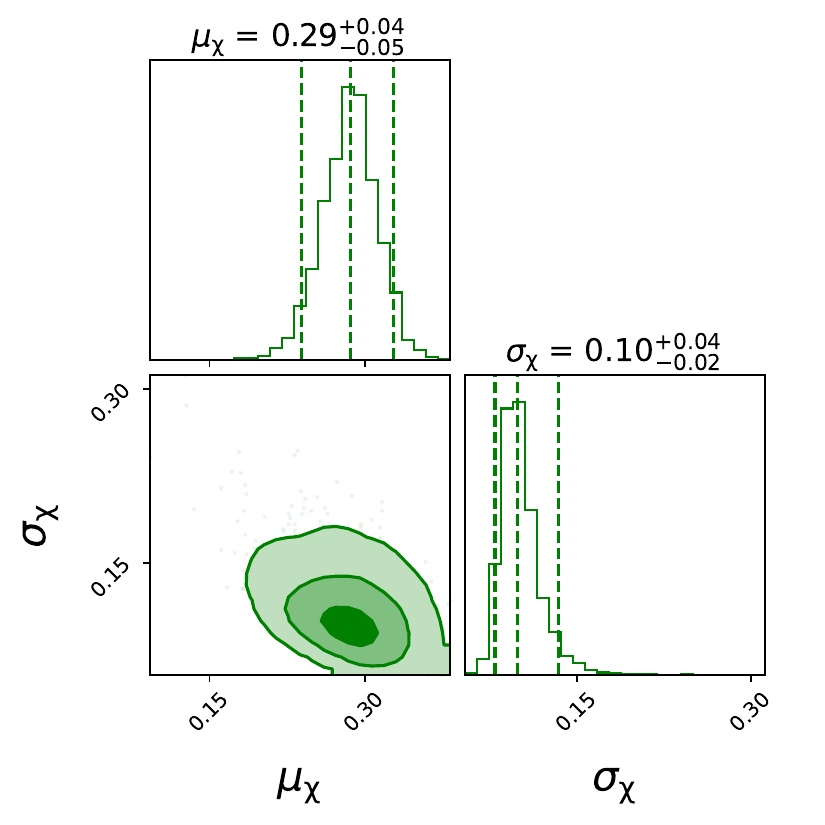}
\caption{Posterior distribution of hyper-parameters for model M1; the dashed lines in the marginal distribution represent the $90\%$ credible intervals. }
\label{fig:one_corner}
\end{figure*}

\begin{figure*}
	\centering  
\includegraphics[width=0.6\linewidth]{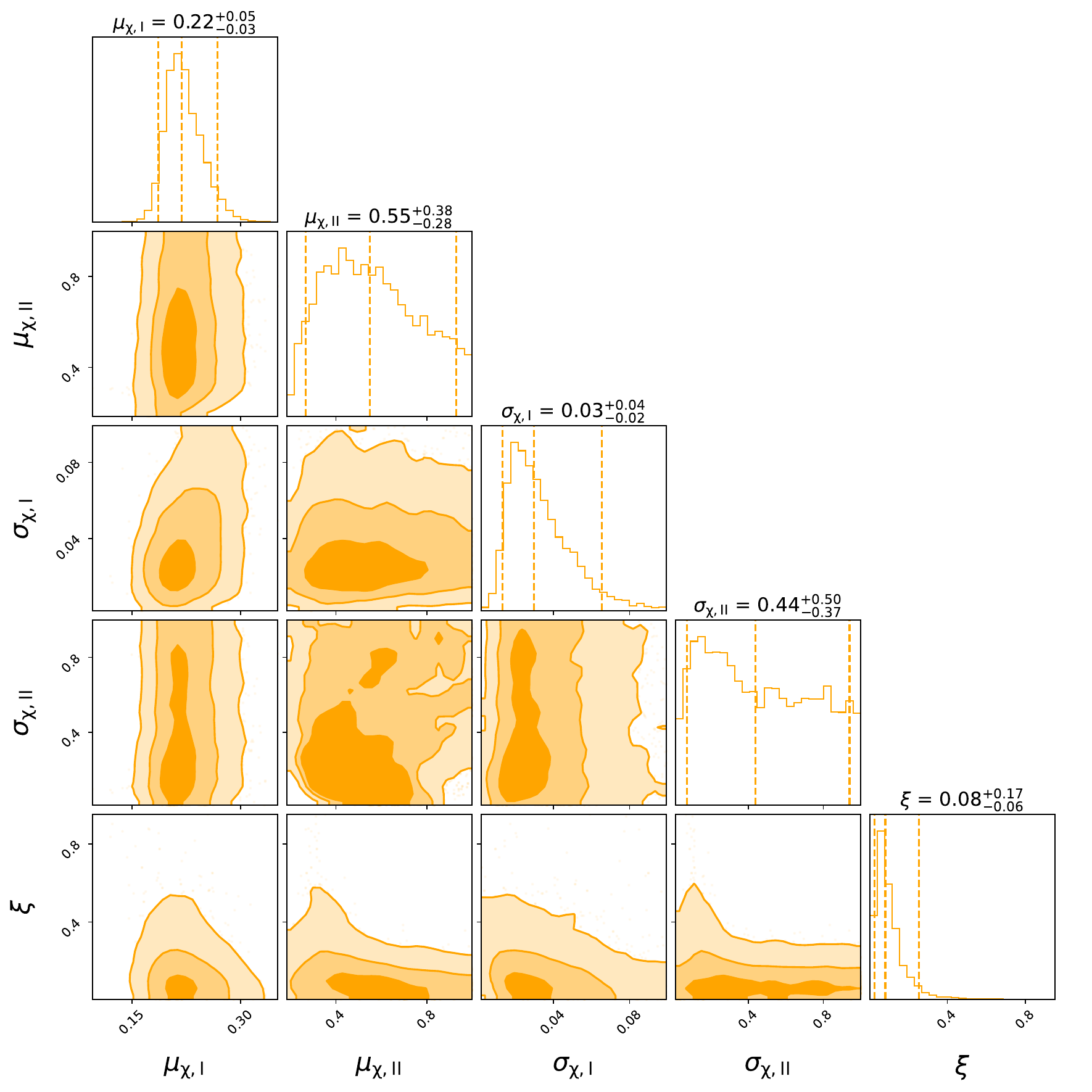}
\caption{The same as Fig.\ref{fig:one_corner}, but for model M2}
\label{fig:two_corner}
\end{figure*}

\begin{figure*}
	\centering  
\includegraphics[width=1.\linewidth]{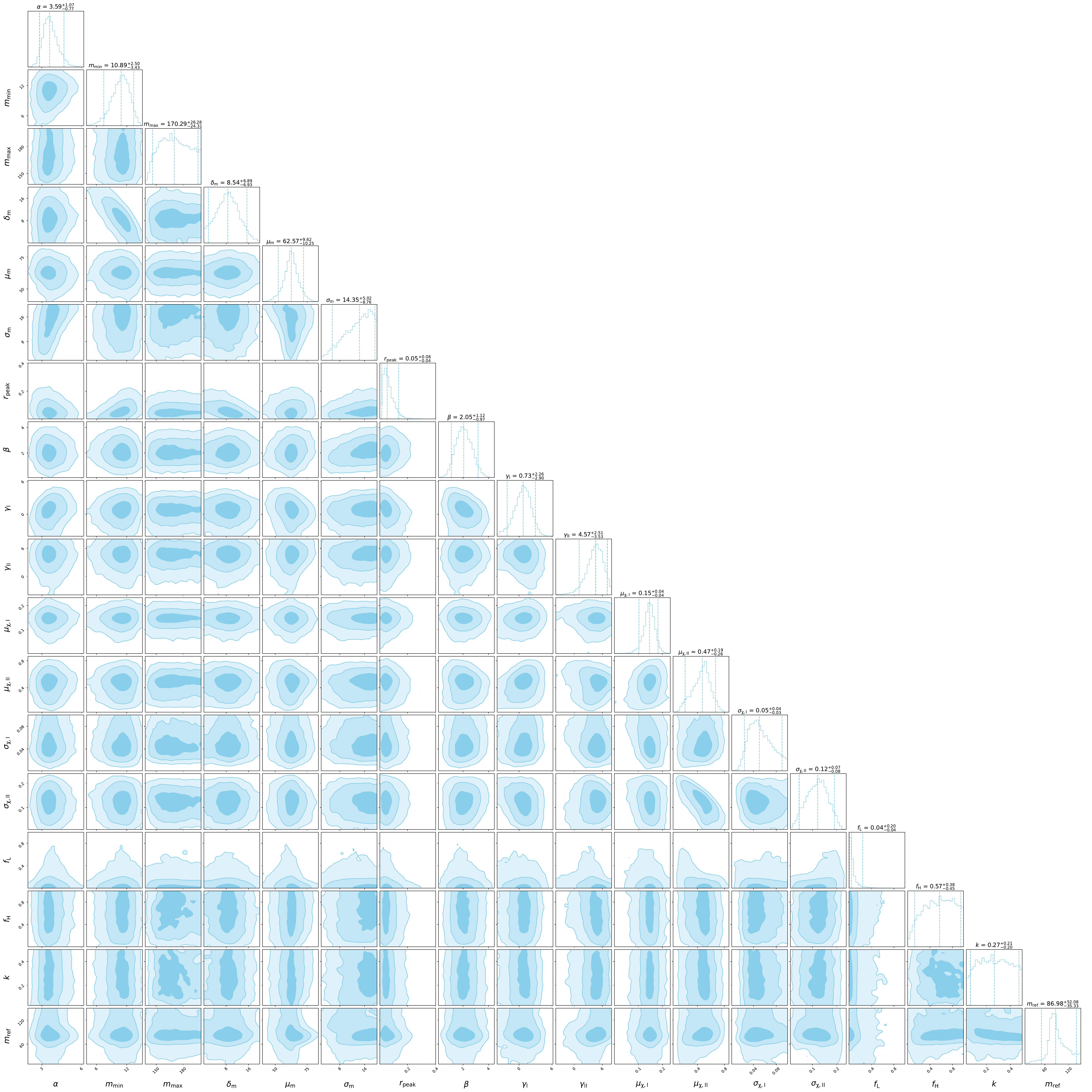}
\caption{Posterior distribution of hyper-parameters for the \textit{JOINT} model as defined in Sec.\ref{sec:joint}.}
\label{fig:full}
\end{figure*}

\begin{figure*}
	\centering  
\includegraphics[width=1\linewidth]{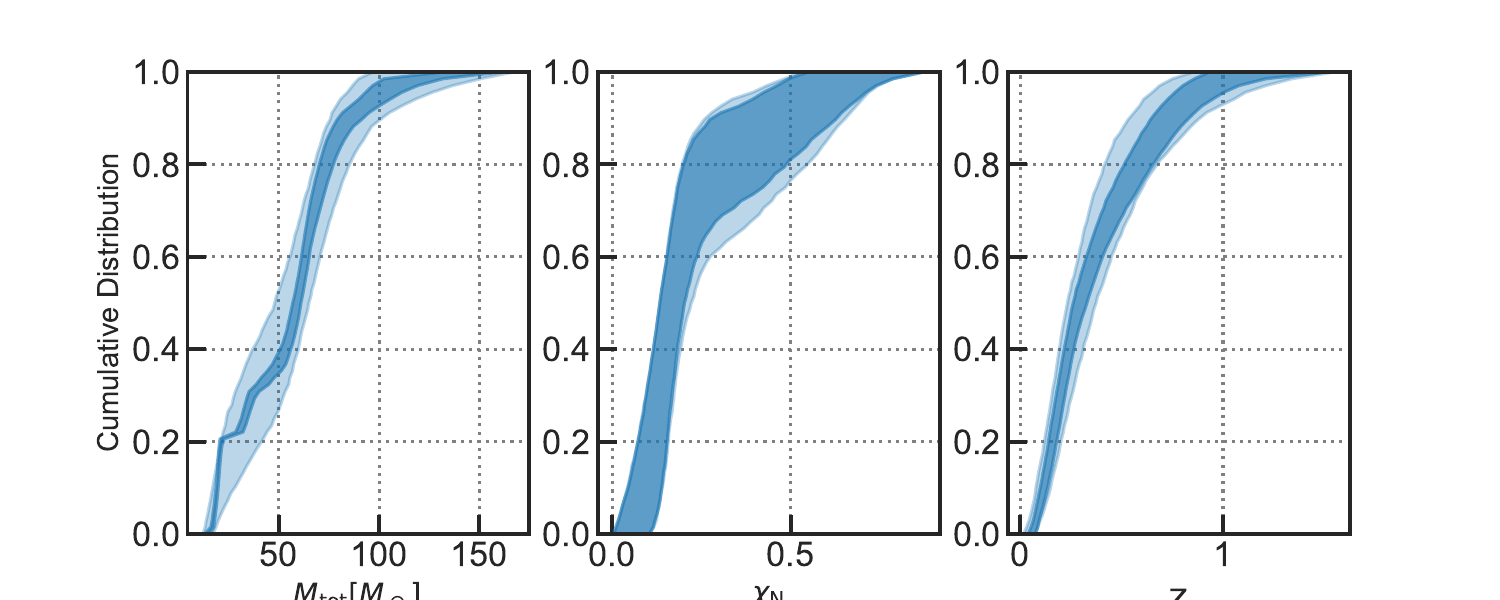}
\caption{Posterior predicted check for the observed total mass distribution (left), the $\chi_{\rm N}$ distribution (middle), and redshift distribution (bottom-right).}
\label{fig:ppcheck}
\end{figure*}

\clearpage

\section{Estimate the Fraction of Systems Disfavor Zero Individual Spin}\label{app:notzero}

\begin{figure*}
	\centering  
\includegraphics[width=0.8\linewidth]{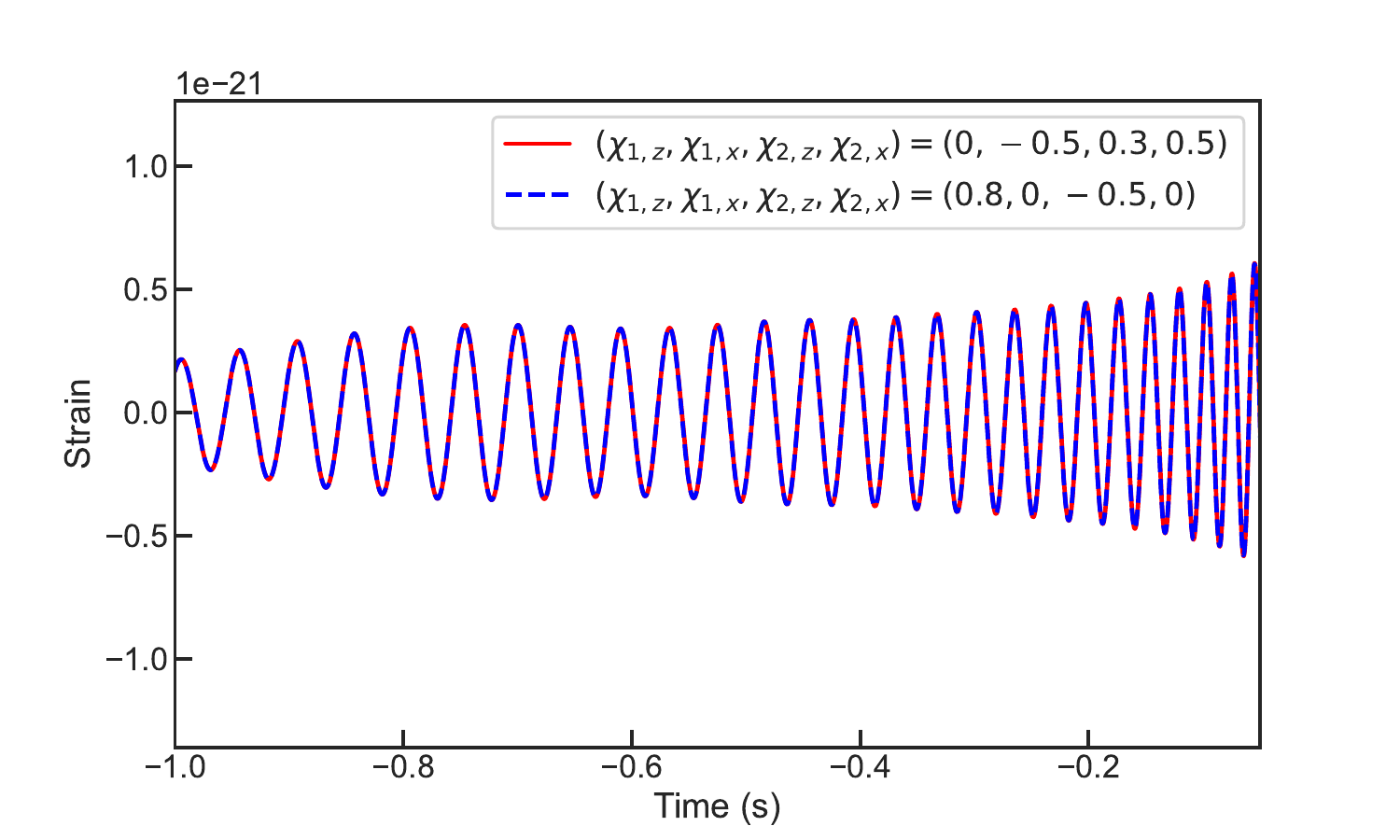}
\caption{Two gravitational waveforms for two $30-30 M_\odot$ BBHs with $\chi_{\rm N}=0.15$ but having different spin projections. The results are generated using the ``IMRPhenomXPHM" model in {\sc PyCBC} \citep{pycbc2019PASP}.}
\label{fig:strain}
\end{figure*}

For two merging black holes with mass ratio close to unity, different combinations of their spin projections are equivalent to each other in producing GW signals as long as they constitute the same $\chi_{\rm eff}$ and $\chi_{\rm p}$. In the case of $q=1$, this also means they should have identical $\chi_{\rm N}$. In Fig.\ref{fig:strain}, we demonstrate two overlapping waveforms from the merger of two $30-30 M_\odot$ BBHs with $\chi_{\rm N}=0.15$ but different individual spin projections. 

In a BBH system, accroding to Eq.~(\ref{eq:3})-(\ref{eq:4}), the maximum $\chi_{\rm N}$ available when only one BH is spinning are $\chi_{\rm N,max2}=q^2/(1+q^2)$ and $\chi_{\rm N,max1}=1/(1+q^2)$ for the cases of zero-spin primary and zero-spin secondary, respectively. For a giving observed event, we count the fraction of posterior samples with $\chi_{\rm N} > \chi_{{\rm N,max}i}$ to estimate the probability that the data disfavors the cases with one zero-spin BH, and the results are listed in the last two columns of Tab.\ref{tab:long_table}. We find that 29 of 69 events have $>50\%$ probability for $\chi_{\rm N} > \chi_{\rm N,max2}$, and 8 of 69 events have $>10\%$ probability for $\chi_{\rm N} > \chi_{\rm N,max1}$. Note that these are only rough estimations for events with mass ratio significantly deviates from unity. We have verified with a more precise analysis considering the combination of both $\chi_{\rm eff}$ and $\chi_{\rm p}$, and find that the numbers for $\chi_{\rm N} > \chi_{\rm N,max2}$ reduced to 13 and remains the same for $\chi_{\rm N} > \chi_{\rm N,max1}$, so the conclusion that ``the net spin of some events are more likely to exceed the maximum value that can be provided by the lighter black hole" still holds.

\begin{longtable}{|c|c|c|c|c|}
\caption{The total mass and dimensionless net spin for events included in this work} \label{tab:long_table} \\
\hline
Event & Total mass & $\chi_{\rm N}$ & $P(\chi_{\rm N}>\chi_{\rm N,max1})$ & $P(\chi_{\rm N}>\chi_{\rm N,max2})$ \\ \hline
\endfirsthead

\multicolumn{5}{c}%
{{\bfseries \tablename\ \thetable{} -- continued}} \\
\hline
Event & Total mass & $\chi_{\rm N}$ & $P(\chi_{\rm N}>\chi_{\rm N,max1})$ & $P(\chi_{\rm N}>\chi_{\rm N,max2})$ \\ \hline
\endhead

\hline \multicolumn{5}{r}{{Continued on next page}} \\ \endfoot

\hline \endlastfoot

GW150914\_095045 & $64.53^{+3.72}_{-3.16}$ & $0.28^{+0.30}_{-0.20}$ & 0.03 & 0.24 \\
GW151012\_095443 & $38.78^{+10.30}_{-4.74}$ & $0.32^{+0.37}_{-0.24}$ & 0.03 & 0.59 \\
GW151226\_033853 & $21.68^{+8.29}_{-1.59}$ & $0.46^{+0.41}_{-0.27}$ & 0.03 & 0.75 \\
GW170104\_101158 & $49.65^{+4.74}_{-3.63}$ & $0.26^{+0.32}_{-0.19}$ & 0.01 & 0.33 \\
GW170608\_020116 & $18.55^{+2.03}_{-0.63}$ & $0.21^{+0.30}_{-0.16}$ & 0.01 & 0.28 \\
GW170729\_185629 & $84.47^{+15.03}_{-10.86}$ & $0.47^{+0.30}_{-0.37}$ & 0.09 & 0.82 \\
GW170809\_082821 & $58.52^{+5.28}_{-3.89}$ & $0.27^{+0.36}_{-0.20}$ & 0.02 & 0.36 \\
GW170814\_103043 & $55.97^{+3.45}_{-2.95}$ & $0.29^{+0.32}_{-0.22}$ & 0.05 & 0.31 \\
GW170818\_022509 & $62.52^{+5.32}_{-4.58}$ & $0.37^{+0.33}_{-0.27}$ & 0.09 & 0.45 \\
GW170823\_131358 & $67.03^{+10.34}_{-7.22}$ & $0.31^{+0.35}_{-0.23}$ & 0.05 & 0.39 \\
GW190408\_181802 & $43.37^{+4.21}_{-3.02}$ & $0.23^{+0.31}_{-0.18}$ & 0.01 & 0.26 \\
GW190412\_053044 & $36.78^{+4.69}_{-4.43}$ & $0.29^{+0.19}_{-0.17}$ & 0.00 & 0.89 \\
GW190413\_134308 & $58.05^{+10.56}_{-7.81}$ & $0.32^{+0.38}_{-0.24}$ & 0.05 & 0.44 \\
GW190421\_213856 & $81.30^{+16.87}_{-11.92}$ & $0.45^{+0.38}_{-0.34}$ & 0.07 & 0.67 \\
GW190503\_185404 & $73.67^{+13.19}_{-9.61}$ & $0.33^{+0.36}_{-0.24}$ & 0.06 & 0.41 \\
GW190512\_180714 & $69.46^{+10.06}_{-8.56}$ & $0.31^{+0.37}_{-0.24}$ & 0.03 & 0.46 \\
GW190513\_205428 & $35.80^{+4.09}_{-3.46}$ & $0.18^{+0.33}_{-0.14}$ & 0.01 & 0.36 \\
GW190517\_055101 & $54.44^{+9.32}_{-6.72}$ & $0.34^{+0.38}_{-0.26}$ & 0.03 & 0.66 \\
GW190519\_153544 & $64.15^{+9.94}_{-9.78}$ & $0.72^{+0.17}_{-0.22}$ & 0.42 & 0.99 \\
GW190521\_030229 & $105.67^{+14.38}_{-13.94}$ & $0.49^{+0.23}_{-0.26}$ & 0.09 & 0.88 \\
GW190521\_074359 & $153.21^{+42.19}_{-16.17}$ & $0.52^{+0.37}_{-0.40}$ & 0.11 & 0.75 \\
GW190527\_092055 & $76.34^{+7.00}_{-5.85}$ & $0.23^{+0.29}_{-0.17}$ & 0.01 & 0.21 \\
GW190602\_175927 & $58.10^{+18.14}_{-8.79}$ & $0.30^{+0.41}_{-0.23}$ & 0.03 & 0.49 \\
GW190620\_030421 & $115.73^{+19.29}_{-14.93}$ & $0.37^{+0.39}_{-0.27}$ & 0.06 & 0.61 \\
GW190630\_185205 & $92.74^{+18.53}_{-13.16}$ & $0.54^{+0.28}_{-0.30}$ & 0.12 & 0.86 \\
GW190701\_203306 & $59.42^{+4.73}_{-4.77}$ & $0.23^{+0.25}_{-0.15}$ & 0.00 & 0.27 \\
GW190706\_222641 & $94.33^{+11.96}_{-9.54}$ & $0.33^{+0.35}_{-0.24}$ & 0.06 & 0.43 \\
GW190707\_093326 & $112.64^{+27.36}_{-16.77}$ & $0.52^{+0.29}_{-0.35}$ & 0.09 & 0.88 \\
GW190708\_232457 & $20.06^{+1.68}_{-1.20}$ & $0.18^{+0.28}_{-0.14}$ & 0.00 & 0.21 \\
GW190720\_000836 & $31.46^{+2.85}_{-2.15}$ & $0.19^{+0.32}_{-0.14}$ & 0.01 & 0.33 \\
GW190727\_060333 & $57.36^{+38.27}_{-11.84}$ & $0.48^{+0.37}_{-0.35}$ & 0.07 & 0.77 \\
GW190728\_064510 & $21.84^{+3.77}_{-2.02}$ & $0.30^{+0.30}_{-0.16}$ & 0.01 & 0.63 \\
GW190803\_022701 & $18.31^{+7.57}_{-1.86}$ & $0.30^{+0.42}_{-0.23}$ & 0.03 & 0.56 \\
GW190828\_063405 & $68.82^{+10.18}_{-7.76}$ & $0.36^{+0.35}_{-0.26}$ & 0.08 & 0.45 \\
GW190828\_065509 & $20.71^{+4.23}_{-1.40}$ & $0.26^{+0.32}_{-0.15}$ & 0.00 & 0.43 \\
GW190910\_112807 & $70.68^{+16.33}_{-11.42}$ & $0.30^{+0.37}_{-0.23}$ & 0.05 & 0.43 \\
GW190915\_235702 & $65.06^{+11.95}_{-8.23}$ & $0.31^{+0.37}_{-0.23}$ & 0.04 & 0.41 \\
GW190924\_021846 & $76.69^{+19.50}_{-13.79}$ & $0.58^{+0.26}_{-0.38}$ & 0.25 & 0.83 \\
GW190925\_232845 & $57.23^{+7.93}_{-4.35}$ & $0.31^{+0.31}_{-0.21}$ & 0.05 & 0.31 \\
GW190929\_012149 & $34.28^{+5.13}_{-4.28}$ & $0.23^{+0.34}_{-0.17}$ & 0.00 & 0.60 \\
GW190930\_133541 & $78.04^{+9.27}_{-9.11}$ & $0.24^{+0.32}_{-0.18}$ & 0.02 & 0.23 \\
GW191105\_143521 & $57.23^{+7.07}_{-5.33}$ & $0.38^{+0.33}_{-0.28}$ & 0.07 & 0.51 \\
GW191109\_010717 & $13.93^{+2.85}_{-0.90}$ & $0.18^{+0.37}_{-0.14}$ & 0.00 & 0.35 \\
GW191127\_050227 & $36.73^{+3.56}_{-2.77}$ & $0.27^{+0.33}_{-0.20}$ & 0.02 & 0.34 \\
GW191129\_134029 & $93.40^{+22.90}_{-15.02}$ & $0.29^{+0.45}_{-0.24}$ & 0.01 & 0.72 \\
GW191204\_171526 & $21.16^{+5.91}_{-1.99}$ & $0.33^{+0.34}_{-0.21}$ & 0.01 & 0.67 \\
GW191215\_223052 & $19.97^{+3.71}_{-1.76}$ & $0.36^{+0.31}_{-0.18}$ & 0.02 & 0.57 \\
GW191216\_213338 & $18.48^{+2.08}_{-1.27}$ & $0.19^{+0.33}_{-0.15}$ & 0.01 & 0.23 \\
GW191222\_033537 & $111.56^{+20.25}_{-15.55}$ & $0.60^{+0.23}_{-0.37}$ & 0.32 & 0.82 \\
GW191230\_180458 & $79.55^{+38.69}_{-22.02}$ & $0.52^{+0.38}_{-0.40}$ & 0.08 & 0.82 \\
GW200112\_155838 & $17.55^{+2.41}_{-1.16}$ & $0.20^{+0.26}_{-0.14}$ & 0.00 & 0.34 \\
GW200128\_022011 & $20.22^{+1.70}_{-0.96}$ & $0.29^{+0.28}_{-0.13}$ & 0.01 & 0.43 \\
GW200129\_065458 & $43.36^{+5.33}_{-4.29}$ & $0.34^{+0.36}_{-0.26}$ & 0.04 & 0.45 \\
GW200202\_154313 & $19.81^{+2.69}_{-0.94}$ & $0.20^{+0.24}_{-0.11}$ & 0.00 & 0.32 \\
GW200208\_130117 & $79.09^{+16.04}_{-10.55}$ & $0.28^{+0.33}_{-0.21}$ & 0.03 & 0.31 \\
GW200209\_085452 & $85.73^{+18.63}_{-11.96}$ & $0.37^{+0.38}_{-0.28}$ & 0.10 & 0.50 \\
GW200219\_094415 & $63.88^{+5.74}_{-4.60}$ & $0.24^{+0.29}_{-0.18}$ & 0.01 & 0.22 \\
GW200224\_222234 & $74.81^{+17.07}_{-12.21}$ & $0.41^{+0.34}_{-0.29}$ & 0.13 & 0.53 \\
GW200225\_060421 & $63.39^{+4.35}_{-3.57}$ & $0.32^{+0.49}_{-0.21}$ & 0.08 & 0.38 \\
GW200302\_015811 & $17.58^{+1.78}_{-0.67}$ & $0.18^{+0.29}_{-0.14}$ & 0.01 & 0.23 \\
GW200311\_115853 & $65.39^{+7.81}_{-6.84}$ & $0.28^{+0.34}_{-0.21}$ & 0.03 & 0.38 \\
GW200316\_215756 & $62.65^{+13.89}_{-9.41}$ & $0.39^{+0.37}_{-0.28}$ & 0.11 & 0.51 \\
GW190413\_052954 & $81.43^{+20.20}_{-14.00}$ & $0.39^{+0.40}_{-0.29}$ & 0.06 & 0.66 \\
GW190719\_215514 & $65.01^{+12.58}_{-8.22}$ & $0.35^{+0.35}_{-0.26}$ & 0.07 & 0.46 \\
GW190725\_174728 & $72.21^{+7.17}_{-5.10}$ & $0.31^{+0.31}_{-0.21}$ & 0.04 & 0.32 \\
GW190731\_140936 & $33.48^{+3.63}_{-2.98}$ & $0.41^{+0.30}_{-0.29}$ & 0.06 & 0.58 \\
GW190805\_211137 & $57.82^{+9.56}_{-6.85}$ & $0.30^{+0.39}_{-0.23}$ & 0.02 & 0.60 \\
GW191103\_012549 & $61.90^{+5.30}_{-4.24}$ & $0.28^{+0.34}_{-0.21}$ & 0.03 & 0.29 \\
GW200216\_220804 & $21.16^{+7.19}_{-1.95}$ & $0.27^{+0.30}_{-0.17}$ & 0.00 & 0.48 \\

\end{longtable}

\end{document}